\documentclass[preprint,12pt]{aastex}
 
\newcommand{\HI}{\ion{H}{1}}
\newcommand{\kms}{km~s$^{-1}$}

\slugcomment{Accepted to ApJ}

\shorttitle{Diffuse Neutral Intragroup Medium in HCGs}
\shortauthors{Borthakur, Yun, \& Verdes-Montenegro}

\begin{document}

\title{Detection of Diffuse Neutral Intragroup Medium in Hickson Compact Groups}

\author{Sanchayeeta Borthakur, Min Su Yun}
\affil{Astronomy Department, University of Massachusetts, Amherst, MA 01003, USA}
\email{sanch@astro.umass.edu, myun@astro.umass.edu}

\author{Lourdes Verdes-Montenegro}
\affil{Instituto de Astrof\'{\i}sica de
Andaluc\'{\i}a, CSIC, Apdo. Correos 3004, E-18080 Granada, Spain}
\email{lourdes@iaa.es}

\begin{abstract}
We present new Green Bank Telescope (GBT) 21 cm neutral hydrogen (\HI) observations of a complete distance limited sample of 22 Hickson Compact Groups (HCGs) with at least four true members. 
We detected an average HI mass of 8~$\times$~10$^{9}$~M$_{\odot}$ (median~=~6~$\times$~10$^{9}$~M$_{\odot}$), which is significantly larger than previous single-dish measurements. Consequently, the HI-deficiencies for these HCGs have been reduced, although not completely eliminated. Spectral comparison of the GBT data with complementary Very Large Array (VLA) data shows significant \HI\ excess in the GBT spectra. 
The observed excess is primarily due to the high surface brightness sensitivity of the GBT detecting diffuse, low column density HI in these groups.
The excess gas forms a faint diffused neutral medium which is an intermediate stage in the evolution of high-surface brightness \HI\ tidal debris in the intra-group medium (IGM) before it is fully ionized.
The excess gas mass fraction, (M(HI)$_{GBT}$~-~M(HI)$_{VLA}$)/M(HI)$_{GBT}$, for our complete sample varies from 5\% to 81\% with an average of 36\% (median~=~30\%). The excess gas mass fraction is highest in slightly \HI\ deficient groups where the tidal debris has had enough time to evolve. We also find the excess gas content increases with the evolutionary phase of the group described in \citet{VM2001}.
 Theoretical calculations indicate that an \HI\ cloud of radius $\ge$~200~pc would survive in an IGM of 2~$\times$~10$^6$~K for more than the typical dynamical lifetime of a group. However, smaller clouds get evaporated and assimilated into the hot IGM in a much shorter timescale.

\end{abstract}


\section{INTRODUCTION}

Most galaxies in the universe are not isolated; instead they are in groups or clusters of galaxies \citep{tully87}. Galaxy groups are an environment where galaxies evolve continuously as they interact with each other and the group potential. Groups also contain a significant fraction of their baryonic content outside the individual galaxies  \citep{fukugita98,fukugita06}. The detection of  a substantial intragroup medium (IGM) in X-ray \citep[see ][]{mulchaey00} provides a conclusive evidence of baryons associated with the group potential. There is also growing evidence that significant changes in galaxy morphology and truncation of star formation take place in groups and cluster outskirts, rather than in dense cluster cores \citep[][and references therein]{dressler80,lewis02,helsdon03,capak07}. Therefore, a full accounting of the gaseous components in galaxy groups and their influence on the evolution of member galaxies should offer an important insight into the evolution of galaxies in general.

Hickson Compact Groups (HCGs) are unique laboratories for studying the effects of multi-galaxy tidal interactions on the morphological and dynamical evolution of galaxies as well as the IGM. HCGs are groups of galaxies showing extreme projected densities and were identified using the following set selection criteria \citep{hick82}: (1) four or more galaxies within 3 magnitudes of the brightest; (2) no other galaxy in the magnitude range of the group members between the group radius\footnote{$R_G$ is the group radius defined by the smallest circle containing the centers of the group members.} $R_G$ and $3R_G$; and (3) mean surface brightness of the group, $\bar{\mu}_G \leq$ 26.0 mag arcsec$^{-2}$ in the red (E) plate of the Palomar Observatory Sky Survey (POSS). A total of 100 compact groups with projected galaxy density ranging from 300 to $10^8$ $h^{-2} Mpc^{-2}$ were cataloged by \citet{hick82}. The projected galaxy densities in these groups are among the highest in the local universe.  Despite the theoretical arguments that these groups cannot last long enough to be real \citep{hern95}, subsequent spectroscopic surveys have shown that the majority are physical groups with low velocity dispersions of $\sim$250 \kms\ \citep{hick92}.   

Morphological analysis has shown that compact groups contain a significantly smaller fraction of late-type (spiral and irregular) galaxies as compared to that seen in the field \citep{roodwill89,kind90}. Moreover, spiral galaxies associated with these groups tend to be deficient\footnote{This is referred to as ``\HI-deficiency'' and is defined as $Def_{HI}\equiv$ log[M(HI)$_{pred}$] $-$ log[M(HI)$_{obs}$] \citep[see ][]{hay84}.} in neutral atomic hydrogen (\HI), containing on an average only 40\% of the expected \HI\ based on their optical morphology and luminosity types \citep[][and references therein]{willrood87,willvan95,huch97, VM2001, stevens04}. These galaxies are also deficient in CO emission indicating low molecular gas content \citep{VM98}. Tidal gas stripping, shock heating, and gas exhaustion by tidally induced star formation are some of the possible explanations for the observed deficiencies. \HI\ deficient groups tend to be brighter in X-rays than the non-deficient groups \citep[][ VM2001 hereafter]{VM2001}. This suggests that the hot IGM facilitates the ionization of tidally stripped cold interstellar medium (ISM). However, the IGM is not always detected in the X-ray, especially in spiral dominated groups, possibly because of a smaller halo mass thus resulting in a lower virial temperature \citep{mul96b,rasmussen08}. For instance, \citet{aracil} detected cooler neutral gas ($<$~10$^5$~K) in the form of a cluster of 13 \HI\ Lyman~$\alpha$ absorption lines spread over 1000~\kms in the IGM of a galaxy groups at z = 0.0635 towards quasar HS~0624+6907.

Atomic gas studies using the Very Large Array (VLA) of HCGs \citep{VM2001,willvan95,VM2005, VM2002, del03, durb08, willyun02} provide insights on the fate of cold gas in HCGs, including ubiquitous tidal features and off-disk gas. Tidal interactions are important in all compact groups, and they produce complex, continuously changing gaseous and stellar structures. Frequent, multiple tidal encounters within the group potential may rapidly disperse \HI\ tidal debris to a diffuse, extended medium. Cold ISM content dictates the current and future star formation activity in galaxies. Examining the Spitzer infrared imaging data, \citet{johnson07} found that the most gas-rich groups also tend to be the most actively star forming. They also found evidence of rapid evolution of galaxy properties in response to the dynamical effects and the level of these activities directly impacts the morphological and spectral evolution of the galaxies. Diffuse emission features with broad ($\ge600$~kms$^{-1}$) line-widths may escape detection when observed with a relatively limited bandwidth and surface brightness sensitivity.
The main aim of this work is to explore whether there exists significant diffuse \HI\ emission previously undetected by the VLA and to determine their quantity, location, and nature. 
 
We have obtained high quality \HI\  spectra for a complete distance limited sample of 22 HCGs along with 4 additional sources using National Radio Astronomy Obsrevatory's (NRAO)\footnote{The National Radio Astronomy Observatory is a facility of the National Science Foundation operated under cooperative agreement by Associated Universities, Inc.} Robert C. Byrd Green Bank Telescope (GBT). The excellent sensitivity and spectral resolution of the GBT provides the opportunity to study the physical state and the fate of missing cold gas in HCGs. The work presented here is part of a broader study of evolution of galaxies in compact group environment through the study of morphology and tidal signatures traced in the 21cm \HI\ line.

A description of our sample along with the details of our GBT observations are presented in \S~2 and the results are presented in \S~3.  We discuss our findings and use comparative studies of the new GBT measurements and the previous VLA measurements to investigate the nature and the distribution of the neutral gas in HCGs in \S~4. Finally, a conclusion is presented in \S~5, summarizing the results and inferences drawn from this work. For this study, we have adopted the value of the Hubble constant, H$_o$ = 75 km s$^{-1}$ Mpc$^{-1}$.

\section{OBSERVATIONS \label{sec:observations}}

\subsection{Sample \label{sec:sample}}
Identified by visual inspection of the Palomar Sky Survey plates and without any redshift information, the original Hickson catalog is fraught with projection effects and selection biases.  Further examination of the catalog incorporating the redshift information has shown that about 1/3 of the groups are false groups, either in projection or part of a more massive group or a cluster \citep[see ][]{hick92,sulentic97}.  Understanding that HCGs are potentially diverse and heterogeneous entities, a new coherent sample of true groups from the Hickson catalog has been constructed for a meaningful statistical analysis.  The criteria we used to define the new complete sample are:
\begin{itemize}
\item having four or more true member galaxies meeting the original Hickson definition, excluding false groups by utilizing redshift information;
\item containing at least one spiral galaxy, to determine \HI\ deficiency meaningfully; and
\item located within a distance of 100 megaparsec (Mpc) so that they can be studied with excellent sensitivity.
\end{itemize} 

These criteria yielded 22 groups which we will refer hereafter as the complete sample: HCG~7, 10, 15, 16, 23, 25, 30, 31, 37, 40, 44, 58, 67, 68, 79, 88, 90, 91, 92, 93, 97, and 100. We have also included four more groups that do not fit the sample criterion but had previous single dish and/or VLA data - HCG~18, 26, 35, and 48 (referred to as the ``additional sources" hereafter). HCG~18 and 48 have less than four group members while HCG~26 and 35 are further than 100~Mpc. Therefore, these four groups will not be included in our statistical studies presented in \S~\ref{sec:trends}. The redshift, size and morphological information of each of these groups and their members are provided in Table~\ref{tbl-Sample}.

Whether HCGs are real groups or chance projections along radially oriented filaments in the large-scale structure \citep{hern95} is a highly debated topic. An independent identification of galaxy groups was performed by \citet{yang07} using the halo occupancy model on the galaxy catalogs from the Sloan Digital Sky Survey (SDSS) and the Two Degree Field (2dF) galaxy redshift survey. Six of our groups, HCG~7, 16, 25, 35, 55, and 88, fall within the sky and redshift coverage of these surveys and all of them have been identified as galaxy groups by \citeauthor{yang07} The SDSS and 2dF surveys are deeper than that used by \citet{hick82} and hence the \citeauthor{yang07} catalog includes additional fainter galaxies associated with these groups. They derived halo masses for these groups, which range between (2 - 115)~$\times$~10$^{12}$~M$_\odot$. The independent identification of all 6 HCGs within the SDSS and 2dF sky coverage lends further support that our sample refinement described above successfully identified real physical groups.

\subsection{GBT \HI\ and OH Observations}

We conducted GBT observations of the 21~cm \HI\ and 18~cm OH transitions for 26 HCGs in May and October 2005 as part of the GBT05A-051 and GBT05C-005 programs.  The data on HCG~67 and HCG~79 were severely corrupted by solar interference, and new observations were obtained in August 2006 as part of the GBT06B-053 program.  We used the dual polarization L-band system that covers the frequency range between 1.15 and 1.73 Gigahertz (GHz). Using 9-level sampling and two IF settings, each with 12.5 MHz total bandwidth, the spectrometer gives 8192 channels at 1.5~kHz (0.3~km s$^{-1}$) resolution, covering a total velocity range of 2500~km s$^{-1}$. We employed the standard position-switching scheme by cycling through the ON-OFF sequence, dwelling for 300 seconds at each position. A position offset of +20\arcmin\ in Right Ascension was adopted for the OFF position so that the presence of any confusing sources in the OFF position could be tracked. We also performed local pointing corrections (LPC) using the observing procedure AutoPeak, which was then automatically applied to the data. Hence, the error in pointing should be less than GBT's global pointing accuracy of 2.7~$^{\prime\prime}$, which is less that 0.5\% of GBT beam size at 1.4~GHz. 

The clean optics and active surface system of the GBT also results in a well-calibrated structure and a stable gain at the 21cm wavelength. The primary flux calibrators 3C~48 (16.5 Jy), 3C~147 (22.5 Jy), and 3C~286 (15.0 Jy) were observed regularly to monitor the instrumental performance, and the antenna gain of 1.65$\pm$0.05 K/Jy is derived from these data. Our regular observation of bright calibration sources has verified the stability of the telescope gain factor. The rms noise after 60 minutes of on-source integration was 1.4 mK when the spectra were smoothed to 50 kHz ($\sim$10 km s$^{-1}$) resolution. The corresponding 5$\sigma$ \HI\ mass detection limit at a distance of 100 Mpc is 1.65$\times 10^8 \frac{\Delta V }{100~kms^{-1}}$~M$_\odot$ where $\Delta V $ is the line-width of the \HI\ emission. 

The GBT data also has very good stability within the bandwidth and the baselines could be fitted using first or second order polynomial for most of the HCGs. All the line-free channels with the exception of 200 channels at each end were used to fit baselines. Since most of the groups were observed near transit, we estimate an attenuation due to opacity of the air to be around 1\% and corrected the GBT spectra accordingly.

The second IF was tuned to search for the 1665~MHz and 1667~MHz OH maser transitions simultaneously with the \HI\ transition.  Dense molecular gas associated with nuclear starbursts or circum-nuclear disks has been detected in OH maser emission in nearby and distant galaxies \citep[see review by ][]{lo05}. Excited primarily through IR pumping, these transitions trace the presence of a substantial quantity of warm and dense molecular material (T = 40-200 K, $n=10^{4-7}$ cm$^{-3}$) heated by an intense radiation field. No OH maser was detected in any of the groups. This result is consistent with the absence of infrared bright starburst systems in these groups \citep{VM98}. Thus, we concentrate on the analysis of the \HI\ emission in the remainder of the paper.


\section{RESULTS \label{sec:results}} 
\subsection {GBT Data \label{sec:GBTdata}}

Among the 26 HCGs observed using the GBT, \HI\ emission was detected in all but one group. The \HI\ masses for the groups are calculated using the relationship M(HI) = 2.36$\times10^5D^2(S\Delta V)~M_\odot$ where $D$ is the luminosity distance to the group in megaparsec (Mpc) and the $S\Delta V$ is the velocity integrated \HI\ flux density in Jy~\kms\ . The derived \HI\ masses within the GBT beam ($\sim$9.1$^{\prime}$) range between $0.5-28.5~\times10^9$~M$_\odot$  (see Table~\ref{tbl-GBT} for individual values including \HI\ mass and linewidth). The velocity range over which the \HI\ line flux is integrated is also presented in Table~\ref{tbl-GBT}. The only undetected group, HCG~35, is the highest redshift group and is not a member of our distance-limited complete sample. The sensitivity of the GBT makes the $3\sigma$ upper limit for this high redshift group ($cz=16249$~\kms) $M(HI) \le 28 \times 10^{9}~M_\odot$ assuming a line width of $\Delta~V = 600$~\kms (average line-width for our sample).  We adopt a systematic uncertainty in calibration and baseline subtraction of 10\%, which is substantially larger than the measured noise level (around 1\%) in most cases. For HCG~40, which was observed under the pilot program GBT05A-051, the antenna gain was not measured. We adopt the same gain factor from our observation under the program GBT05C-005. The reduced and calibrated spectra are shown in Figure~\ref{spectra}.

Visual comparison of our GBT spectra with previous single dish measurements by \citet{willrood87} and \citet{huch97} confirm our calibration accuracy. A detailed comparison is not discussed here. The large aperture, clean optics, and the low noise receiver of the GBT lead to superior calibration, sensitivity, and baseline stability compared with the earlier single dish measurements. This makes the GBT an excellent instrument for detecting the diffuse neutral component of the IGM, which may be very faint, extended, or  broad in velocity ($\Delta V=1000$ \kms). Also, the noise level achieved in our observations after an hour of integration is about 5 times lower than that of \citet{huch97}. The extended nature of the \HI\ emission in these groups also makes a meaningful comparison with other single dish measurements with different beam sizes difficult.  Instead, we focus our comparative analysis on the \HI\ data obtained using the GBT and the VLA.


\subsection{Re-evaluation of HI-deficiency for the Groups \label{sec:HIdef}}

HI deficiency is a measure of the deviation of the observed \HI\ mass from the predicted \HI\ mass for a galaxy based on its optical luminosity and morphology \citep[see][Table 5, page 791]{hay84}. It is defined as  $Def_{HI}\equiv$ log[M(HI)$_{predicted}$] $-$ log[M(HI)$_{observed}$]. The predicted \HI\ mass for a galaxy group is estimated as the sum of the predicted \HI\ masses for individual galaxies (VM2001). 

 The values of predicted \HI\ masses used in our analysis are taken from VM2001, which used optical luminosities and morphologies from \citet{hick89}. HCG~90 was not included in the VM2001 sample and thus we estimate the predicted \HI\ mass for this group following the same procedure as VM2001.  In most cases the GBT beam covered all the galaxies in a group, with the exception of HCG 10, 44, 68, \& 93 where we correct the predicted \HI\ mass to account for the GBT beam coverage. We assume that the predicted \HI\ mass in individual galaxies is distributed uniformly inside the optical extent of the galaxy and hence the correction in the predicted mass translates to the fraction of the optical galaxy covered by the GBT beam. The observed \HI\ masses, the predicted \HI\ masses within the GBT beam and their corresponding \HI\ deficiencies are provided in Table~\ref{tbl-GBT}. 

\begin{figure}
\figurenum{2}

\includegraphics[scale=0.65,angle=-90]{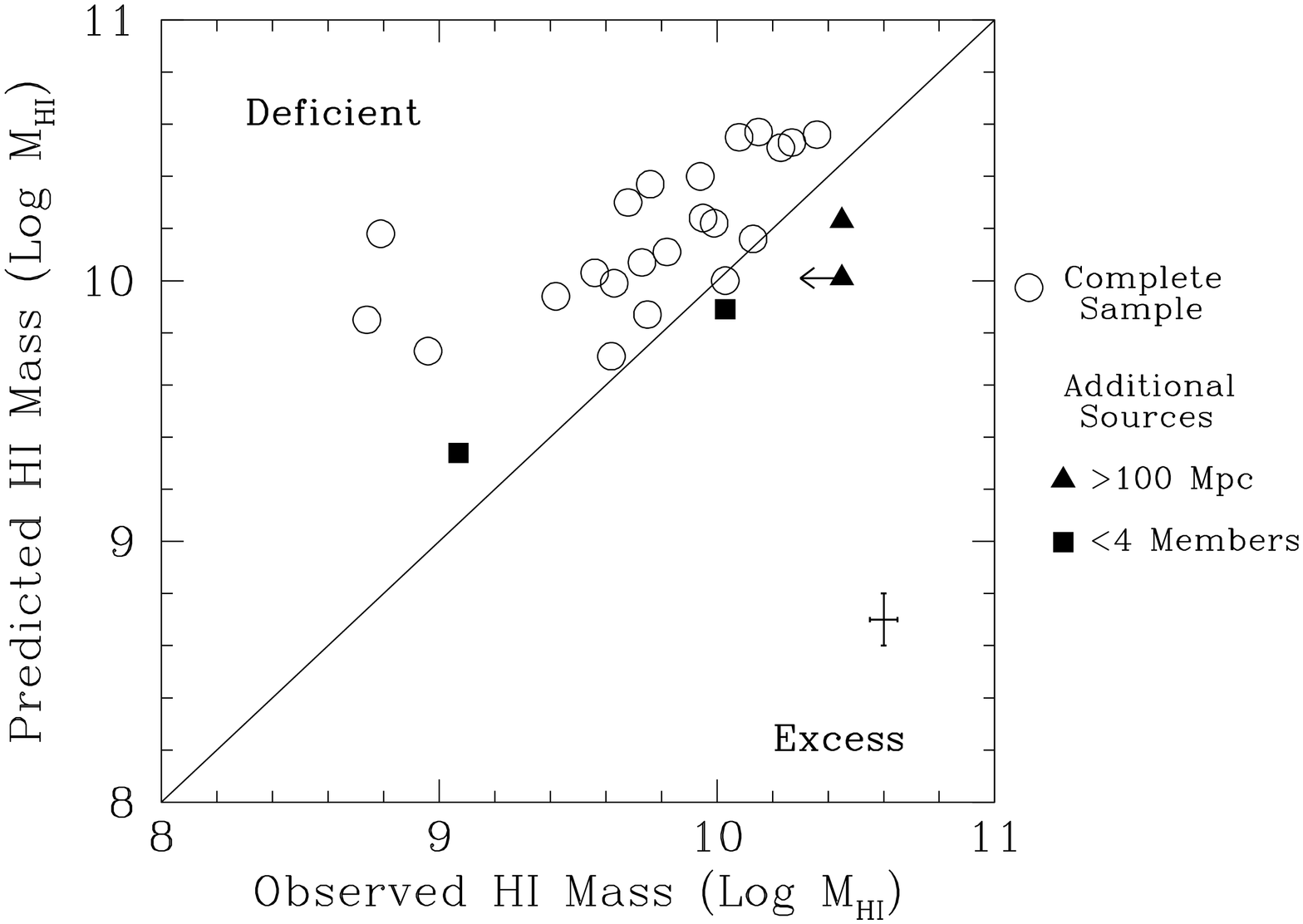}

{\caption{The \HI\ mass observed using the GBT is compared with the predicted \HI\ mass for each of the 22 HCGs from our complete sample as well as the 4 additional sources. The diagonal line drawn indicates \HI\ normalcy.  A clear majority of the groups appear on the \HI\ deficient side. Typical uncertainties in both masses are shown using error bars on the bottom right corner. The additional sources are \HI\ normal. No \HI\ was detected in HCG~35 and consequently an uppper limit is presented. 
 \label{mass_obs_pred} }}
\end{figure}

A comparison of the predicted \HI\ mass and the observed \HI\ mass for our entire sample is shown in Figure~\ref{mass_obs_pred}. We find that most of the groups are \HI\ deficient. While the predicted \HI\ masses for most of the groups are similar, ranging from 9.34 to 10.57 dex;  the observed \HI\ masses vary over two orders of magnitudes due to the range in \HI\ deficiency. The highest redshift group HCG~35 (part of our additional sources) can be seen with an arrow indicating the observed mass as a 3~$sigma$ upper limit. \HI\ was detected in the other three groups part of our additional sources and all of them lie in the  \HI\ normal region of the plot.

The \HI\ content and spatial distribution in the groups is dependent on the evolutionary history of individual galaxies. Any transformation of neutral hydrogen as a result of tidal interaction within the group environment is expected to have some correlation with the group's \HI\ deficiency. Therefore, we have divided the HCGs from our complete sample into three classes based on their predicted \HI\ mass and the GBT observed \HI\ mass:
\begin{itemize}
\item normal \HI\ content: observed mass more than 2/3 of the predicted value (HCG 23, 25, 79, \& 68);
\item slightly \HI\ deficient: observed mass ranges between 2/3 to 1/3 of the predicted value (HCG 10, 15, 16, 31, 37, 40, 58, 88, 91, 92, 97, \& 100); and
\item highly \HI\ deficient: observed mass less than 1/3 of the predicted value (HCG~7, 30, 44, 67, 90, \& 93).
\end{itemize} 
These distinctions are based on the observed scatter in the \HI\ content for each Hubble type \citep[$\sim$0.2 in dex; see ][]{hay84}. This classification enables us to examine quantitatively the hypothesis that the \HI\ content of each group indicates their age or evolutionary stage.

\begin{figure}
\figurenum{3}

\includegraphics[angle=-90,scale=0.65]{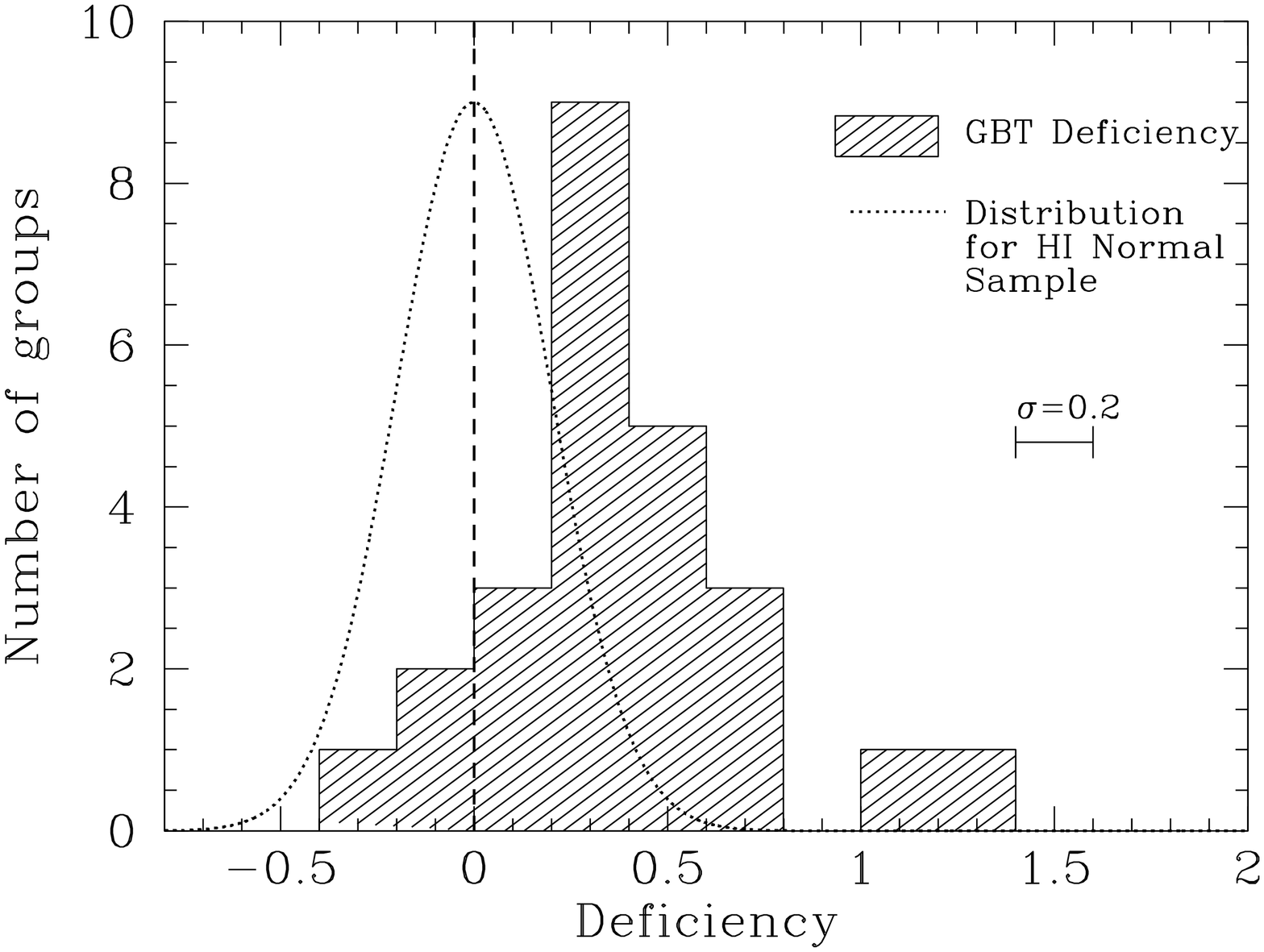}\\
{\caption{ Histogram showing HI-deficiency evaluated using our GBT measurements. The dashed line represents zero deficiency and the dotted line represents the expected distribution for an \HI\ normal sample (see \S~\ref{sec:HIdef}). Typical uncertainty in HI-deficiency is estimated to be about 0.2 in the log scale. HCG~35 is not included in this plot since no \HI\ was detected and hence reliable \HI\ deficiency value could not be estimated.
\label{plot_his}}}
\end{figure}

The \HI\ deficiencies estimated by VM2001 using previous single dish measurements are also presented in Table~\ref{tbl-GBT} for comparison. For 80\% of the groups, the GBT detected more \HI\ than previous single dish observations using NRAO's 91m telescope even though the beam size for these observations was 10.8$^{\prime}$  \citep[][VM2001]{willrood87}. For three slightly deficient groups and two highly deficient groups the previous  single dish values are larger than that measured using the GBT.  The most likely source of this difference is the extended nature of the \HI\ distribution which was captured better by the 10.8$^{\prime}$ beam as opposed to the  9.1$^{\prime}$ GBT beam. The histogram of \HI\ deficiency evaluated from GBT data is shown in Figure~\ref{plot_his}. The peak of the distribution lies in the deficiency range of 0.2 to 0.4. The dotted curve represents the expected distribution of \HI\ deficiency for a \HI\ normal sample. This  shows that the uncertainty in deficiency estimation cannot account for the observed distribution and therefore, confirms the existence of \HI\ deficiency in HCGs.

Even though these highly sensitive measurements decrease the \HI\ deficiency of HCGs considerably, however, it does not elevate the total \HI\ content of these groups to the normal level. The addition of hot IGM traced in X-ray may in some cases. \citet{rasmussen08} analyzed new and archival X-ray data from Chandra and XMM-Newton telescopes for eight HCGs, and their X-ray results add a few times $10^{10}$ to 10$^{11}M_\odot$ of hot IGM gas to the total gas content in some of the X-ray bright groups. This result suggests a direction for solving the \HI\ deficiency problem by accounting for colder tidally stripped material that might have been ionized in the IGM. However, it is crucial to account for all sources of hot gas in galaxy groups like accretion of gas from large-scale structures by the group potential, feedback from constitute galaxies as well as the deposition of cold ISM in the IGM as a result of tidal activities. It is also clear that the tidal debris is not the major contributor to the hot IGM gas mass as the missing \HI\ mass (M( \HI)$_{pred}$ - M( \HI)$_{obs}$) for X-ray bright groups like HCG~15, 37, 40, \& 97 is on an average an order of magnitude smaller that the hot gas mass \citep{rasmussen08}.

\subsection {Comparison with the VLA Data \label{sec:comparison}}

 It is a common practice to compare the flux recovered by an interferometer with a single dish measurement in order to characterize the limitations of the resulting aperture synthesis images.
For instance, an interferometer such as the VLA is a spatial filter and is thus insensitive to diffuse, extended structures $\ge15^{\prime}$ in the D-array\footnote{For a uniformly weighted, untapered map produced from a full 12 hour synthesis observation of a source which passes near the zenith, the largest angular scale that could be detected by the VLA in D-array is 15$^{\prime}$. However, since most of the observations are shorter than 12 hours and used natural weight the maximum angular scale should be smaller than 15$^{\prime}$} . Although the collecting area of the GBT and the VLA are comparable, the VLA  has a much lower surface brightness sensitivity than the GBT. In addition, the GBT has a larger instantaneous bandwidth ($\Delta~V~\sim~2500$~\kms) that allows identification of emission features with a much broader ($\ga~500$~\kms) line-width. In comparison, the bandwidth of the VLA is less than half of the GBT, and broad spectral features can be easily missed by the VLA due to limited spectral coverage.

The VLA data used for comparison come primarily from our own imaging survey of HCGs \citep{VM2001,willyun02,VM2005,bor,Yunprep}.  Additional comparison data for six groups are constructed by our own reduction of the archival data that were published previously \citep{willvan95,will98,barnes01,heid05}. Table~\ref{tbl-VLA} summarizes the details of the VLA data used for comparison. The flux density sensitivity of the comparison VLA data is similar in most cases ($\sim$ 0.5~mJy/beam), but the angular and spectral resolutions vary slightly.

Comparisons of the new GBT spectra with the spectra derived from the VLA observations for our complete sample as well as for two of the additional sources are shown in Figure~\ref{comparisons}. We define the  difference between the VLA and the GBT spectra as the excess \HI\ mass 
 \begin{equation}{\label{eq-excess}}
 M_{excess} = M_{GBT} - M_{VLA}
\end{equation}
We make a quantitative characterization of the excess gas as a fraction of the total \HI\ in a group and define it as the excess gas mass fraction, $f$
\begin{equation}{\label{eq-f}}
f=\frac{M_{GBT}-M_{VLA}}{M_{GBT}} \equiv  \frac{M_{excess}}{M_{GBT}} 
\end{equation}
 In groups where the excess \HI\ mass dominates, the value of this ratio is greater than 50\%. Similarly, in groups where this ratio is smaller than 50\%, the primarily \HI\ component is the high surface brightness \HI\ detected by the VLA.

\begin{figure}
\figurenum{5}
\includegraphics[angle=-90,scale=0.6]{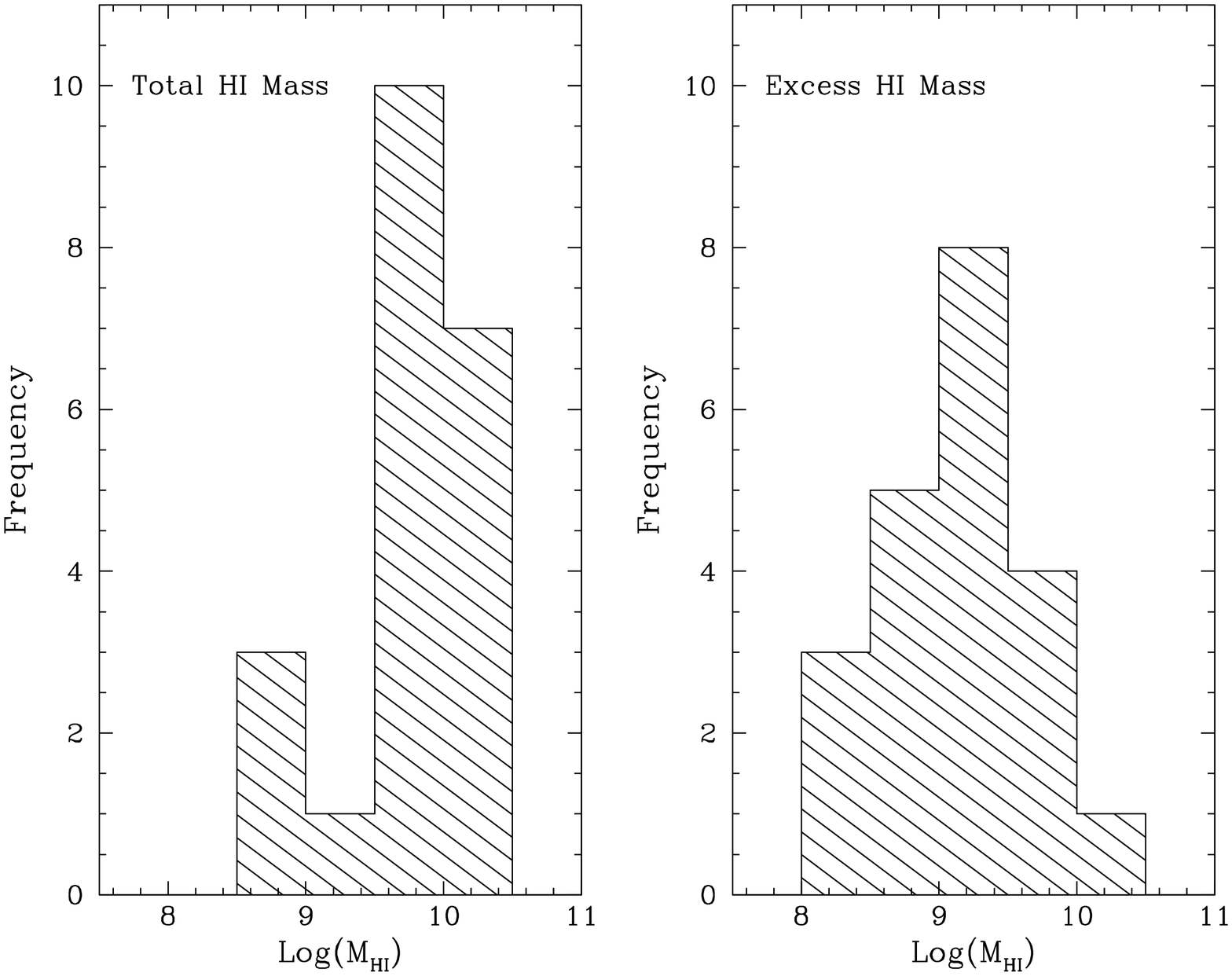}\\
{\caption{Histogram showing the distribution of total HI mass detected by the GBT (left) and excess HI mass (right) for our complete sample with the exception of HCG~40. On an average the excess gas is more than a third of the total \HI\ detected by the GBT, although individual values vary from 5\% to 81\% within the sample.  \label{mass_his}}}
\end{figure}

Figure~\ref{mass_his} is a histogram of the total \HI\ mass and the excess \HI\ mass in our sample. The excess gas mass fraction,$f$ for our complete sample varies between 5\% and 81\% with an average value of $36\%$ (Table~\ref{tbl-filling_factor}). In order to ensure that this comparison is fair, we multiply the original VLA map with the GBT beam pattern and then summing the \HI\ flux in each channel to construct the corresponding VLA spectra. Therefore, the VLA spectra used in the comparison are for regions identical to those observed by the GBT. Furthermore, the quantitative comparison of the \HI\ mass is limited to the velocity range that is common to both spectra although the GBT detected much broader emission features in some cases. Consequently, the values mentioned above are lower limits of the \HI\ missed by the VLA. The GBT data is also much more sensitive than the VLA data and this easily noticeable  in HCGs with faint \HI\ emission such as HCG~15, 37, 90, and 97.

Extra steps were taken to ensure consistency in the calibration as reliable cross-calibration is critical for quantitative comparisons. To ensure excellent calibration with respect to the VLA, we observed the same three VLA primary flux calibrators 3C~48, 3C~147, and 3C~286 throughout the observing run.  By taking these extra calibration steps, we estimate that the calibration between the GBT and the VLA spectra should agree within 10\%. The excellent agreement between the GBT and the VLA spectra for groups dominated by high surface brightness (HSB) features, such as HCG~7, 10, 16, and 67, nicely demonstrates the success of our cross-calibration. Even in groups where there is an overall difference in total flux such as  HCG~26 and 79, parts of the spectra match well within the statistical errors. The fact that the total flux measured by the GBT {\it always} agrees or exceeds the VLA measurement (except in the case of HCG~40 where we have a large calibration uncertainty) but is never smaller rules out the possibility of relative calibration accuracy varying wildly. Therefore, calibration errors are not responsible for the dramatic differences seen in HCG~31,~44, and 100, and the excess emission in the GBT spectra is real. The excess emission is also correlated with the group properties (see \S~\ref{sec:trends}), and this is a strong indication that the measured excesses are real. One may plausibly suspect a continuum subtraction error for missing broad spectral features in the VLA data. However, the absence of negative flux values in the VLA images, which is expected from overdoing continuum subtraction, rules out significant errors in VLA data reduction.

Three broad categories on the spectral distribution of the excess gas have emerged from our comparison of the GBT and the VLA spectra: (I) identical line widths but with a large line flux difference; (II) broad emission wings (50-200 \kms) associated with VLA detected spectral features; and (III) broad ($>400$ \kms) emission features outside the VLA \HI\ line-width. The most commonly seen is the category I, and nearly all groups show this trait to a degree.  The identical line widths is a strong indication that the additional emission detected by the GBT is associated with the structure traced by the VLA.  It is possible that additional, independent \HI\ systems sharing the same velocity range are present, and a few distinct features with a narrow line-width were seen in the GBT spectra. 
 
\begin{figure}
\figurenum{6}
\includegraphics[angle=-0,scale=0.6]{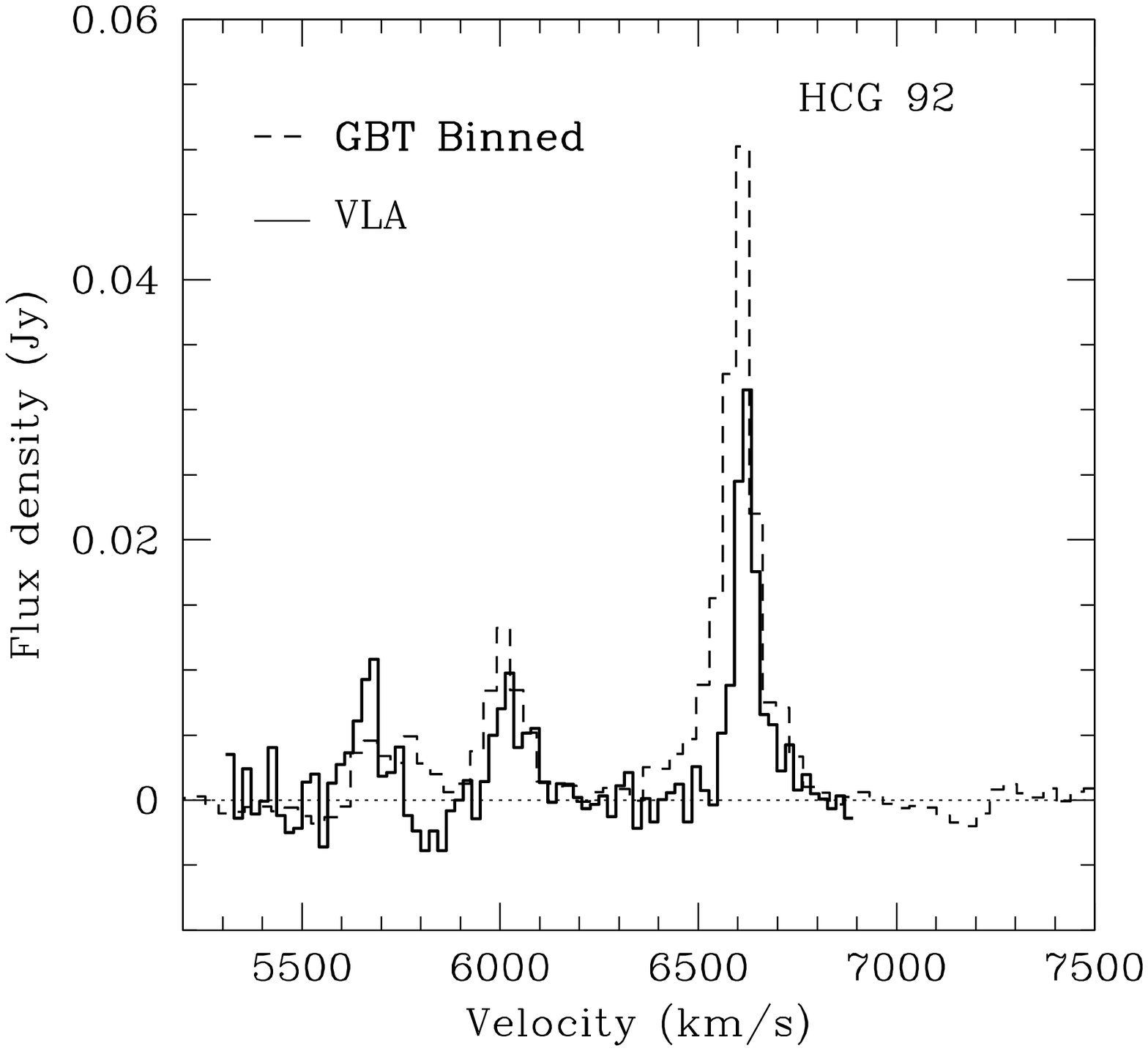}\\ 
{\caption{ An overlay of the VLA spectrum (solid black) and the GBT spectrum in (black dash) obtained by binning GBT data to VLA spectral resolution for HCG~92. The high velocity wings associated with the peak at 6620~\kms extends from 6200 to 6550~\kms.  \label{h92_binned}}}
\end{figure}

The second category with broad emission wings is also relatively common.  HCG~92 (``Stefan's Quintet'') and HCG~100 represent two outstanding examples of this type, characterized by broad \HI\ emission wings in excess over the velocity range detected by the VLA. The wing features in HCG~92 cover the velocity range 6300 to 6530 \kms and 5750 to 5900~\kms where as in HCG~100 the wing feature can be seen in the velocity range from 5450 to 5650 \kms. The excess can be seen in the GBT data even when it is binned to match the resolution of the VLA spectra (see Figure~\ref{h92_binned}). A collision involving an intruder at a velocity in excess of 1000 \kms\ is currently occurring in HCG~92 \citep[see ][ and references therein]{appleton06}, and the broad \HI\ wings are plausibly associated with high velocity shocks in this case. The VLA imaging of HCG~100 shows a cloud of \HI\ surrounding most of the member galaxies and extending far beyond the group diameter in the VLA image \citep{bor}.  The VLA spectrum is clearly missing the broad wings on both sides of the main \HI\ peak.  Other groups showing similar broad emission wings are HCG~ 31, 44, and 79.

The third category is the broad, faint emission features unseen in the VLA spectra.  In HCG~15, we detected 56\% of total \HI\ in this system as excess \HI\  emission spread over a velocity range of more than 1000~\kms.  The VLA channel maps of HCG~15 show \HI\ emission associated with a single galaxy \citep{bor} spread over no more than 250~\kms\ .  The VLA not only missed the broad \HI\ emission because of a smaller bandwidth, but also could not detect fainter emission due to the limitation in its surface brightness sensitivity.  The VLA observations of HCG~30 and 37 have failed to detect {\it any} high surface brightness \HI\ emission \citep{Yunprep} while their GBT spectra show faint, broad emission covering the entire redshift range of the member galaxies (see Figure~\ref{comparisons}). These observations suggest that the {\it entire} \HI\ emission in these groups is in a diffuse component.  In addition, HCG~44, 58, 79, 90, and 97 also show evidence of faint, broad emission not detected by the VLA. 

 HCG 31 is one of the most interesting cases where the comparison of \HI\ spectra shows an enormous difference between the GBT and the VLA, confirming the presence of a substantial diffuse neutral IGM with much larger \HI\ mass than was previously believed. The peak flux and the line profile of the GBT spectrum matches very well with spectra published by \citet{willrood87}, thus confirming our calibration.  Another interesting fact that stands out in the comparison plot for HCG~31 in Figure~\ref{comparisons} is that discrete features seen in the VLA spectrum are not seen in the GBT spectrum.  In fact, the GBT spectrum is much smoother, and the velocity range between the discrete peaks at 4000~km s$^{-1}$ and 4120~km s$^{-1}$ is filled in.  It is likely that the diffuse \HI\ emission spread over a velocity range $\Delta V \sim200$ \kms\ accounts for the majority of the GBT spectrum.  The diffuse \HI\ is likely associated with features much larger in angular extent than group size, possibly in \HI\ filaments being accreted by the group, similar to those seen in the VLA image reported previously by \cite{VM2005}.  Additional observations are underway for an in-depth investigation of the nature of the excess \HI\ detected by GBT.

\section{DISCUSSION \label{sec:discussion}}

\HI content in HCGs varies enormously from one group to another. VM2001 found a broad range of \HI\  deficiency with a mean \HI\ content of 24\% of the expected amount ($Def_{HI}=0.62\pm0.09$) for the individual galaxies and about 40\% ($Def_{HI}=0.40\pm0.07$) for the groups.  The \HI\ distribution in these groups is also quite diverse, ranging from groups dominated by normal galaxies with distinct individual \HI\ disks to groups with galaxies submerged in a single common \HI\ cloud.  Evidence for strong tidal interactions is frequently seen in the \HI\ distribution imaged by the VLA. Such interactions continuously shape the gas distribution and morphology. Consequently, the \HI\ content and morphology reflects the evolutionary history of a group \citep{VM2000,VM2001}.

A new result from our new GBT survey is that the excess \HI\ emission is quite common among HCGs. Here we attempt to understand the distribution and nature of the excess gas through a quantitative analysis of the VLA and the GBT spectra and group properties. 

\subsection{Geometry and Kinematics of the Intragroup \HI \label{sec:geometry}}

\begin{figure}
\figurenum{7}
\includegraphics[angle=-0,scale=0.6]{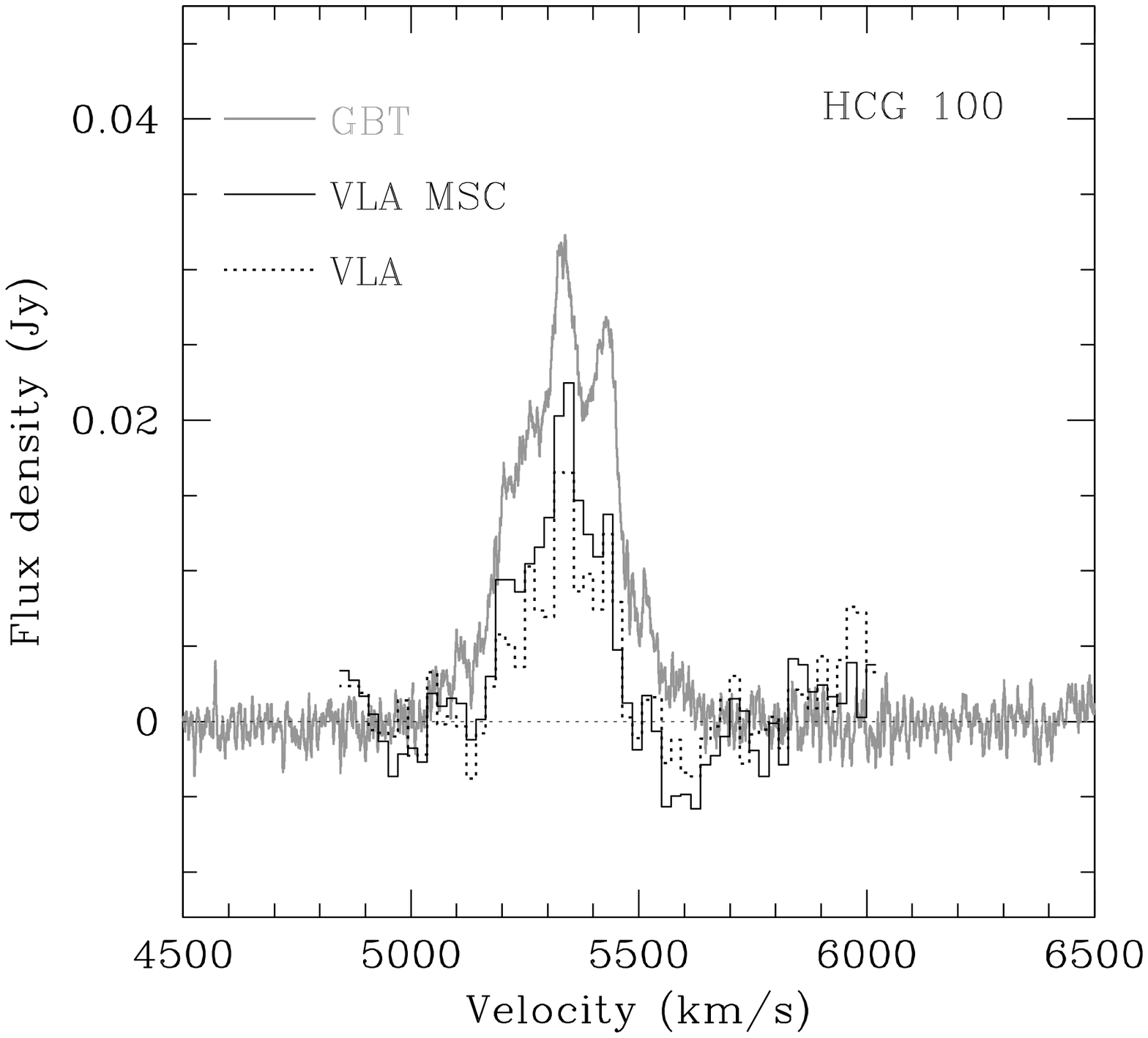}\\ 
{\caption{ An overlay of the VLA spectrum in solid black obtained using the Multi-scale CLEAN (MSC) algorithm in AIPS over the GBT and the VLA spectra for HCG~100. The VLA MSC spectrum recovers 20\% more flux than the VLA spectrum although it does not fully recover the GBT flux. This confirms that there is flux at shorter baselines or larger scales in the VLA data, although the VLA misses the remainder of the GBT flux due to the absence of shorter baselines (or zero baseline) in the data. This translates to a spatial scale of $\sim$15$^{\prime}$ for the \HI\ structures on the plane of the sky. \label{h100_msclean}}}
\end{figure}

The amount of \HI\ detected by the GBT exceeds the VLA detected amount by a factor of two or more in HCG~15, 31, 37, 44, 68, 92, 97, and 100, suggesting that the {\it majority} of \HI\ in these groups is associated with extended diffuse structure, undetected by the VLA. We experimented with the VLA data to look for signs of a spatially extended \HI\ component using multi-scale CLEAN in the IMAGR procedure of the Astronomical Image Processing Software (AIPS) which enhances flux associated with larger angular scales (or shorter baselines) \citep{cornwell08,rich08}. Figure~\ref{h100_msclean} shows an overlay of the multi-scale CLEANed VLA spectra of HCG~100 over its GBT and the VLA spectra.The VLA multi-scale CLEAN spectrum was obtained by using three Gaussian source model widths- 0, 200 and 300 arcsecs, in deconvolving the images. Even in this case, not all of the GBT flux is recovered, and nearly all ``recovered'' features are faint extensions of features already seen in the standard reduction map \citep[details in][]{bor}.

Thus, the missing gas in the VLA images may exist as either: (1) low column density diffuse medium which can only be detected by GBT due to its superb surface brightness sensitivity; (2) spatially extended sheets, filaments or extended disks that are missed by the VLA because of spatial filtering; or (3) as a combination of both.  A physical parameter that provides a quantitative way of distinguishing between these paradigms is the filling factor of the excess gas.  We characterize the filling factor as the fraction of the GBT beam the excess gas would cover if it was distributed uniformly with column density corresponding to the noise level (3~$\sigma_{N(HI),VLA}$) in VLA maps. Thus, the filling factor A$_{ff}$ is computed as   

 \begin{equation}{\label{eq-filling}}
A_{ff} = \frac{M_{excess}}{ 3~ \sigma_{N(HI),VLA}~\Omega_{GBT}~M_{H}} \equiv \frac{M_{excess}}{M_{3\sigma VLA}}
\end{equation}
where $\Omega_{GBT}$ is the area of the GBT beam in cm$^2$ and M$_{H}$ is the mass of an hydrogen atom.

Table~\ref{tbl-filling_factor} summarizes the excess \HI\ masses, VLA noise properties and the filling factors for each of the groups. The 3$\sigma$ column density (seventh column of table~\ref{tbl-filling_factor}) was derived assuming that the features are spread over 2 channels thus implying an intrinsic line-width of 42~\kms\ . The derived filling factor covers a wide range of values, from as little as 3\% in HCG~79 to 566\% in HCG~68.  For most of the HCGs, $A_{ff}$ is much smaller than 100\%, which supports the presence of a faint diffuse \HI\ medium that could have been missed by the VLA due to its lower surface brightness sensitivity.  For groups with $A_{ff}$ $\ga100$\%, spatial filtering by the VLA leading to a significant loss of flux is a highly plausible explanation. This is corroborated by the fact that in all four cases where the filling factor is greater than 100\%, (HCG~44, 68, 91, \& 100), VLA maps show cloud-like structures or extended disks covering around 8-10 $^{\prime}$ (nearly the entire GBT beam).  Furthermore, striking similarities in the spectral profile between the VLA and the GBT spectra in these five systems strongly advocate that the excess gas is dynamically similar to the \HI\ structures mapped by the VLA. 

The excess gas is often associated with a broad; smooth spectral feature rather than discrete peaks.  The dominant spectral feature in spectra of \HI\ deficient groups like HCG~15, 90, and 97 is a faint broad feature. The GBT also detected broad wing-like features in many other groups including HCG~92, 79, and 58.  A noteworthy characteristic is that these broad emission tends to fill the gaps between peaks in the spectrum, resulting in a much smoother spectrum such as in HCG~31 (also see HCG~26, 67, \& 100). This suggests the presence of a diffuse fainter \HI\ component distributed over a large velocity range. The probable origin of this \HI\ component is the evolution of tidally stripped ISM in the hot IGM (see \S~\ref{sec:lifetime_origin}). A detailed discussion of the fate of tidally stripped ISM clouds in the hot IGM is presented in the next section.

In summary, we conclude that the excess gas is spread over a large velocity range and is smoothly distributed as a diffuse, low surface brightness medium. The broad spectral signatures specifically rule out cold discrete filaments. The filling factor values suggest that the excess gas is extended and must occupy at-least a significant fraction (median = 40\% ) of the GBT beam. Consequently, the excess gas exists as a faint diffuse neutral intragroup medium in most HCGs. In a few groups where  $A_{ff} > 1$,  large HSB \HI\ structures that were spatially resolved out by the VLA may also contributes to the excess. A combination of the diffuse IGM and some HSB features is likely in groups exhibiting large line-flux differences along with broad wing-like spectral features (e.g., HCG~100). These groups may also represent a transitional phase occurring between groups with clumps of tidally stripped \HI\ in their IGM and groups with a faint diffuse extended \HI\ component distributed in their IGM. 

\subsection{Lifetime for \HI\ Structures in IGM \label{sec:lifetime}}

Compact groups are evolving structures. The dynamical time ($t_{dyn}$) for a typical group with a diameter of 100~kpc and velocity dispersion of 250~\kms\ is around 400~Myr. During this time, galaxies experience multiple tidal interactions with each other, and a considerable amount of ISM may be removed from individual galaxies and deposited in the group environment. This is evident in groups like HCG~16 which show multiple tidal tails (VM2001) and also in groups like HCG~92 where the \HI\ has been tidally removed from their parent galaxies and deposited as large \HI\ clouds in the IGM \citep{willyun02}. The nature and fate of neutral gas in such extreme environments are determined by the interplay of heating and cooling in these systems. The fate of tidally stripped debris can provide clues about possible causes of \HI\ deficiency commonly observed among galaxies in such dense environments.

In this section, we attempt to check theoretically if the neutral tidal debris should survive for a considerable amount of time for us to be able to detect it in systems like the HCGs. In other words, should we expect to find a surviving component of the tidal debris deposited in the IGM? In order to do so, we investigate the timescales and conditions for the survival of neutral clouds in the hot IGM. For a spiral dominated galaxy group, \citet{mul96a} calculated the expected virial temperature of the IGM to be around 2~$\times$~10$^6$~K. Any cold \HI\ cloud/filament will be conductively heated by the hot IGM. This can ionize the cloud and incorporate it into the hot IGM. In addition, a neutral cloud will expand into the IGM due to its intrinsic velocity dispersion. This will eventually lower its column density and make it more susceptible to ionization by the ultra-violet radiation background. Based on the timescales of these processes, we can derive the expected lifetime of a typical tidally stripped cloud. Since we have ample evidence of diffuse neutral IGM in HCGs, we can explore the parameter space for physical conditions that will ensure the survival of the \HI\ clouds over the dynamical time-scale of the groups. This will help in constraining the properties of the \HI\ clouds as well as the IGM, which in most galaxy groups escape detection by being too cold to emit in the X-ray.

\subsubsection{Conductive Heating Lifetime \label{sec:lifetime_cond}}

We begin by investigating the effects of conductive heating on a cold neutral cloud embedded in hot IGM. This process is not instantaneous, and the evaporation timescale depends on the physical properties of both the \HI\ cloud as well as the IGM. Conductive heating can take place in two paradigms, classical or saturated, depending on the ratio of the mean free path of the electrons to the temperature scale-height of the hot IGM \citep{cowie77}. For a spherical neutral cloud with radius R$_{cl}$ (in parsecs) and density n$_{cl}$ (in cm$^{-3}$) floating in the hot IGM at a temperature of T$_{IGM}$ (in Kelvin) and density $n_{IGM}$ (in cm$^{-3}$), the ratio of classical heat flux to saturated heat flux is given as \citep[][eq. 11 ]{vollmer01} and ,

\begin{equation}{\label{eq-sigma}}
\sigma_{o}  \simeq (T_{IGM}/1.5\times10^7)^2/(n_{IGM}\times R_{cl}) 
\end{equation}

The classical evaporation is applicable for systems where $\sigma_o  \ll 1$, where as saturated evaporation will be applicable for systems with $\sigma_o  \gg 1$. The evaporation timescale in the classical paradigm is given by \citep[][eq. 22]{cowie77}
\begin{equation}{\label{eq-tc}}
t^{c}_{ev}=3.3\times 10^{20}~~\bar{n}_{cl}~~R^{2}_{cl}~~T^{-5/2}_{IGM}~~\frac{ln \Lambda}{30} ~~yrs
\end{equation}
 where $ln~\Lambda = 29.7 + ln~n^{-1/2}_{IGM} (T_{IGM}/10^6~~K)$. For a fixed IGM temperature and cloud mass, 
  \begin{equation}{\label{eq-tc-cor}}
t^{c}_{ev}~\propto~~\bar{n}_{cl}~~R_{cl}^{2}~~\propto~\frac{M_{cl}}{R_{cl}}
\end{equation}
 This implies that a smaller and consequently denser cloud will much survive longer than a less dense cloud of the same mass.

In the saturated paradigm, the evaporation timescale is given by  \citep[][eq. 64]{cowie77}
 \begin{equation}{\label{eq-ts}}
t^{s}_{ev} \sim 10^6~~ \bar{n}_{cl}~~ n^{-1}_{IGM} ~~  R_{cl}  ~~ T_{IGM}^{-1/2}~~yrs
\end{equation}
From Eq.~\ref{eq-tc} and \ref{eq-ts}, it is clear that the lifetime of the cloud increases as its density and/or radius increases in both paradigms, although at different rates. Also the lifetime is inversely related to the temperature of the IGM.

Many of these physical quantities are poorly understood for compact groups, but we can make educated guesses based on our understanding of tidal interactions, ISM phases, and the intergalactic medium. From multi-galaxy interactions seen in HCG 92, \citet{willyun02} found \HI\ structure associated with HCG~92A to be as small as 7.3~$\times$~4.4~kpc$^{2}$. Thus for our calculations we assume a lower limit of tidally stripped neutral cloud to be around 1~kpc. In a recent study of jet-IGM interactions, \citet{freeland08} derived IGM density between 9~$\times$~10$^{-4}$ and 4~$\times$~10$^{-3}$~cm$^{-3}$. Using X-ray data, \citet{rasmussen08} derived IGM densities for HCGs that ranged from 6~$\times$~10$^{-5}$ to nearly 4~$\times$~10$^{-2}$~cm$^{-3}$. Adopting $n_{IGM}$=10$^{-3}$~cm$^{-3}$,  $T_{IGM}$= 2~$\times$~10$^6$~K, and $R_{cl}$= 1~kpc, we find  $\sigma_{o}$= 0.0177 $\ll$ 1 and the classical paradigm is applicable. 

Tidally stripped \HI\ clouds have been part of the ISM of the individual galaxies and had similar densities and ionization fractions as the \HI\ emitting clouds typically found in the ISM. From \citet{mckee_ost77}, the lowest density phase of ISM with ionization fraction significantly less than 1 is the warm ionized medium (WIM). The density and the ionization fraction in this phase is 0.25~cm$^{-3}$ and 0.68 respectively. For phases with lower ionization fraction than 0.68, the density of the ISM will be much higher. Using \HI\ cloud density of 0.25~cm$^{-3}$ and the above-mentioned values for the ISM and IGM parameters, we derive an evaporation timescale of $\sim$ 15~Gyrs, which is  $\gg$ $t_{dyn}$. In the classical paradigm, the evaporation timescale is strongly correlated with the size of the cloud. The lower limit of \HI\ cloud size that can survive the IGM for more than $t_{dyn}$ is $\sim$200~pc. For instance, the evaporation timescale for a smaller cloud with radius 100~pc is  $\sim$~0.15~Gyrs.

\subsubsubsection{Constraining the physical parameters of the IGM \label{sec:lifetime_const}}

Our GBT detections, combined with information from VLA images, confirm that \HI\ in HCGs survives long enough to be seen frequently in their harsh environments. Using this crucial piece of information, we can constraint the physical conditions of the \HI\ clouds as well as the hot IGM by finding the region in the parameter space where the evaporation timescale is greater than the dynamical timescale of the groups. Throughout our analysis we will assumed the \HI\ clouds to be stationary. If the proper motion of the \HI\ clouds could be estimated more stronger constraints could be achieved. Unfortunately with the present state of observational data such estimation is not possible. 

\begin{figure}
\figurenum{8}
\includegraphics[angle=0,scale=0.64]{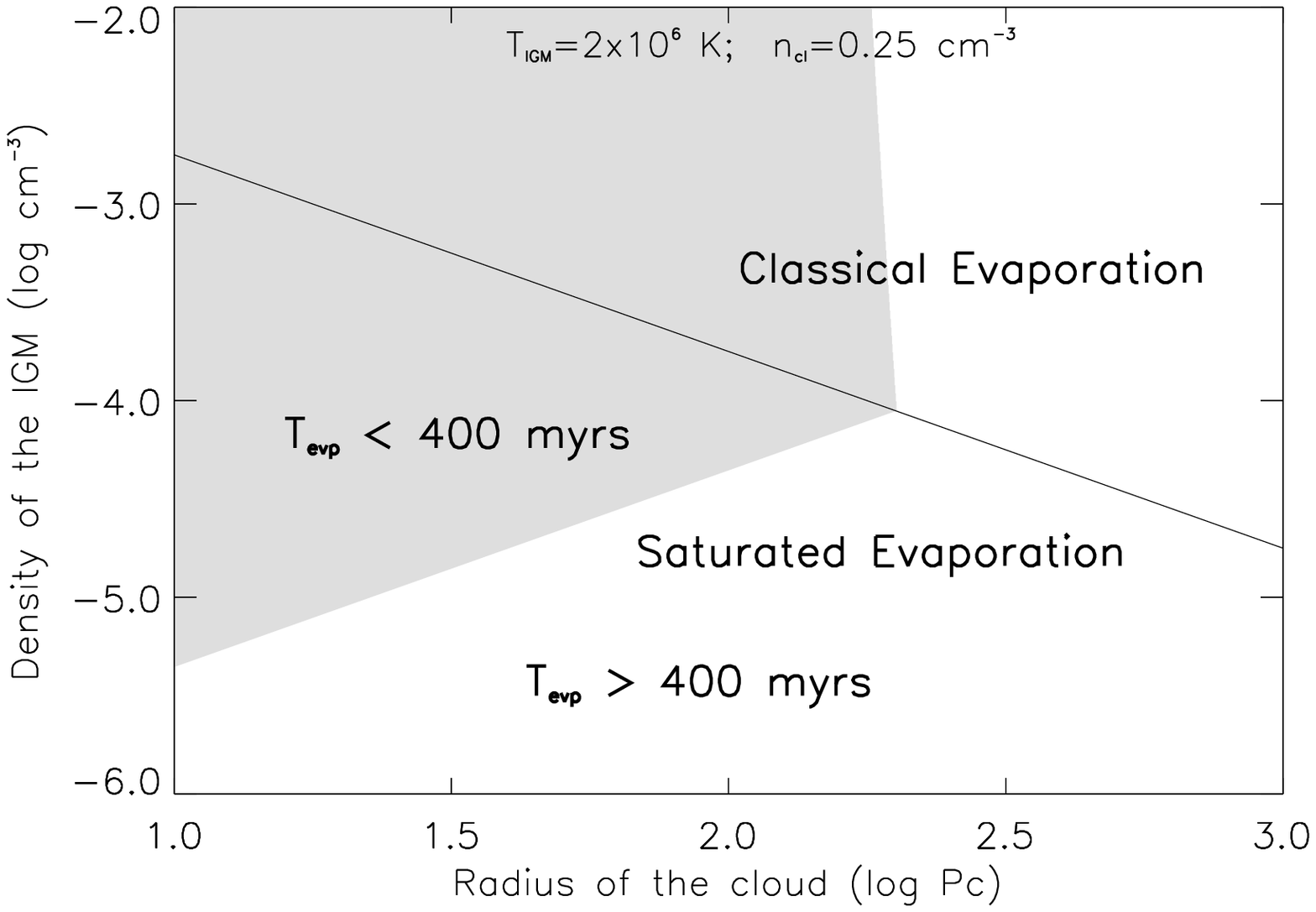}\\
{\caption{Plot shows cloud evaporation timescale regimes as a function of cloud radius and density of the IGM. The values for the temperature of the IGM, T$_{IGM}$ and the density of the cloud, n$_{cl}$, are fixed and are shown on the top. The solid line divides the region of saturated evaporation from that of the classical evaporation. The grey and white areas represent the range of parameters for which the evaporation timescale is shorter and longer than the dynamical time (400~Myrs for a typical HCG) respectively.  \label{parameter}}}
\end{figure}

Figure~\ref{parameter} shows the cloud evaporation time scale as a function of density of the IGM and the radius of the cloud. The solid line divides the parameter space into the classical and saturated evaporation regimes. The white region represents the range for which the evaporation timescale is more than 400~Myrs (typical dynamical time) where as the grey region is the range of parameters for which a neutral cloud would not survive the heating and evaporate within a single crossing time. The timescale increases or decreases  as the parameter values moves away from the solid line and follows the proportionality seen in equation~\ref{eq-tc-cor} and \ref{eq-ts}. {\it It is evident that for clouds of radius $\ge$200~pc with an average density $\ge$ 0.25~cm$^{-3}$ are robust against
evaporation for an IGM with T$_{IGM}$ = 2$\times$10$^6$~K,  irrespective of $n_{IGM}$.}

The large timescale for heating can also be understood in the context of the critical radius, which is defined as the cloud size at which radiative cooling balances heating by thermal conduction. For an IGM with T$_{IGM} \gtrsim $~10$^5$~K, the critical radius as formulated by \citet[][eq. 14 ]{mckee77} is 
 \begin{equation}{\label{eq-rc}}
 R_c \approx 4.8\times10^5~T^2_{IGM} / n_{IGM}~~ cm.
 \end{equation}
The critical radius for a cold cloud in a hot ($T_{IGM}=2\times10^6~K $) medium is $0.6 \leqslant R_c \leqslant 6$~kpc when the density of the IGM is  $10^{-3} \geqslant n_{IGM} \geqslant 10^{-4}$~cm$^{-3}$. Thus, large \HI\ structures with R$_{cl}~\ge$~R$_c$ should survive for more than $t_{dyn}$ (400 Myrs) due to a balance between conductive heating and radiative cooling.

\begin{figure}
 \figurenum{9}

\begin{tabular}{c c}
\includegraphics[scale=0.39,angle=0]{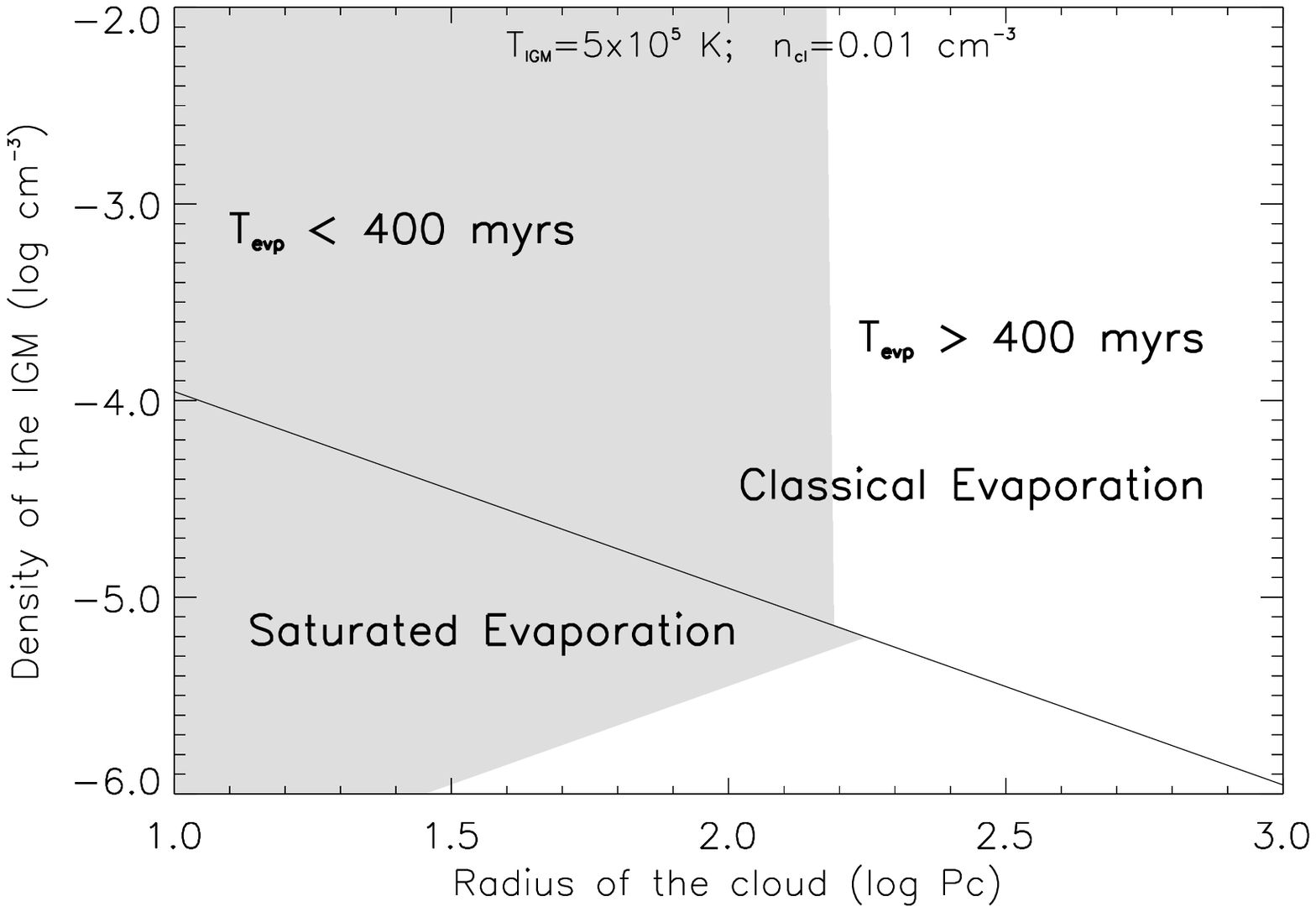} &
\includegraphics[scale=0.39,angle=0]{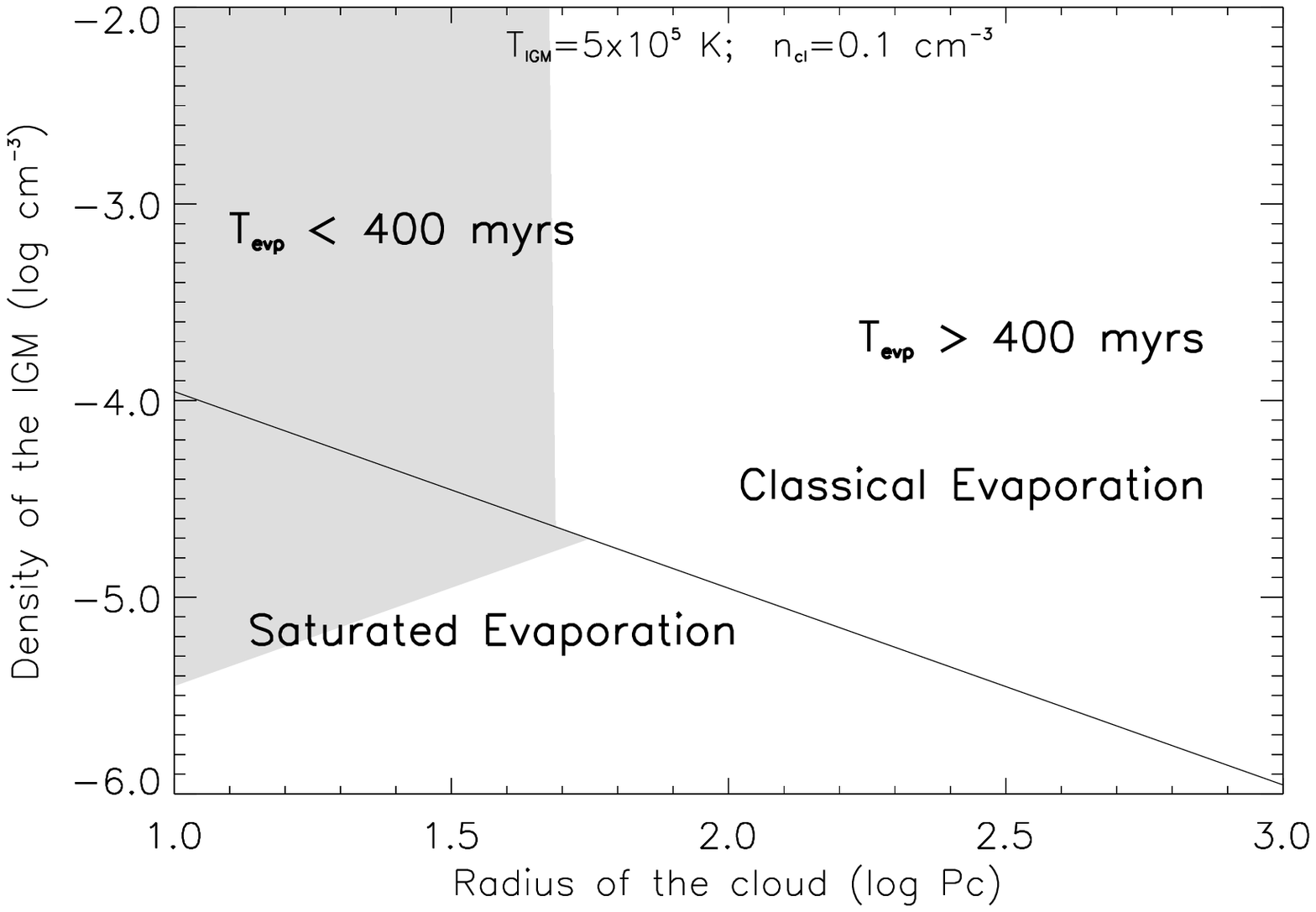} \\
\includegraphics[scale=0.39,angle=0]{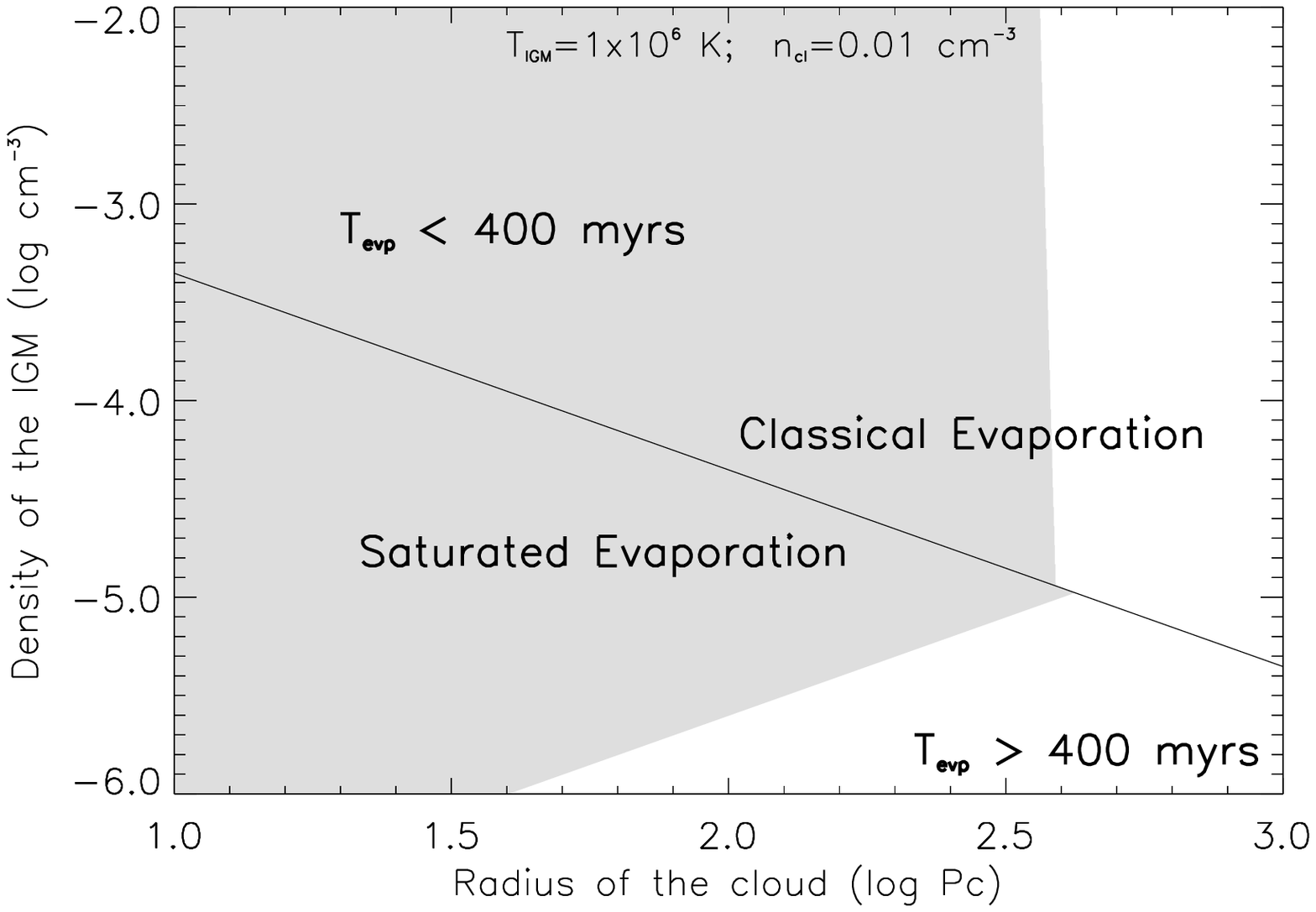} &
\includegraphics[scale=0.39,angle=0]{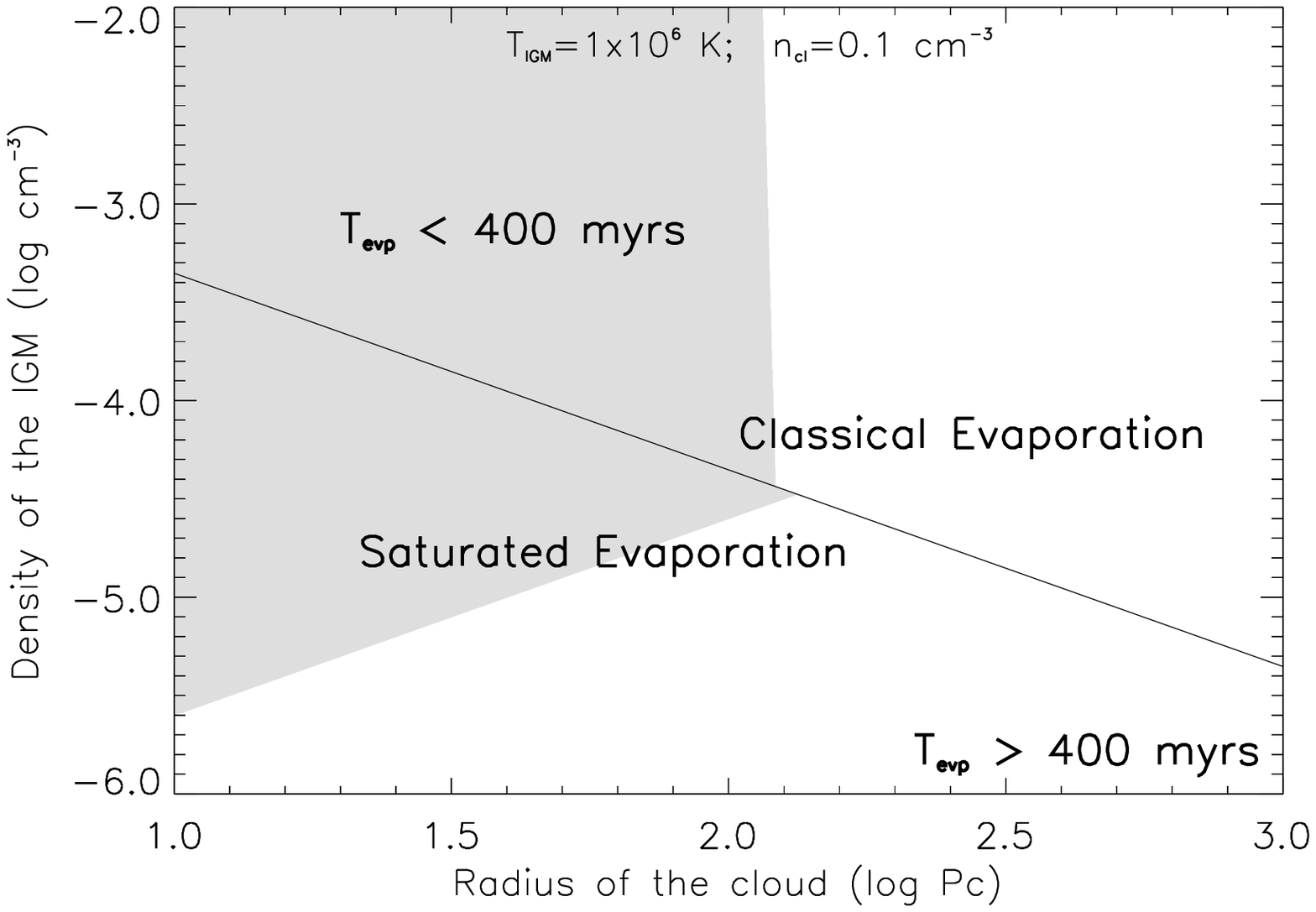} \\
\includegraphics[scale=0.39,angle=0]{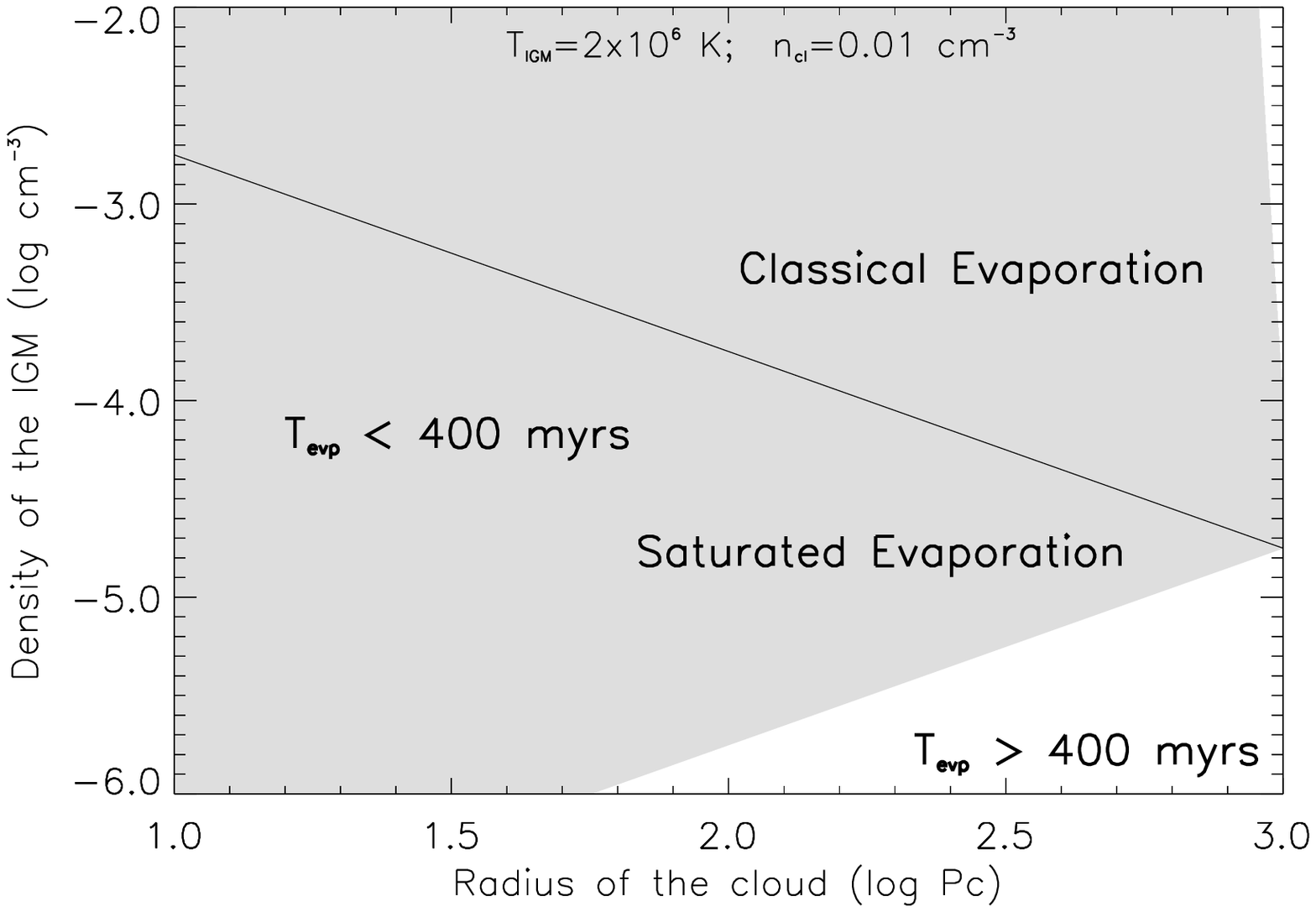} &
\includegraphics[scale=0.39,angle=0]{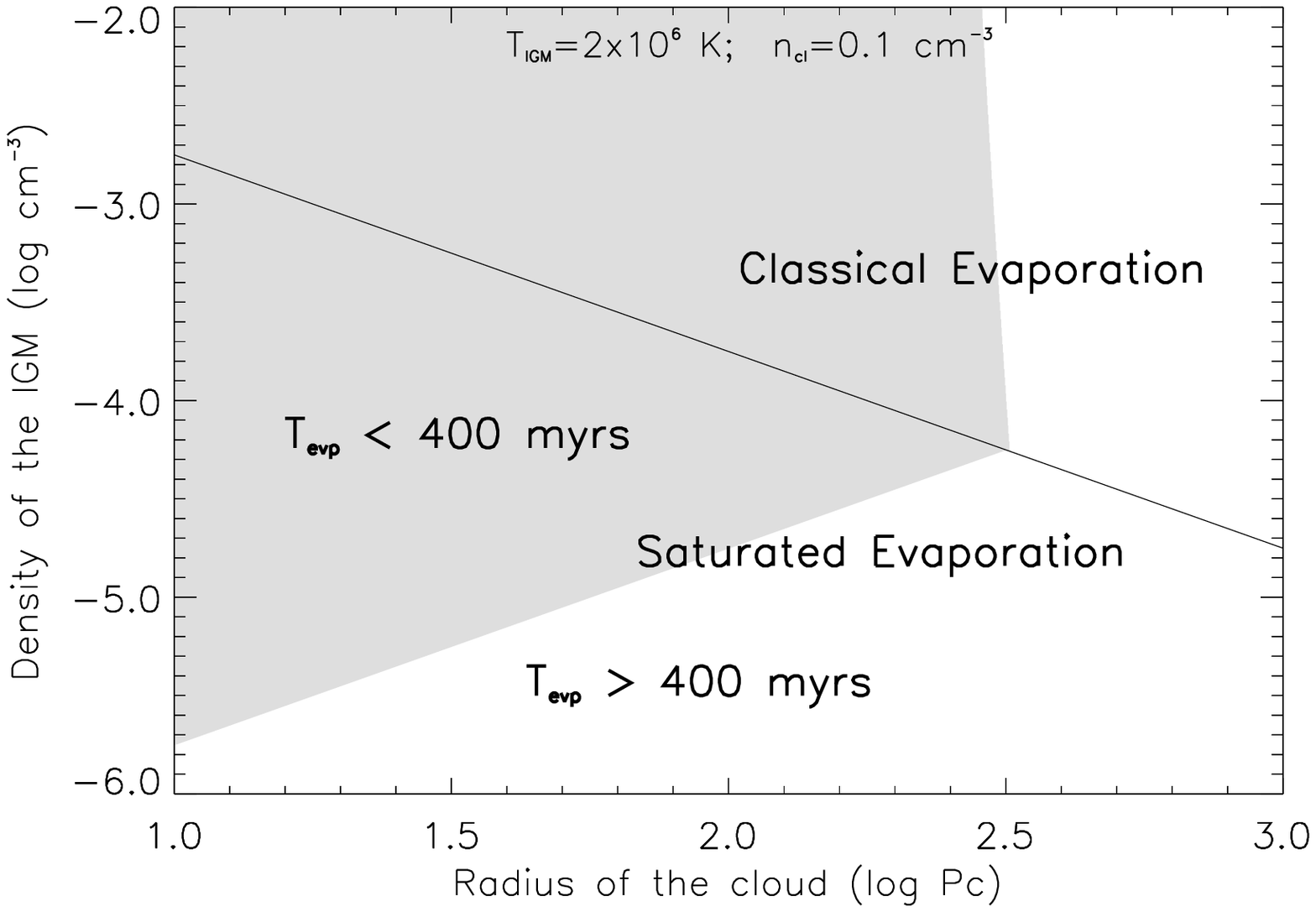} \\
\includegraphics[scale=0.39,angle=0]{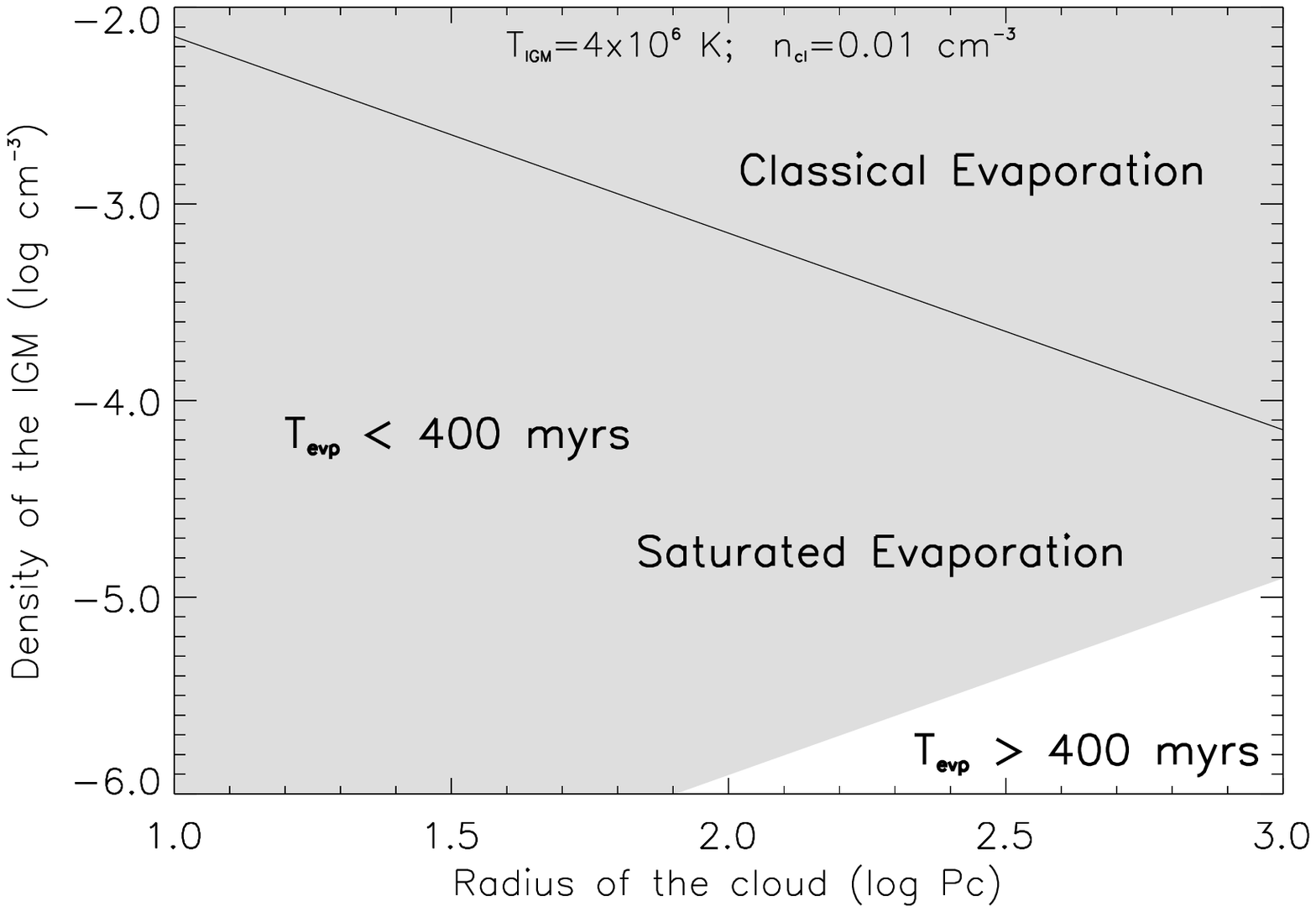} &
\includegraphics[scale=0.39,angle=0]{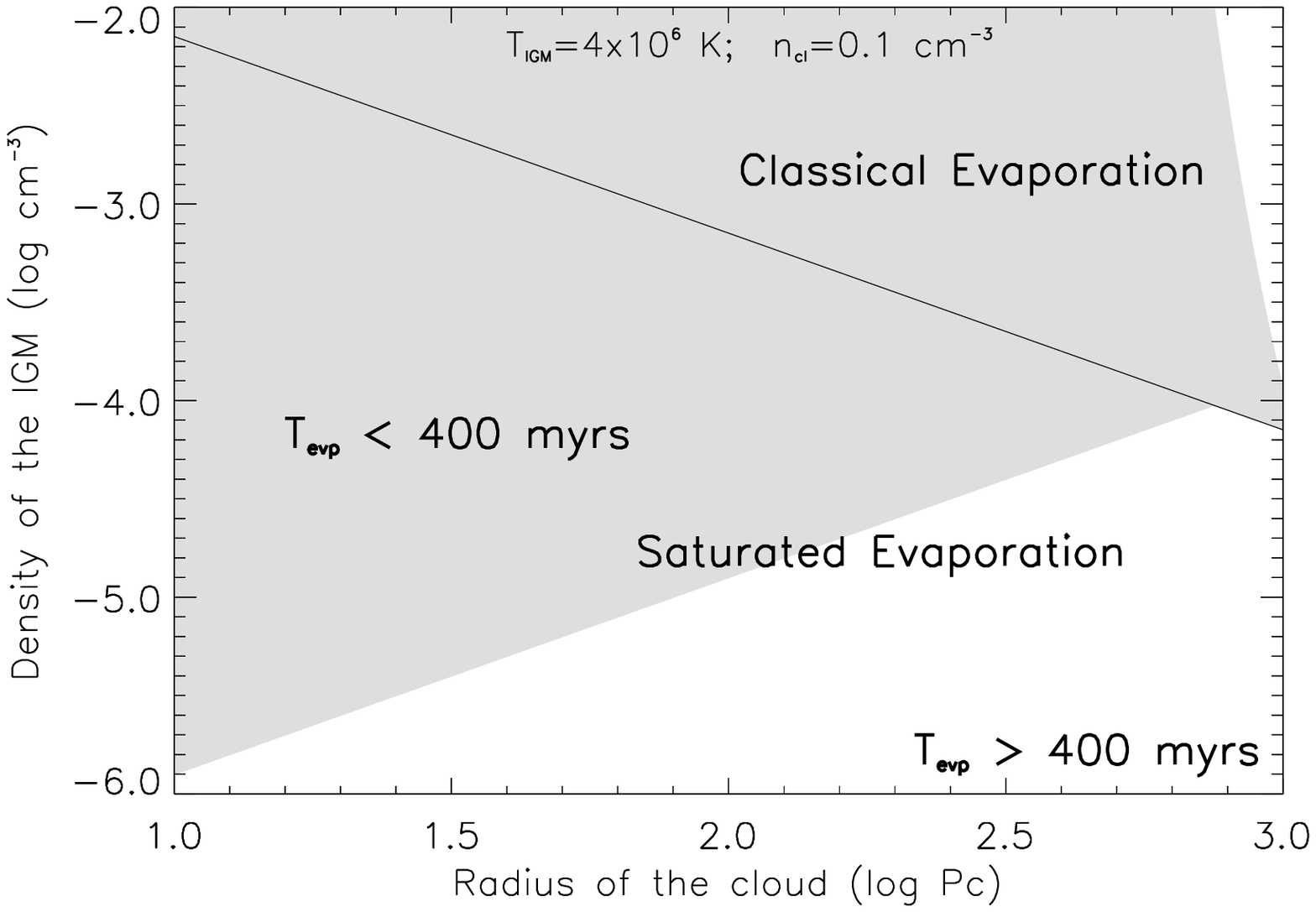} \\

\end{tabular}
{\caption{Plots showing cloud evaporation timescale regimes as a function of cloud radius and density of the IGM for different sets of IGM temperature and cloud density. The T$_{IGM}$ and n$_{cl}$ values are printed on the top of each plot. All notations are same as Figure~\ref{parameter}. \label{lifetimes}}}
\end{figure}

We have also explored the dependence of $t_{ev}$ for a small range of IGM temperatures and HI cloud densities. Figure~\ref{lifetimes} shows the parameter for eight different combinations of IGM temperature and cloud density. The plots have been arranged with increasing temperature from top to the bottom and increasing density from left to right. These plots clearly show that evaporation timescales decrease with increase in the IGM temperature and with the decrease in the density of the cloud, as expected from Eqs~\ref{eq-tc} and \ref{eq-ts}. Given that we detected 21~cm emitting cold gas in most of the groups, we can conclude that the IGM temperature is less than 4~$\times$~10$^6$~K unless the density of the hot IGM is unusually low. Also, the density of the \HI\ clouds, responsible for excess emission in the GBT spectra, is most likely higher than 0.01~cm$^{-3}$ unless the IGM is substantially cooler than 2~$\times$~10$^6$~K. This follows from the fact that the clouds with density lower than 0.01~cm$^{-3}$ would not survive long enough to be detected in an IGM corresponding to typical virial temperature of 2~$\times$~10$^6$~K. Since the relevant parameter space is mostly in the classical regime, the variation of IGM temperature has a much stronger effect on the timescale than the density of the cloud. On the other hand, the density of the IGM effects the timescale only in the saturated evaporation regime and consequently, we cannot draw a strong constraint on the  n$_{IGM}$ from our data. 

\subsubsection{Lifetime Based on Expansion of the \HI\ structures \label{sec:lifetime_exp}}

In addition to heating, the tidal \HI\ debris should also experience expansion as a result of their intrinsic velocity dispersion unless they are bound by their self-gravity. Such an expansion of an \HI\ cloud would lower its column density, making it prone to ionization by the intergalactic UV radiation field. The 21~cm \HI\ studies by \citet{corbelli89,corbelli93}, and others have shown that a minimum column density of $2\times 10^{19}$~cm$^{-2}$ is required for an \HI\ cloud to shield itself from the background UV radiation field. Therefore, the expansion timescale, which refers to the time an \HI\ cloud would take to reach the above mentioned column density, is important in estimating the lifetime of an expanding tidal structure. As a cloud expands its column density decreases which can be expressed in terms of the initial column density ($N_i$) and initial radius $R_i$ as 
\begin{equation}
\frac{N}{N_i}= \frac{Area_i}{Area}= \frac{(R_i)^2} {(R_i~+~v~t_{exp})^2}
\end{equation}

Assuming that the tidally stripped cloud in question is a disk/cylinder with $R_i$~=1~kpc and $N_i ~=~3\times 10^{21}$~cm$^{-2}$, expanding at $v~=~ 20$~\kms . The expansion timescale for this cloud to reach the above mentioned threshold value is around 500~Myrs. In this time the cloud expands more than 100 times its initial area (or 1000 times its initial volume), thereby creating a faint diffuse neutral IGM.

\subsection{Possible Origins of the Diffuse \HI\ Emission \label{sec:lifetime_origin}}

Based on the observed deficiencies as well as complex \HI\ tidal structures seen in VLA imaging of the HCGs, the excess gas mass can be attributed to evolving tidal debris deposited by the individual galaxies. The fact that the excess gas does not render the HCGs superabundant in \HI\ supports the tidal debris origin of the excess gas. Furthermore, the velocity distribution of the excess gas is extremely similar in line shape to that of the VLA detected high surface brightness \HI\ associated with individual galaxies (catagory I) in more than 80\% of the groups. The excess gas content also increases with the evolutionary phase (VM2001) of the group (discussed in detail in \S~\ref{sec:Dep_evol}), which supports the idea of evolution of tidal debris into this component. Similar detections of tidal \HI\ structures by \citet{dav04} and \citet{osvan05} in an even more extreme environment of the Virgo intercluster medium also lends support the existence and preservation of tidally stripped neutral gas for $>~10^8$ years. 

A potentially important process for the IGM in some of the X-ray bright compact groups such as HCG~97 and HCG~15 is the precipitation of cold gas from the hot IGM through the radiative cooling \citep[so-called ``cooling flow''][]{fabian94}. Unless a constant source of energy is present to reheat the gas back to the IGM temperature, hot gas is expected to cool down in the centers of groups and clusters. Calculations based on work by \citet{narayan01} suggest that the transfer of heat from hot IGM to the condensing cores would take about 0.5~Gyr for a group of size 100~kpc. Such a conduction time-scale for the IGM suggests the possibility of a neutral IGM resulting from cooling of the X-ray bright IGM at the centers of these groups. However, 
most of the HCGs used in our study are not X-ray bright and therefore, cooling flow is not likely to be the dominant mechanism responsible for the existence of the neutral gas in our sample. Therefore, we can safely conclude that the majority of the excess gas detected by the GBT is not a result of condensation of the hot IGM. 

In a handful of groups where the GBT beam covered a region much larger than the group size, accretion of cold gas from the large-scale structure \citep{keres05} is a possibility. For most of the HCGs we have surveyed the central region of the groups where the accreted cold gas, if any, is expect to get shock heated and ionized thus eliminating the possibility of its survival for ubiquitous detection in 21~cm emission. Therefore, the most plausible origin of the excess gas found in the central regions of the HCGs is \HI\ tidal debris deposited in the IGM as individual galaxies interacted with each other and the group potential. In this scenario, no additional cold gas contribution from accretion or cooling flows (for X-ray bright HCGs) is required, although minor contributions from these processes cannot be ruled out.

\subsection{Excess \HI\ and Group Properties  \label{sec:trends}}

M$_{excess}$ is substantial in most of our groups and has important implications on our understanding of the group medium and its evolution. As discussed earlier, this excess gas most likely exists as an extended diffuse medium that is being created and maintained by depositing tidally stripped ISM from individual galaxies into the IGM. In this section, we explore the distribution of the excess component using excess gas mass fraction in terms of the group radius, the \HI\ deficiency, the evolutionary stage as well as Hubble type of the central galaxy of the HCGs. For our statistical analysis, we use a sub-sample of our complete sample for which VLA D-array \HI\ data was available for comparison. The additional sources - HCG~26 and HCG~48 are not included. HCG~40, although a part of our complete sample, is also excluded from the analysis due to the uncertainty in the calibration of the GBT data. In addition, VLA  spectra has more flux than in the GBT spectra for HCG~40 and following equation~\ref{eq-f}, $f_{HCG40}~<$~0. Hence, our statistical analysis sample consists of - HCG~7, 10, 15, 16, 23, 25, 30, 31, 37, 44, 58, 67, 68, 79, 88, 90, 91, 92, 93, 97, and 100.

\subsubsection{Radial Dependence with Group Diameter \label{sec:groupdiameter}} 

\begin{figure}
 \figurenum{10}
\includegraphics[angle=-90,scale=0.6]{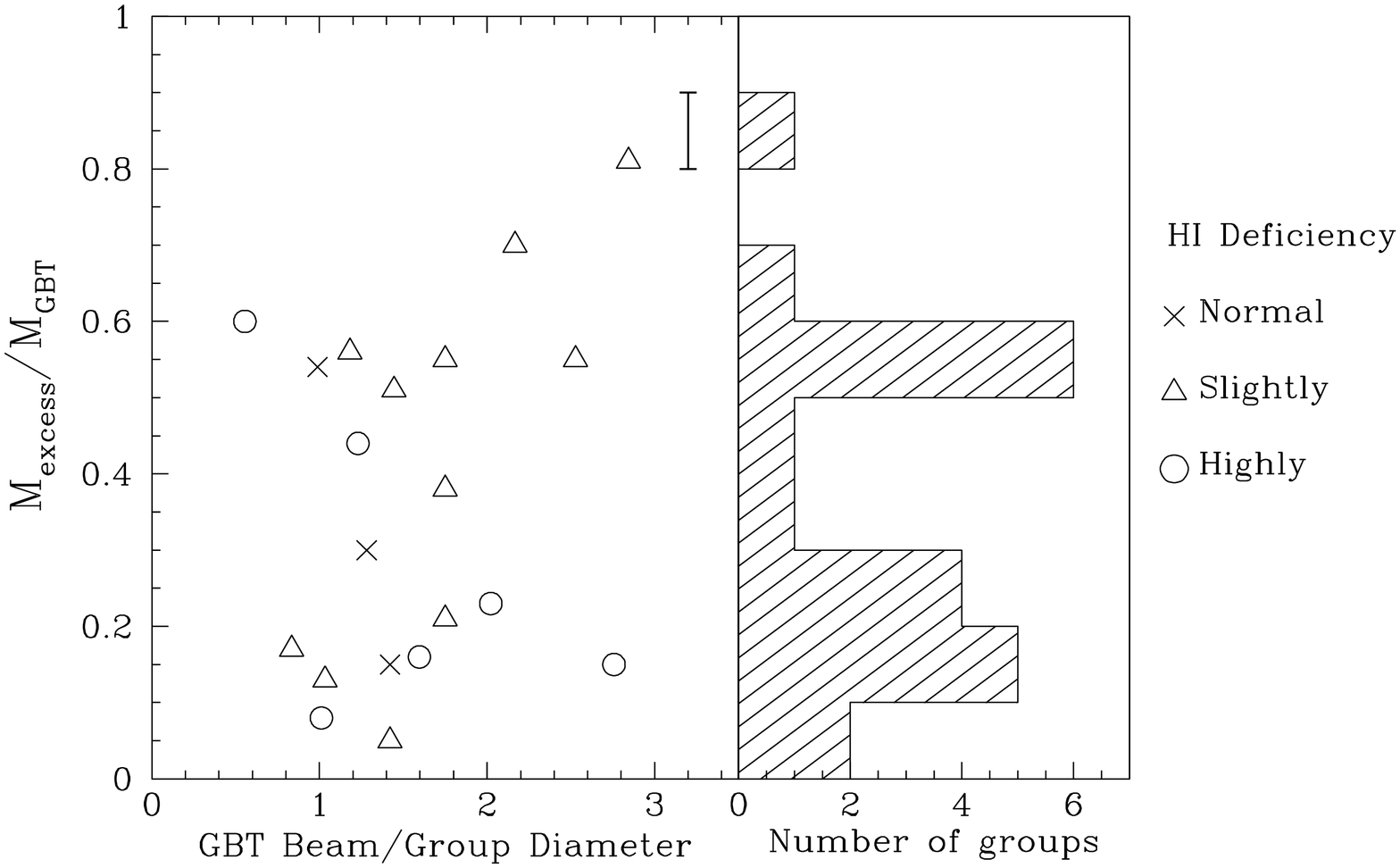}\\
{\caption{The diffuse gas fraction is plotted as a function of the GBT beam size (in terms of the group radius). On the right is a histogram HCGs is shown as a function of the diffuse gas fraction.
The plots do not include the extremely compact group HCG~79 (GBT beam/group diameter $\sim$7) with  21\% of excess gas.  
\label{diff_beam_ratio}}}
\end{figure}

HCGs represent a diverse set of galaxy groups with a range of sizes, number of members, \HI\ content and evolutionary phases. In our sample, the angular diameter of the groups varied between 1.3$^{\prime}$ to 16.4$^{\prime}$, which translates to 0.14 to 1.80 times the GBT beam at 1.4~GHz. The excess gas mass fraction is shown as a function of the group diameter sampled by the GBT beam in Figure~\ref{diff_beam_ratio}. For groups where  $f~\ge 0.5 $, the GBT beam covers the inner region up to a maximum of 3 times the group diameter. For instance in HCG~44, where the excess gas mass fraction is 60\%, the GBT beam covers only the inner one-third of the group area (group diameter is 16.4$^{\prime}$). In other words, the diffuse excess gas is the major contributor to the total \HI\ mass in the central region of this system. This indicates that the distribution of the diffuse component is within the central region of the groups along with the individual galaxies.  The majority of groups are either lack any diffuse component or
are dominated by it with few groups lying in between as seen in Figure~\ref{diff_beam_ratio}. In fact, this offers the first indication that the $f$ distribution is clustered, with one peak near $f \sim 0.2$ and a second peak near $f \sim 0.6$, irrespective of the fraction of group diameter covered by the GBT.  There are only two systems with 0.3~$\le~f~\le$~0.5. This suggests that the transition from the low diffuse gas phase to high diffuse gas is not gradual. Instead, the transition  may be quite rapid and the change in the ratio indicates an evolution in the group property.

\begin{figure}
\figurenum{11}
\includegraphics[angle=-90,scale=0.6]{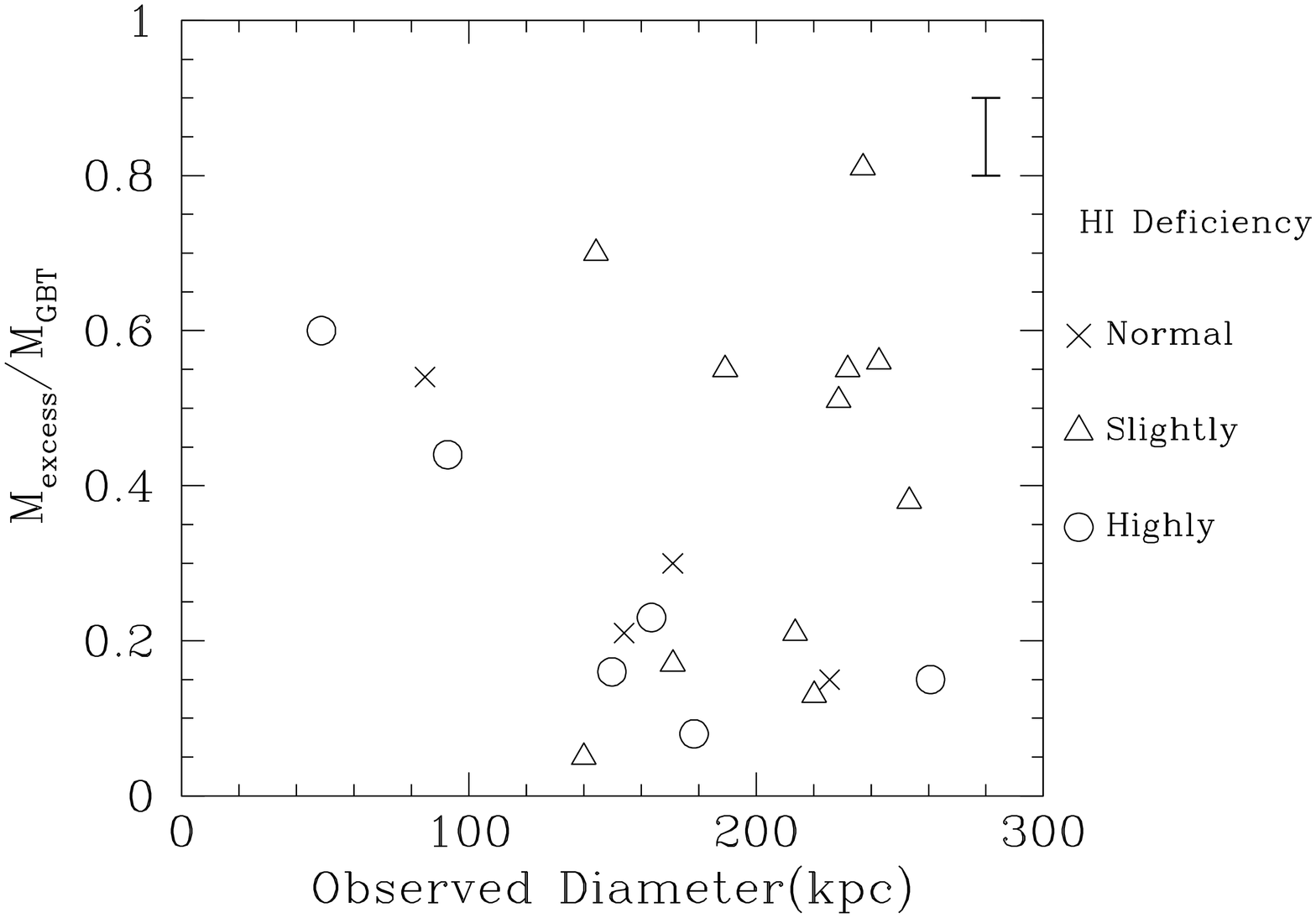}\\
{\caption{The diffuse gas fraction is plotted as a function of the GBT beam size in kiloparsecs. 
\label{diff_size}}}
\end{figure}

In terms of the physical dimension, the GBT beam coverage varies from 40~kpc to 250~kpc. Figure~\ref{diff_size} shows the same data as Figure~\ref{diff_beam_ratio} but in physical units (kilo parsec). The trends seen are essentially identical to those presented in Figure~\ref{diff_beam_ratio}, but a potentially important and new insight is offered: {\it groups where the surveyed region is smaller than 100~kpc show a rather significant fraction of excess gas, $f > 0.4 $}.  While the statistics are poor, such a trend is expected only in scenarios where the group radius actually reflects the physical proximity of the member galaxies.  Specifically, such a trend should not exist in apparent groups identified by chance projection only \citep[e.g][]{hern95}.

\subsubsection{Dependence on HI-deficiency \label{sec:Dep_HIdef}}

 Our HCG sample can be divided into three categories on the basis of their \HI\ deficiencies, as normal \HI\ content, slightly \HI\ deficient, and highly \HI\ deficient (see \S~\ref{sec:HIdef}). The original definition of \HI\ deficiency was based on the \HI\ content of individual galaxies. However, in HCGs the \HI\  distribution varies enormously from the distribution of the optical galaxies, and it is often not possible to associate \HI\ structures with individual galaxies. Therefore, we extend the definition of \HI\ deficiency to groups by evaluating the ratio of the sum of the predicted \HI\ mass for individual members to the observed \HI\ mass associated with the entire group. 

The excess \HI\ fraction for the three deficiency classes of HCGs is summarized in Table~\ref{hi_def}. The mean (median) excess gas mass fraction, $f$, for the normal \HI\ content, slightly \HI\ deficient, and highly \HI\ deficient groups are 30~\% (26\%), 42~\% (51\%), and 28~\% (20\%) respectively. The comparison of \HI\ deficiency for the groups is also shown graphically on the top panel of Figure~\ref{3histograms}. Most of the groups with a high fraction of excess gas are slightly \HI\ deficient groups whereas most of the normal and highly deficient groups have much smaller fraction of excess gas. The scatter within each class is large which makes the average distribution ineffective in deducing any statistically significant conclusion.

\begin{figure}
\figurenum{13}
\includegraphics[scale=0.7,angle=-90]{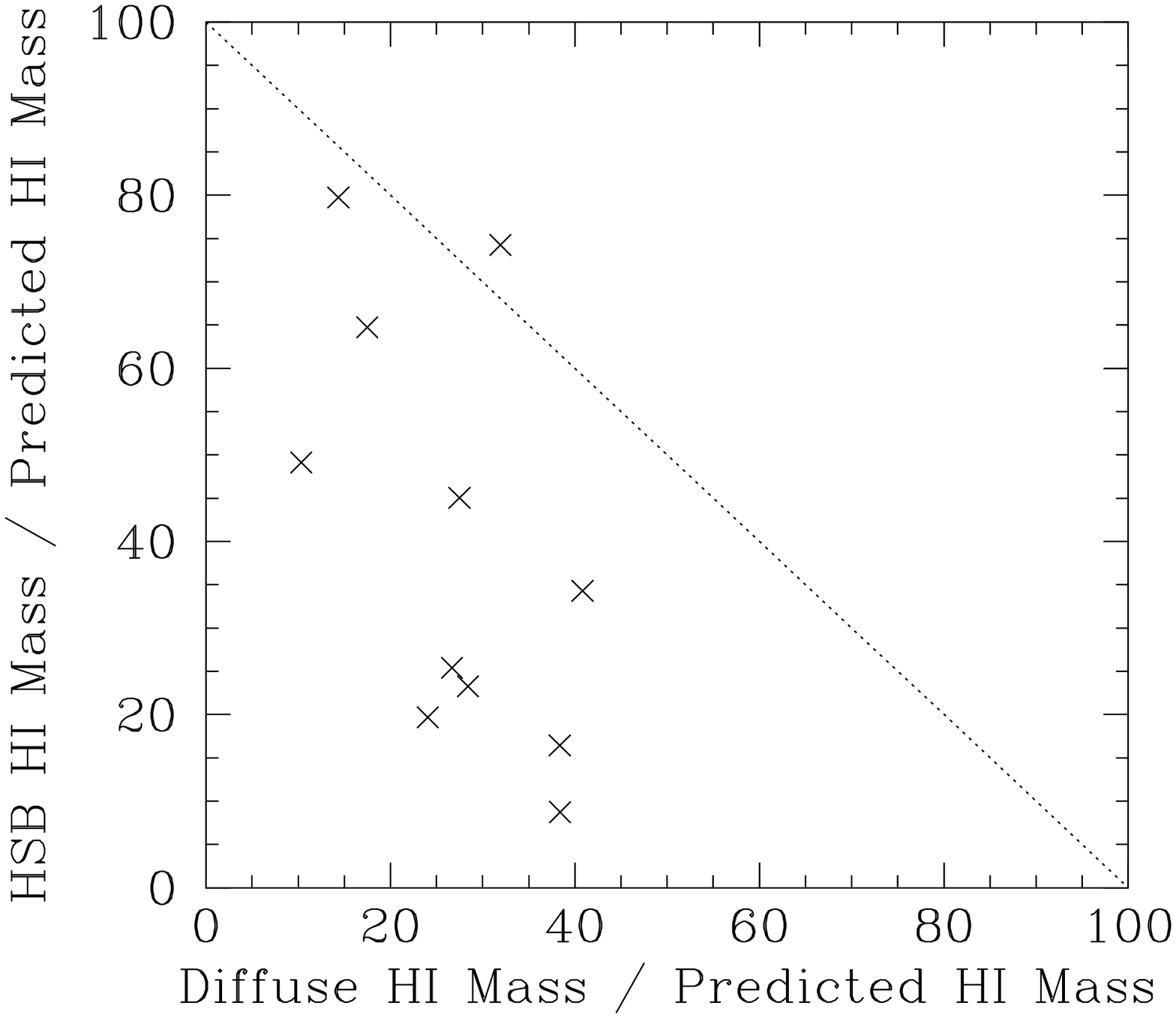}
{\caption{Variation of diffuse gas content and the high surface brightness (HSB) VLA detected \HI\ in terms of the predicted \HI\ mass for 11 HCGs with \HI\ deficiency $\le$ 0.4. The dotted line marks the case of \HI\ normalcy. 
\label{hsb_diff} }}
\end{figure}

On the other hand, for groups with \HI\ deficiency~$<$~0.4 ($\equiv$~2~$\sigma$ in predicted \HI\ mass), one would expect the two tidally stripped \HI\ components - diffuse \HI\ detected only by the GBT and the high surface brightness (HSB) \HI\ detected by the VLA, to constitute the total \HI\ in these systems. Figure~\ref{hsb_diff} shows a comparison of the excess gas content and the HSB \HI\ content in terms of the predicted mass for a sub sample of 11 HCGs, where \HI\ deficiency~$<$~0.4. In general, the fraction of diffuse excess gas increases with the increase in the \HI\ deficiency. The Spearman's (rho) rank correlation coefficient for this subsample is $-$0.523919 with significance of its deviation from zero of 0.098. This provides an indication that HSB structures in the IGM are evolving into the diffuse \HI\ component before being ionized and assimilated into the hot IGM. 

In terms of the evolution and fate of the neutral tidal debris, the diffuse gas can be understood as an intermediate stage between HSB \HI\ and ionized hydrogen. In this picture, the stripped \HI\ from the galaxies begin their life as tidal features (VM2001) in the IGM, which evolves into a diffuse faint cold gas component that was detected with the GBT. After a certain threshold, depending on the physical parameters of the IGM and the diffuse \HI\ clouds (see \S~\ref{sec:lifetime}), the hot IGM ionizes and assimilates the diffuse component. Thus, in the early stages of evolution, groups show an increasing fraction of diffuse gas component with increasing deviation from \HI\ normalcy. Towards the later stages of evolution, groups continue to show strong \HI\ deficiency as the diffuse \HI\ component is transformed into the ionized IGM and cold neutral gas is truly lost from these systems. Since the tidal interactions do not continuously deposit constant amount of neutral debris in the IGM, no correlation between the total HSB component and the total diffuse gas component is expected or observed for highly evolved systems.

\subsubsection{Dependence on Group Evolutionary Phase \label{sec:Dep_evol}}
 
 HCGs represent the full range of formative and evolutionary stages of galaxy groups.  VM2001 proposed an evolutionary sequence in HCGs based on their \HI\ morphology and distribution. According to this model, during the first phase of evolution (Phase 1), the \HI\ distribution and kinematics are relatively unperturbed, and more than 90\% of the \HI\ mass is found in the disks of individual galaxies with the remaining gas found in incipient tidal tails. In the second stage of evolution (Phase 2), groups still retain a significant amount of \HI\ in the disks, but 30\% to 60\% of the total \HI\ mass forms tidal features. In the final phase, most of the \HI\ is stripped from the disks and is found in tails produced by tidal interactions or not detected at all (Phase 3a). A slightly less common phase is the Phase 3b where the \HI\ forms a single large cloud containing all galaxies with a continuous velocity gradient and a single peak line profile.
 
 A statistical comparison of the excess gas mass fraction, $f$, for our sample based on their evolutionary stage is shown in Table \ref{evolution}.   For this analysis we do not distinguish between Phases 3a and 3b. The mean (median) excess gas mass fraction for groups in the evolutionary phases 1, 2, and 3 are 21~\% (21~\%), 30~\% (17~\%), and 48~\% (54~\%), respectively. Dispersion in excess \HI\ fractionamong the members in a phase may be understood in terms of a continuous evolutionary sequence in HCGs where the three phases are broadly (and somewhat arbitrarily) divided to quantify the gradual change. It is apparent that in the initial phases of evolution there is much less diffuse neutral IGM than in the last phase, where the neutral IGM is well developed. For instance, all the groups in the Phase~3, with the exception of HCG~93, have a considerably large fraction of diffuse \HI\ . HCG~93 may be a special case where the addition of a new \HI\ rich member  (93B) reduces the fractional significance of the diffuse neutral \HI\ in the IGM.  
 
\begin{figure}
\figurenum{12}
\includegraphics[scale=0.7,angle=0]{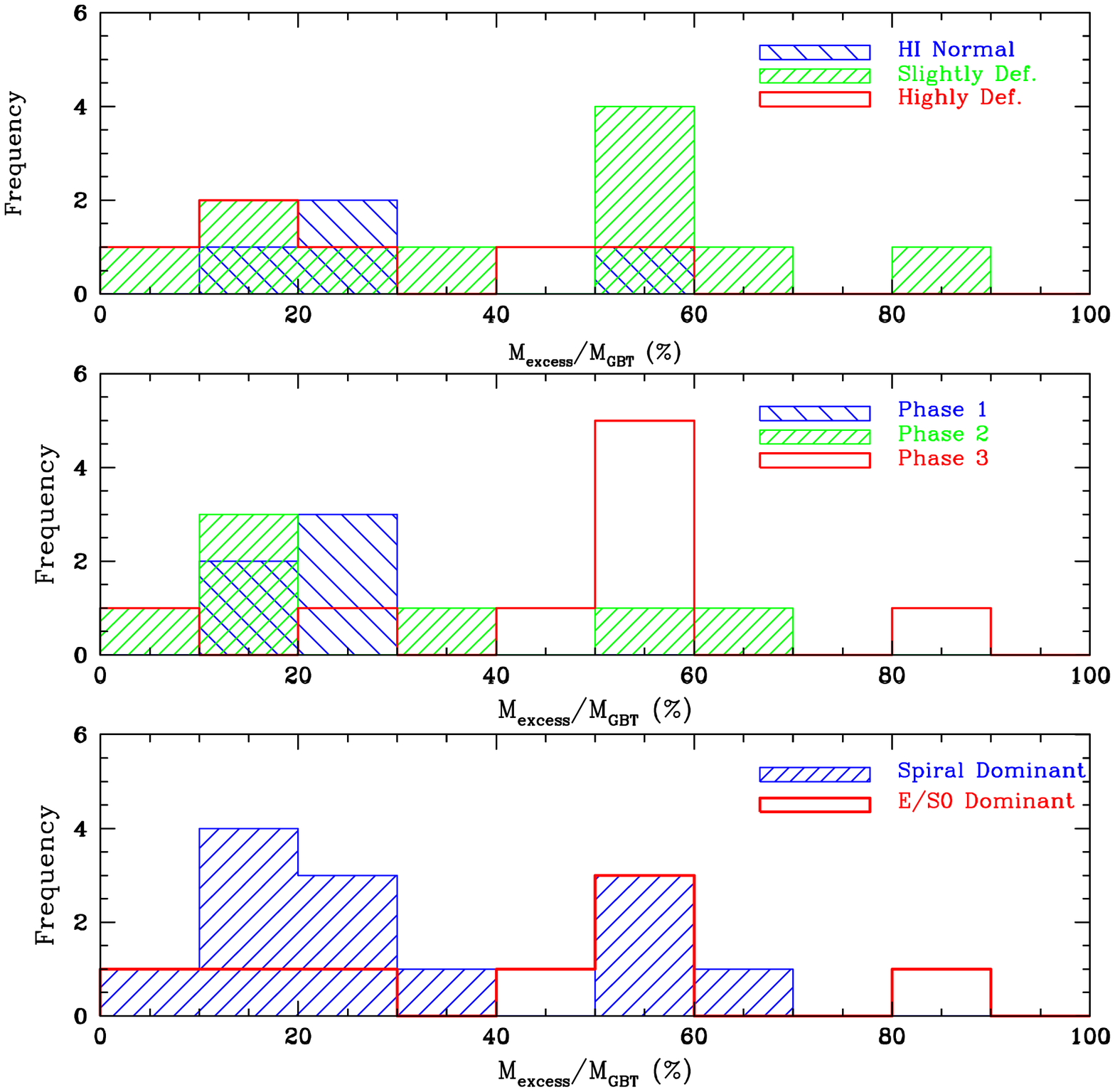}
{\caption{Comparisons of the diffuse gas mass fraction for our complete sample (except HCG~40) in terms of \HI\  deficiency (top), group evolutionary phase (middle), and the dominant group galaxy type (bottom).  The bimodal distribution of the diffuse gas mass fraction is clearly seen in these histograms.
\label{3histograms} }}
\end{figure}

A graphical comparison of excess gas mass fraction shown in the middle panel of Figure~\ref{3histograms} demonstrates the clustering of Phases 1 and 2 towards lower values and Phase 3 HCGs towards higher values of the excess gas mass fraction. As discussed in \S~\ref{sec:groupdiameter}, this clustering may indicate a rapid transition for groups dominated by \HI\ mostly in the ISM of individual galaxies to a phase where most of the \HI\ is found in the IGM. We emphasize that the group evolutionary phase is determined from the \HI\ {\it morphology} as mapped by the VLA while the excess gas mass fraction comes from the {\it integral total line flux} in the GBT spectra. One should not {\it a priori} expect any correlation between these two completely independent parameters, except for an underlying physical mechanism that governs them both. Groups with $f~<~40\%$ in Figure~\ref{3histograms} consist mainly of HCGs in Phases 1 \& 2 while those with$f~>~40\%$) are almost entirely from Phase 3. In the group evolutionary scenario, this trend indicates that the evolutionary Phases 1 \& 2 are much closer in timescale and in terms of the neutral IGM build-up process, and the evolutionary phase 3 is indeed more distinct.  Again, the small sample size and a large dispersion within each sub-sample is an important limitation.

\subsubsection{Central Dominant Galaxy and Diffuse Intragroup \HI  \label{sec:Dep_domgal}}

In the cold dark matter (CDM) paradigm, dark matter halo properties such as mass and angular momentum determine the galaxy properties.  Specifically, for a given massive halo, the nature of the central galaxy is thought to be determined by the merger history of the halo.  By examining the dependence of the cross-correlation between galaxies and galaxy groups on group properties, \citet{yang06} found that the dark matter halo bias of galaxy groups decreases as the star formation rate of the central galaxy increases. These authors interpret this finding in terms of the halo bias dependence on halo formation time \citep{gao05}, in the sense that halos that were assembled earlier are more strongly biased with a redder central galaxy.  In other words, on an average galaxy groups with a red central dominant galaxy have formed much earlier than groups with a blue central dominant galaxy even if their present halo masses are the same. 

An excess of early type galaxies in HCGs have been reported by \citet{hick88}. \citet{sulentic00} and VM2001 suggested that some of the lenticulars and S0 galaxies were spirals that were stripped of their \HI\ disks as a result of galaxy interactions. We have examined whether the excess \HI\ content has any dependence on the Hubble type of the group central dominant galaxy.  If the earlier Hubble type of the central dominant galaxy is an indicator of an earlier formation time or equivalently an advanced stage of evolution for the group, greater excess \HI\ content is expected, provided the diffuse intragroup gas can survive long enough compared to the mass assembly time.  We find the mean and median for the excess gas mass fraction is 32~\% and 23~\% for the spiral dominated HCGs and the mean and the median for non-spiral dominated HCGs are 42~\% and 49~\%, respectively (see Table~\ref{hubbletype}). The graphical comparison shown on the bottom panel in Figure~\ref{3histograms} shows the distribution of the spiral and non-spiral dominated groups. Most of the spiral dominated groups have lower excess gas mass fractions where as the non-spiral dominated groups cover the entire range. However, there is a some overlap between the two sub-samples. Based on the Analysis of Variance test\footnote{An extension of Students t-test to more than two samples.}, one can conclude that these two sub-samples are drawn from different distributions with more than 75\% confidence. The weakness of the distinction between the two classes may indicate that the IGM evolution timescale is shorter than the mass assembly time \citep[ $\sim$ 4 Gyrs see][]{barnes89}.  

Another interesting observation based on the VLA imaging of the non-spiral dominated systems \citep{bor}  is that most of the \HI\ mass is contributed by a single \HI\ rich spiral galaxy, except for in the case of HCG~97 and HCG~37, where no \HI\ was observed in the VLA imaging. In all the cases, the \HI\ rich spiral galaxy is relatively undisturbed and shows no sign of tidal interaction. Interactions in galaxies can be seen in terms of disturbed \HI\ morphology even when they are well away from the center of the cluster \citep{chung07}. The most plausible explanation for a galaxy with an undisturbed \HI\ distribution in a compact group is that it is new to the group and has not yet passed through the group center. So for all these systems, the VLA detected \HI\ includes both the tidally stripped HSB gas as well as the \HI\ mass of the undisturbed galaxy, thus diluting the excess gas fraction's ability to trace evolution of tidal debris. This may have caused the excess gas mass fraction in the non-spirals to span a much wider range as seen in Figure~\ref{3histograms}.

\section{CONCLUSION  \label{sec:Concl}}

A total of 26 HCGs have been observed in the 21~cm \HI\ emission with the 100-meter GBT. We detected \HI\ in almost all the compact groups with the exception of HCG~35. The derived \HI\ masses ranged between $0.5-28.5~\times~10^{9}$~M$_\odot $.  In HCG~40 and HCG~48, we detected 3 to 5 times the previously detected \HI\ (compiled by VM2001) reducing their \HI\ deficiencies significantly.

For our distance limited complete sample of 22 HCGs, we compared our GBT spectra with the corresponding VLA spectra for the region covered by the GBT beam.  Strong differences between the two spectra can be seen for some groups while others showed very little or no differences at all. The total flux recovered by the GBT was always equal to or larger than that by the VLA (with the exception of HCG~40) as expected from correct calibration. Although the GBT had larger line flux, the shape of the two \HI\ profiles for most of the groups are quite similar. There are fine differences in the form of broad faint emission and wing-like emission features that were only detected by the GBT. We have taken extra steps to ensure that these differences are not a result of calibration errors. Perfect match between the VLA and the GBT spectra of HCG~16, 93, and others supports our claim of calibration accuracy. By comparing the instrumental properties and data analysis techniques for the two instruments, we infer that this excess \HI\ is in the form of a faint diffuse extended neutral IGM. In a few groups, spatial filtration of extended \HI\ structures of angular size $>$12-15$^{\prime}$ by the VLA is also plausible.

Owing to their high spatial density and low velocity dispersion, tidal interactions are ubiquitous in HCGs. This results in a substantial deposition of cold gas from the ISM of individual galaxies to the hot IGM. The detection of neutral IGM in HCGs confirms that the cold tidally stripped filaments and clouds survive the conductive heating and expansion for longer than a typical group dynamical timescale ($>$~400~Myrs). Based on our calculations, we constrain the radius of an \HI\ cloud to be $\ge$ 200~pc to survive over the dynamical time of the group. Furthermore, the presence of neutral medium also constrains the temperature of the IGM to be lower than 4~$\times$~10$^6$~K. We also conclude that the diffuse neutral gas content should increase with interactions between member galaxies and hence is related to the age of the group.

We characterize the excess gas content in terms of the total \HI\ content in the galaxy-group. The excess gas mass fraction, $f$= M$_{excess}$/M$_{GBT}$,  varies from almost none to 81\%. We discovered a much larger fraction of excess gas (median=51~\%) in slightly \HI\ deficient groups as compared to \HI\ normal groups ( median=26~\%) or highly \HI\ deficient groups (median=20~\%). For groups with \HI\ deficiency less than 0.4 ($\equiv 2~\sigma$ of deficiency estimation), the excess gas fraction increases with increasing deviation from \HI\ normalcy. The opposite trend is seen in the case of high-surface brightness \HI\ detected by the VLA. We conclude that the diffuse neutral gas represents an intermediate stage in the evolution of tidal debris in the hot IGM between HSB \HI\ and ionized hydrogen.

We also examined the excess gas content in terms of group evolution based on the scenario proposed by VM2001. According to this evolutionary scenario, the tidal interactions increasingly deposit ISM from individual galaxies into the IGM as a group evolves and hence based on the \HI\ morphology one can predict the stage of evolution. The tidal debris, which survives the hot IGM, will evolve into the diffuse neutral IGM and is expected to be a substantial fraction of the total \HI\ content. Thus the excess \HI\ content is expected to be correlated with the evolutionary phase. From our analysis we found that this is indeed the case. The groups in the final phase of evolution had a larger amount of excess \HI\ (average of 48\%) as compared to groups in phase 1 (21\%) and phase 2 (30\%).  Furthermore, we find a clustering in the distribution with only two systems showing the transition phase. This suggests that the transition is rapid and hints at the presence of a physical threshold like a characteristic age or an interaction threshold aiding the transition. We also examined the dependence of excess gas content on the Hubble type of the central dominant galaxy in the group. While spiral dominated groups tend to be have a lower excess gas mass fraction, the non-spiral dominated groups have a wide range of values covering the entire range.

The existence of cold neutral IGM in HCGs offers a strong support for the survival of the tidally stripped ISM from individual galaxies. A significant amount of \HI\ is tidally stripped from the galaxies as they interact with the group potential and other members and ultimately ends up in the diffuse IGM. Thus, the excess gas content gives valuable information on the evolutionary history of groups. In addition, the neutral gas distribution also holds the key to understanding current and future star formation in groups, which in-turn can considerably change the morphology and spectral properties of the galaxies. Cooler regions of the IGM might form clumps with sufficient densities to fall back into the galaxies, and providing them with fresh fuel for star-formation. The presence of neutral IGM also raises the possibility of non-disk star formation. Therefore, this discovery opens up a host of unanswered questions and we hope that a GBT-VLA combined map will provide clues to some of these questions. Sensitive X-ray masses and maps along with a UV absorption study of the IGM would be the next crucial step in bringing out the full picture.

\acknowledgments

This manuscript has benefited from insightful discussions and comments by J. Hibbard, J. Kwan, F. Lockman, A. Minter, H. Mo, T. Ponman, J. Rasmussen, T. Tripp, J. van Gorkom, and K. C. Xu. 
The authors are grateful to the observatory staff at the GBT who made these observations possible.
Support for this work was provided by the NSF through award GSSP06-0002 from the NRAO. LVM is partially supported by DGI Grant AYA 2005-07516-C02-01 and Junta de Andaluca (Spain). This research has made use of the NASA/IPAC Extragalactic Database (NED) operated by the Jet Propulsion Laboratory, California Institute of Technology, under contract with the National Aeronautics and Space Administration.

{\it Facilities:} \facility{GBT ()}, \facility{VLA ()}

\clearpage

{\rotate
\begin{figure}
\figurenum{1}
\epsscale{2}
\includegraphics[angle=-90,scale=0.65]{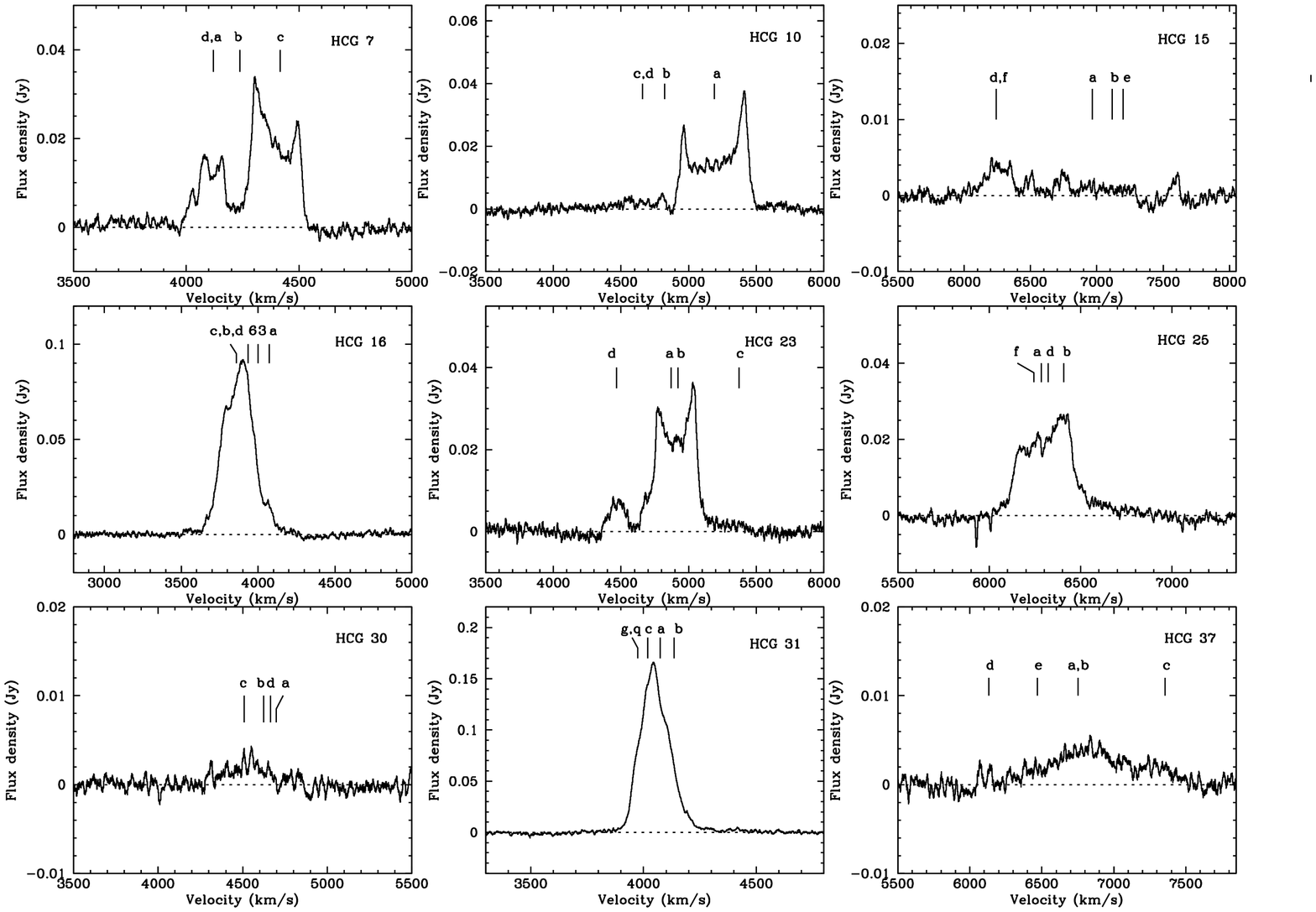}
{\caption{\HI\ profile of the 26 Hickson Compact Groups observed with the 100 meter diameter GBT.  
The redshifts of individual members are marked with short vertical lines on the spectra. \label{spectra}}}
\end{figure}

\clearpage

\begin{figure}
\figurenum{1}
\includegraphics[angle=-90,scale=0.65]{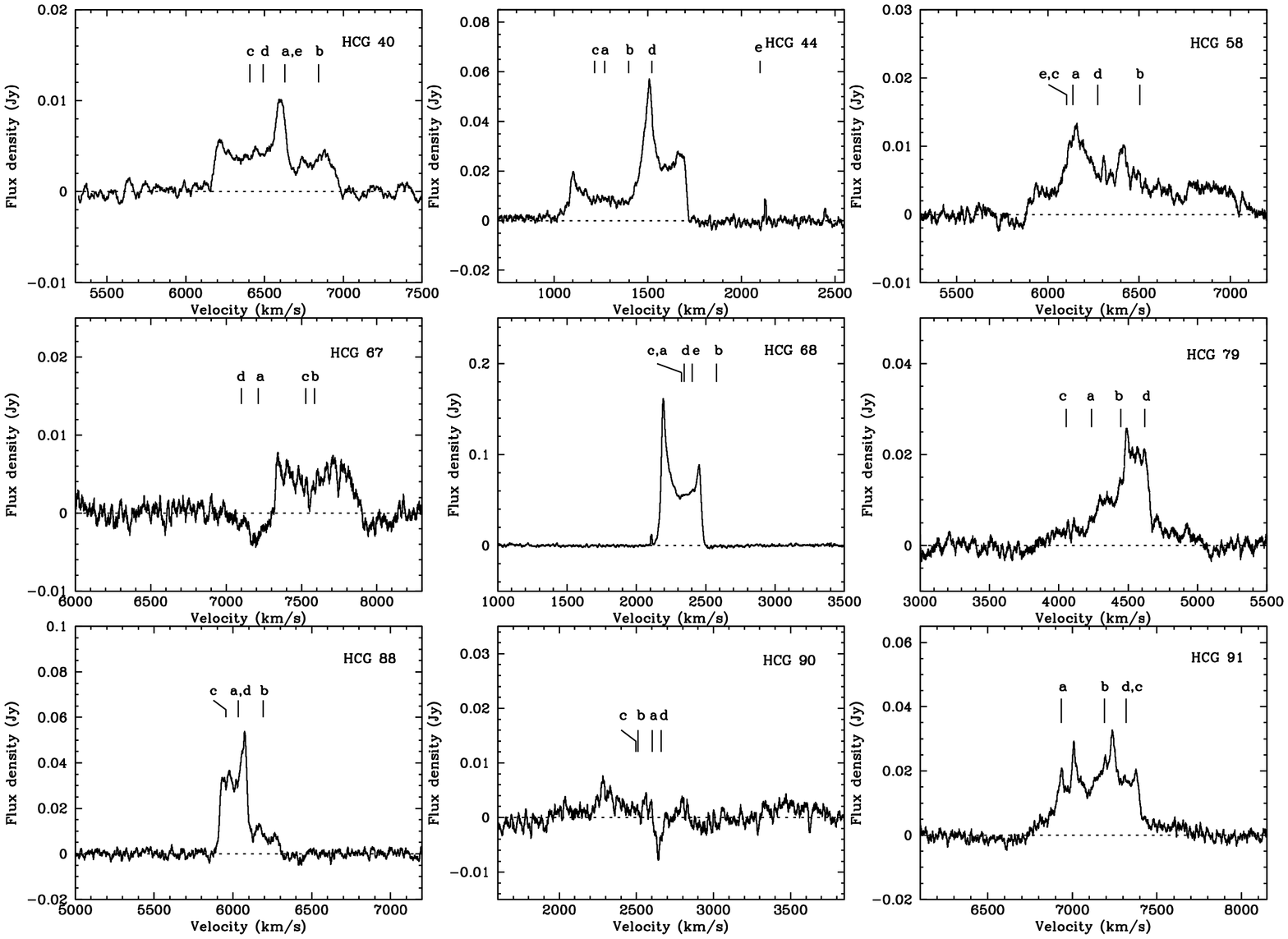}
{\caption{continued. }}

\end{figure}

\clearpage

\begin{figure}
\figurenum{1}
\includegraphics[angle=-90,scale=0.65]{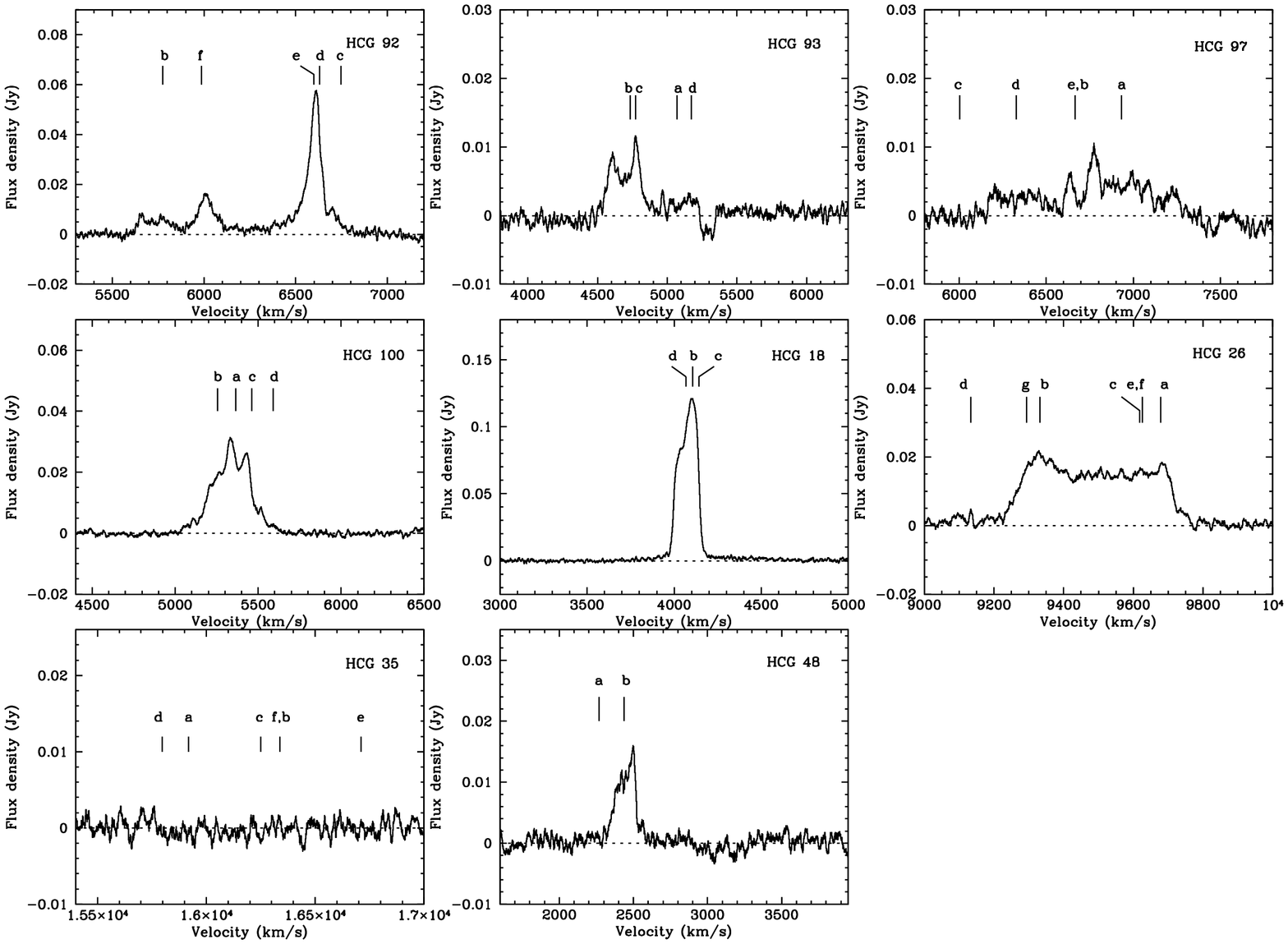}
{\caption{continued. }}

\end{figure}
}

\clearpage

\clearpage

\clearpage

 \begin{figure}
 \figurenum{4}
\begin{tabular}{c c}
\includegraphics[scale=0.35,angle=-0]{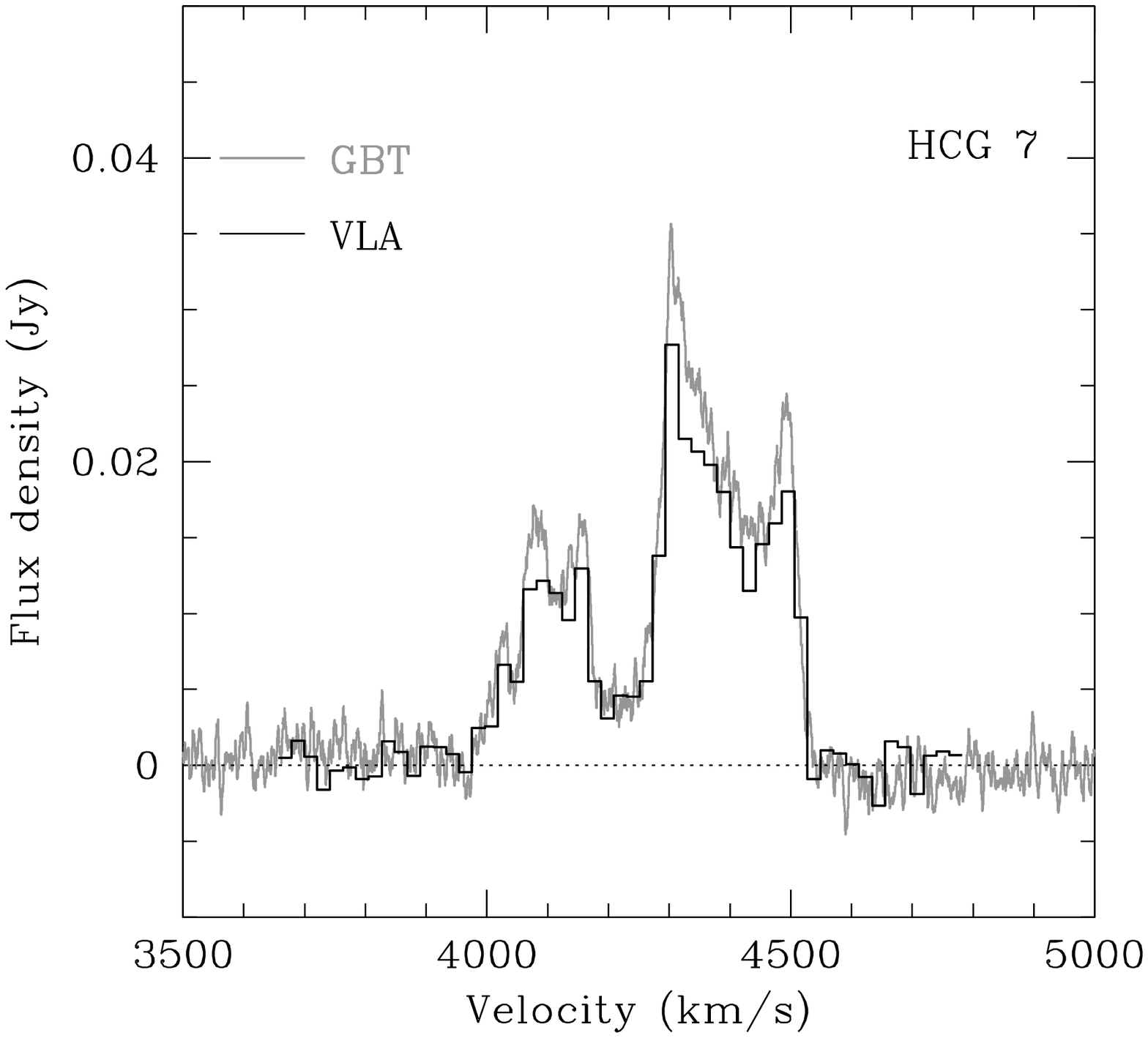} & \includegraphics[scale=0.35,angle=-0]{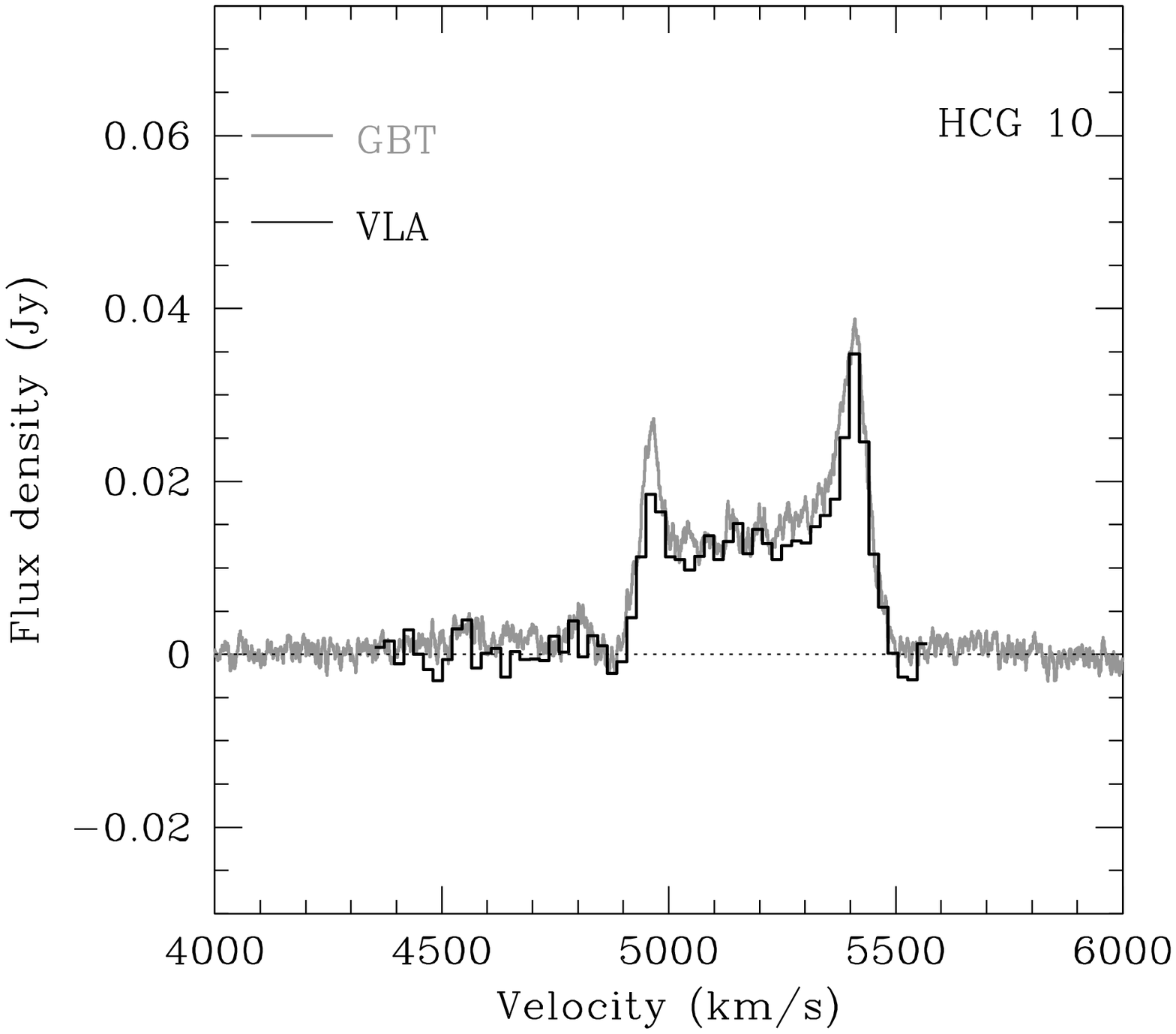} \\
\includegraphics[scale=0.35,angle=-0]{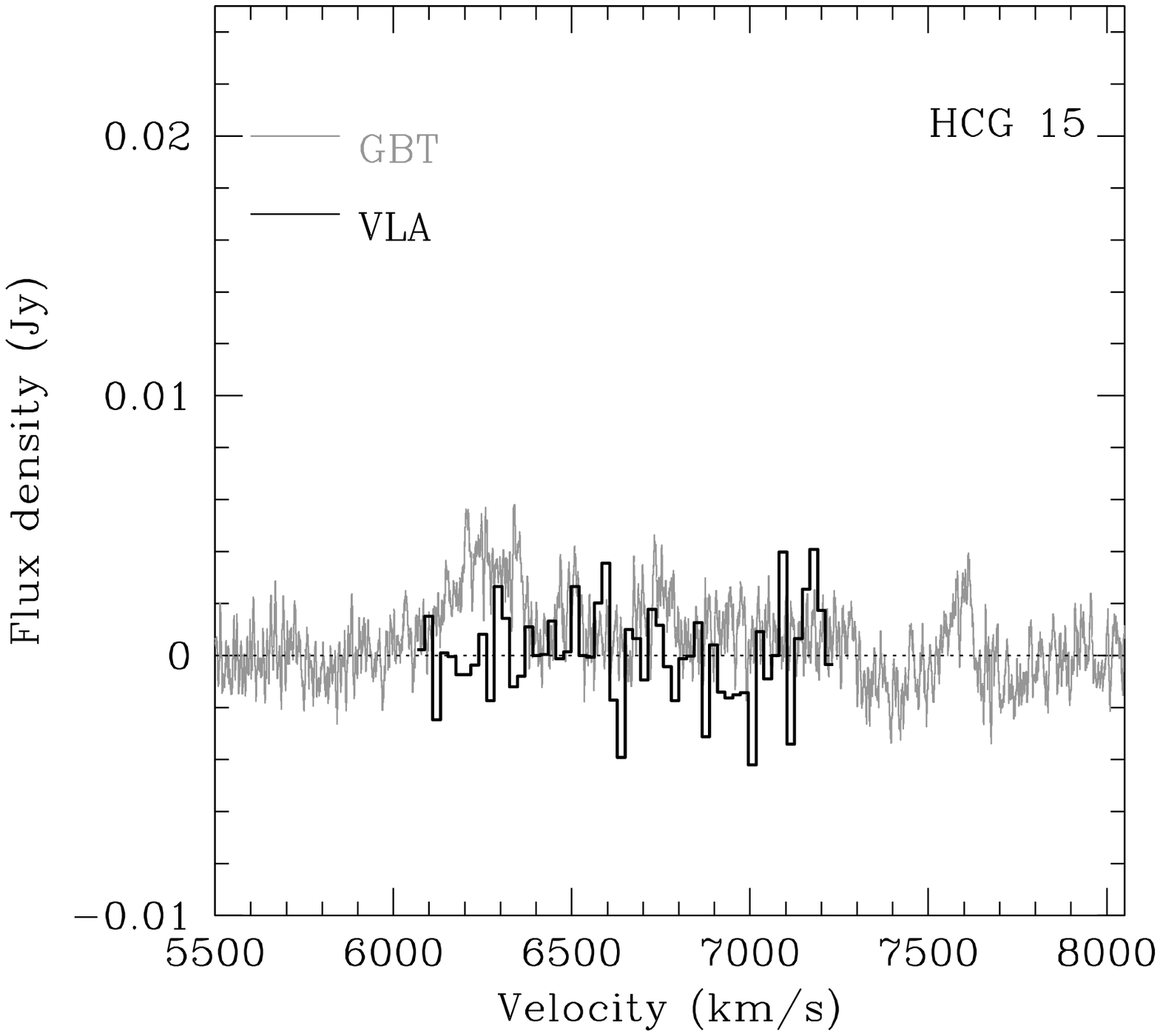} & \includegraphics[scale=0.35,angle=-0]{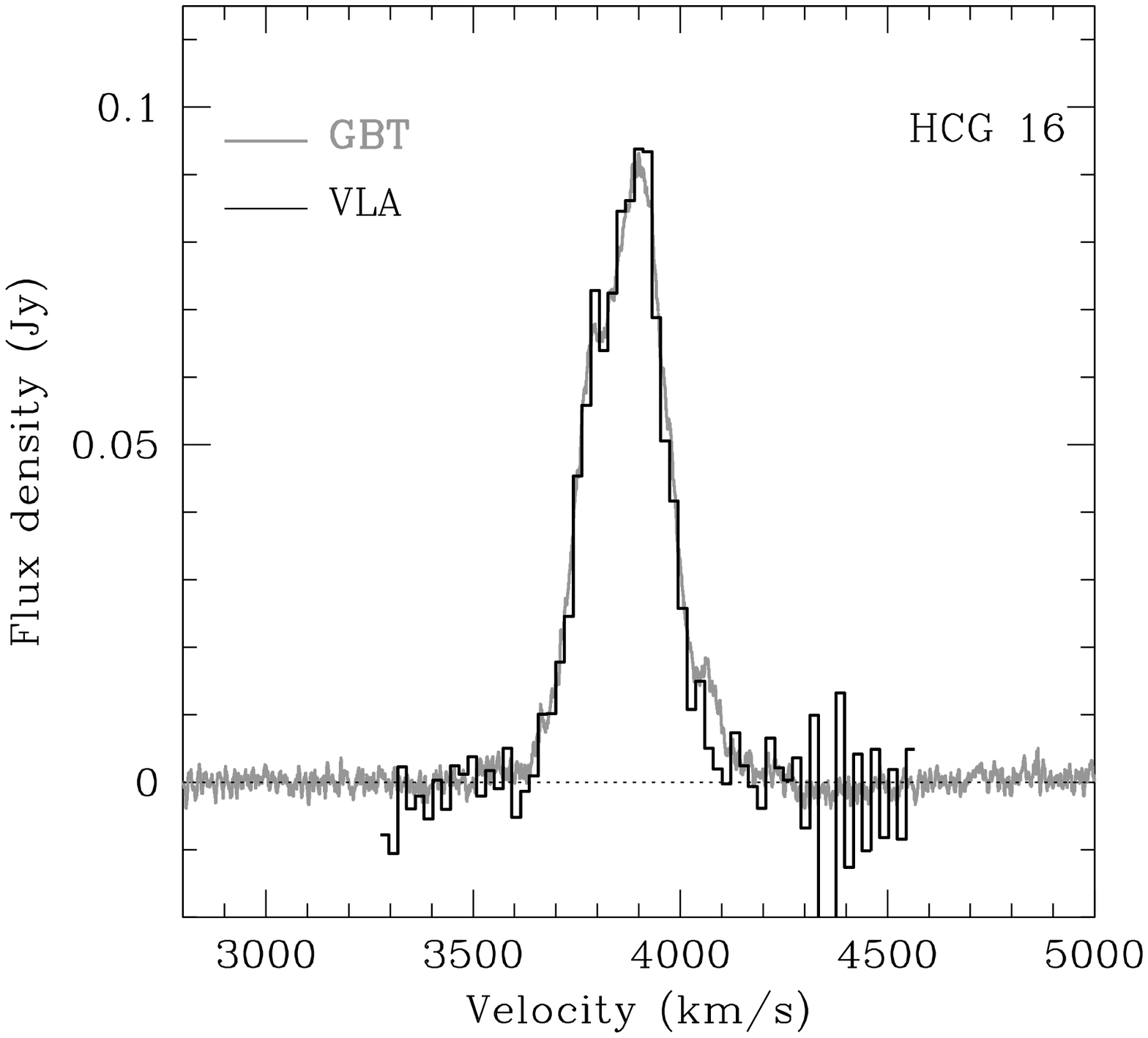} \\
\includegraphics[scale=0.35,angle=-0]{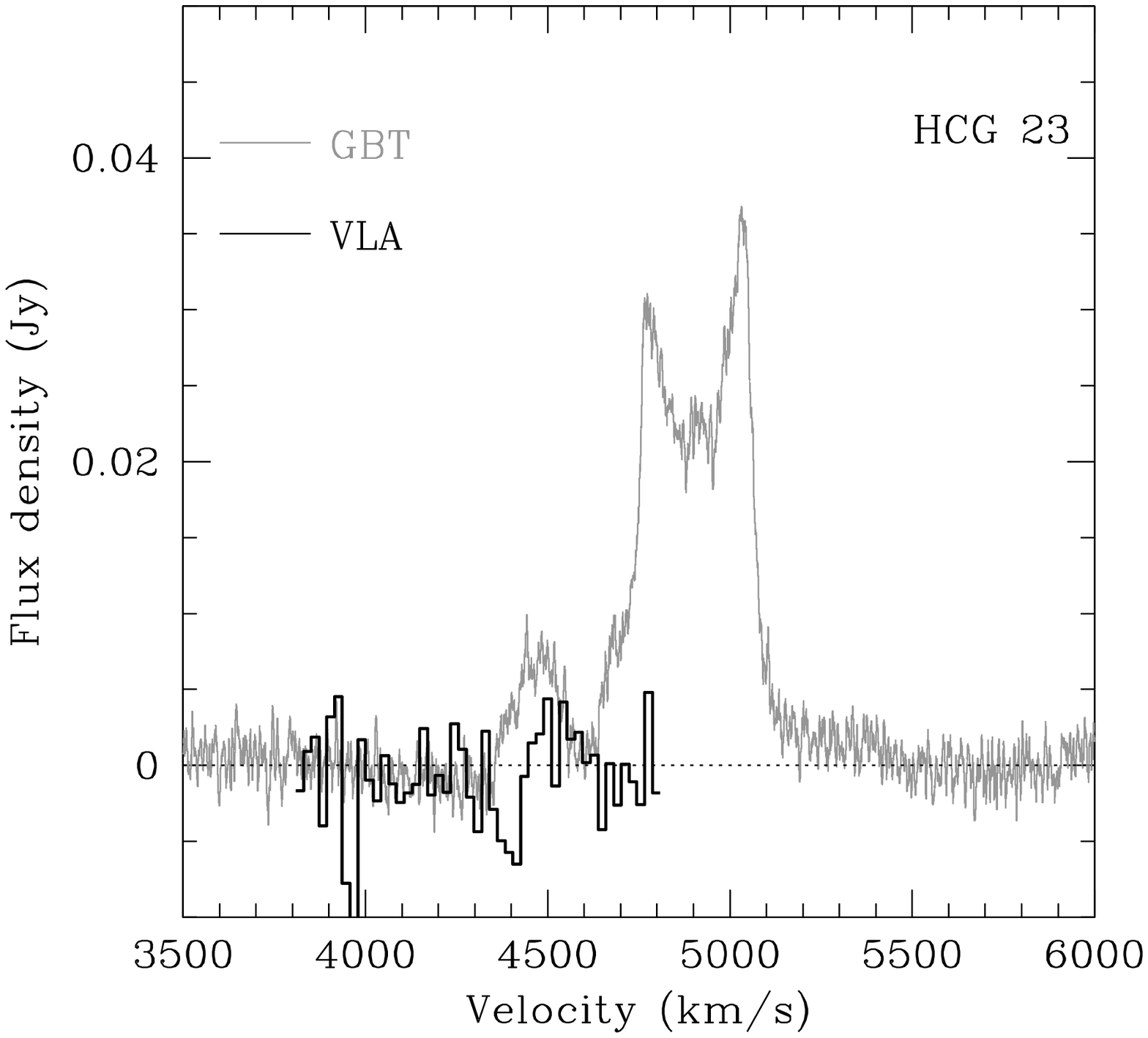} & \includegraphics[scale=0.35,angle=-0]{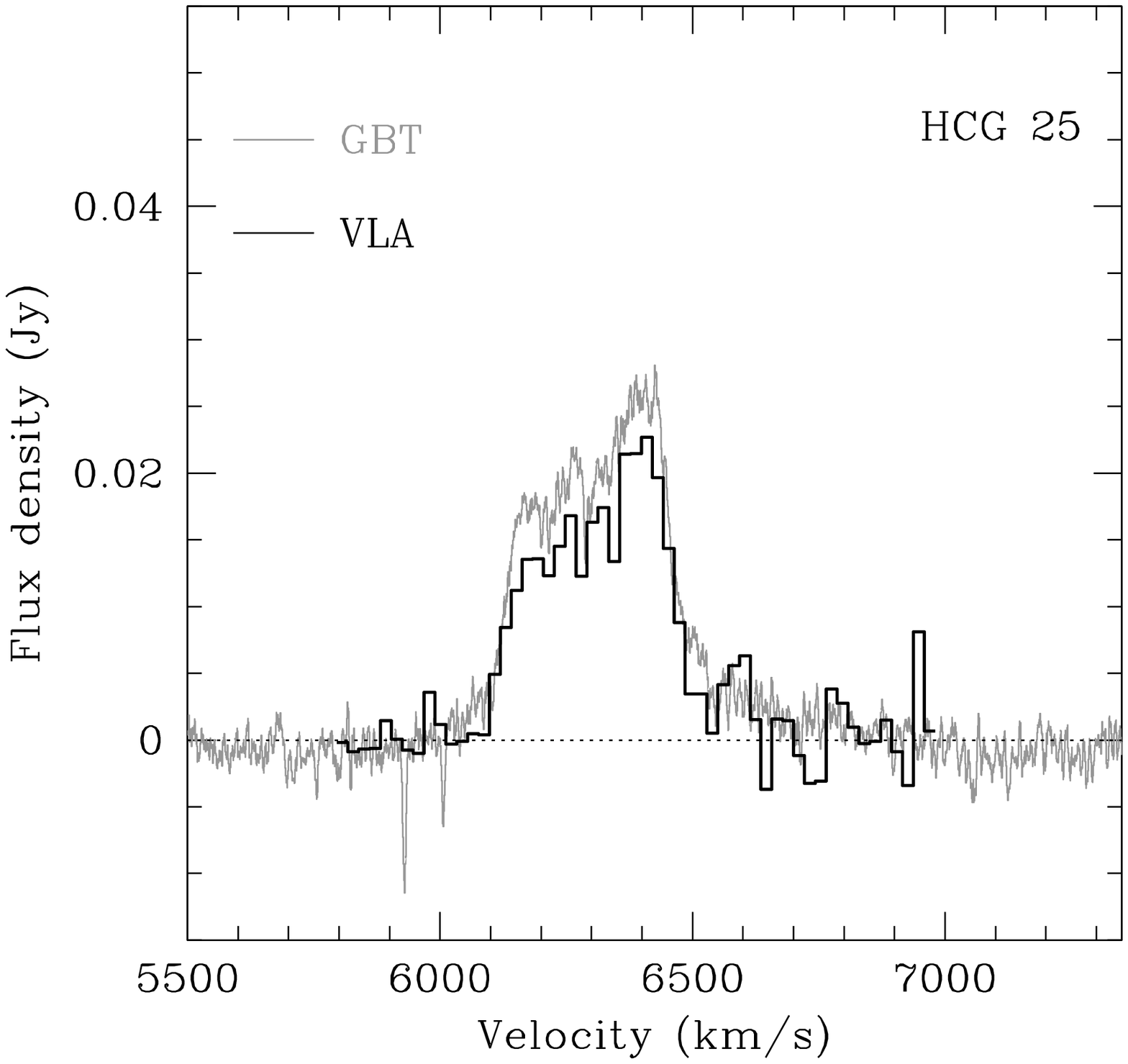} \\
\end{tabular}
{\caption{Comparison of the GBT spectrum (gray line) with VLA spectrum (dark line) for our complete sample of 22 groups as well as two of the addition sources- HCG~26 and 48 in the last two panels. The VLA spectra presented here were obtained by superimposing the GBT beam on the VLA image (see \S~\ref{sec:comparison}). Details of the VLA data are summarized in Table~\ref{tbl-VLA}.\label{comparisons}}}
\end{figure}
\clearpage

 \begin{figure}
 \figurenum{4}
\begin{tabular}{c c}

\includegraphics[scale=0.35,angle=-0]{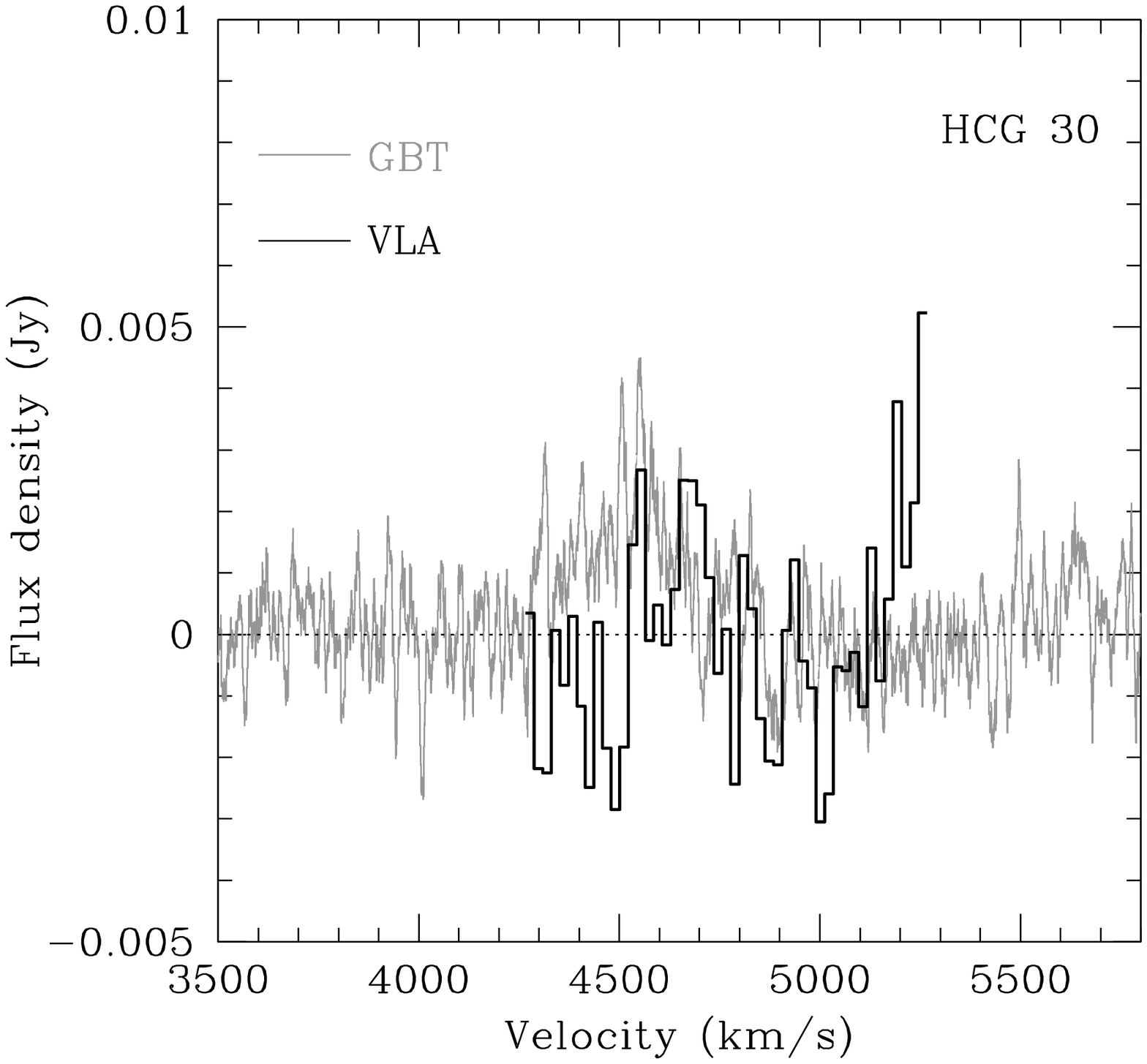} & \includegraphics[scale=0.35,angle=-0]{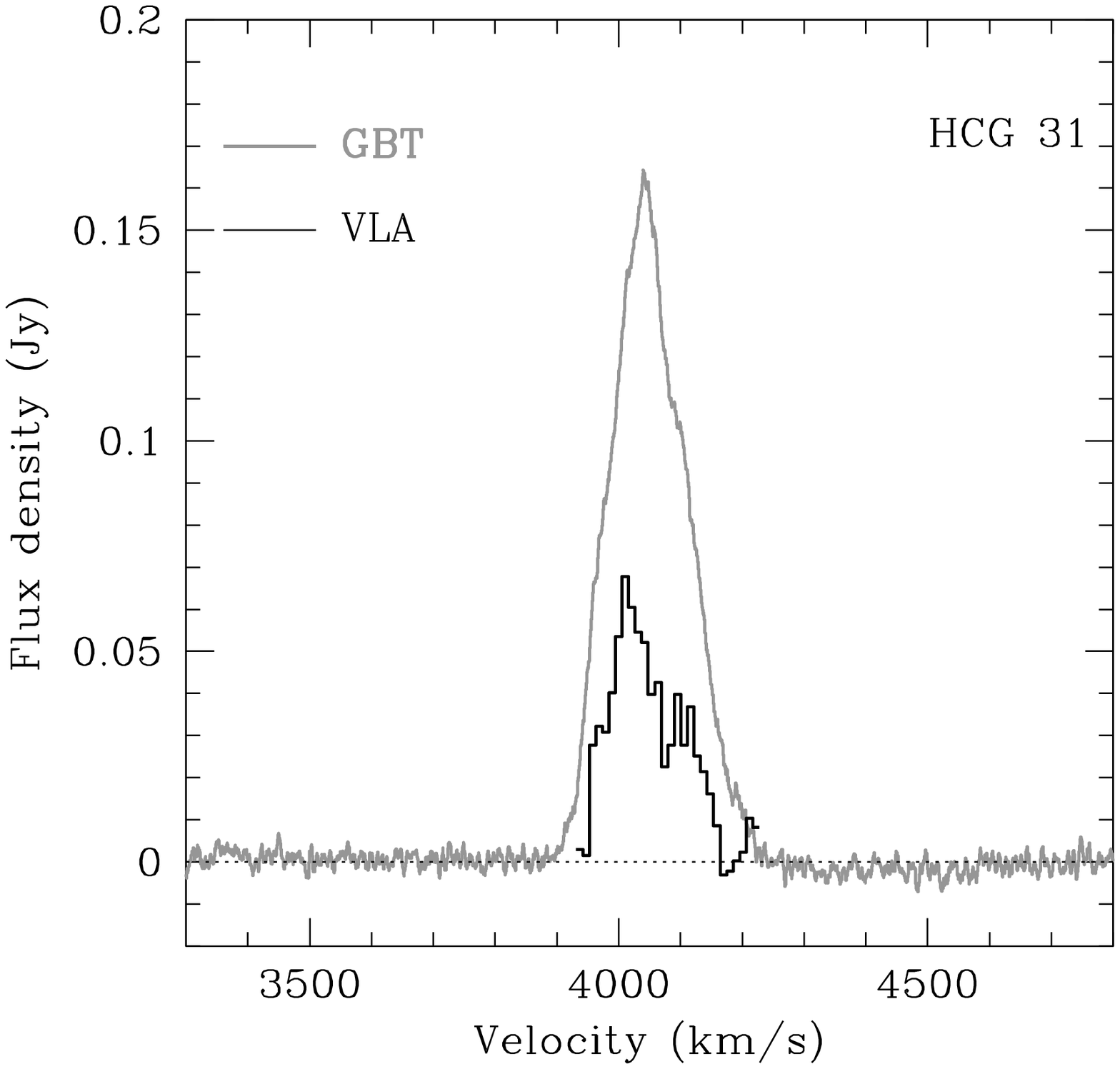} \\
\includegraphics[scale=0.35,angle=-0]{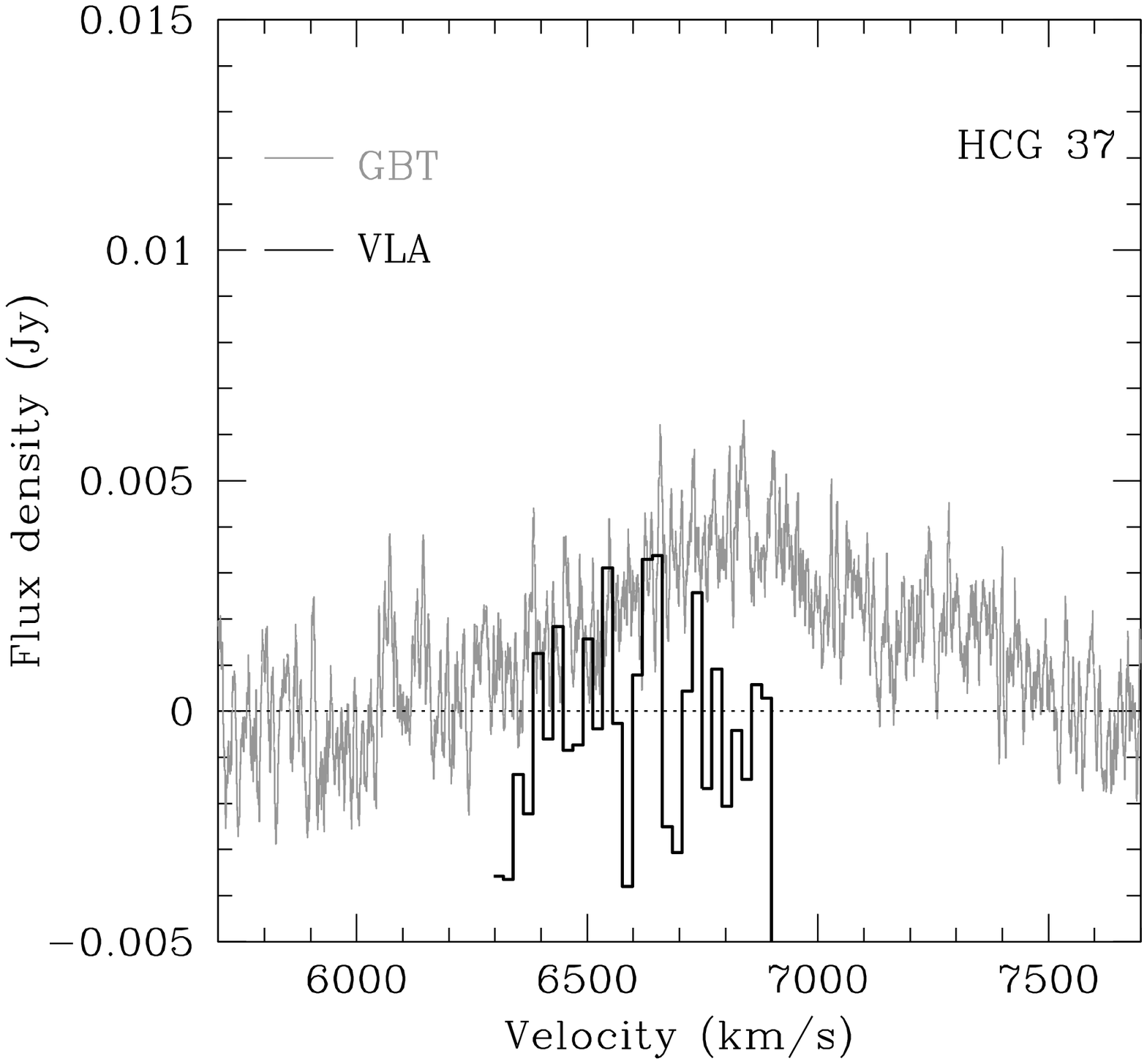} & \includegraphics[scale=0.35,angle=-0]{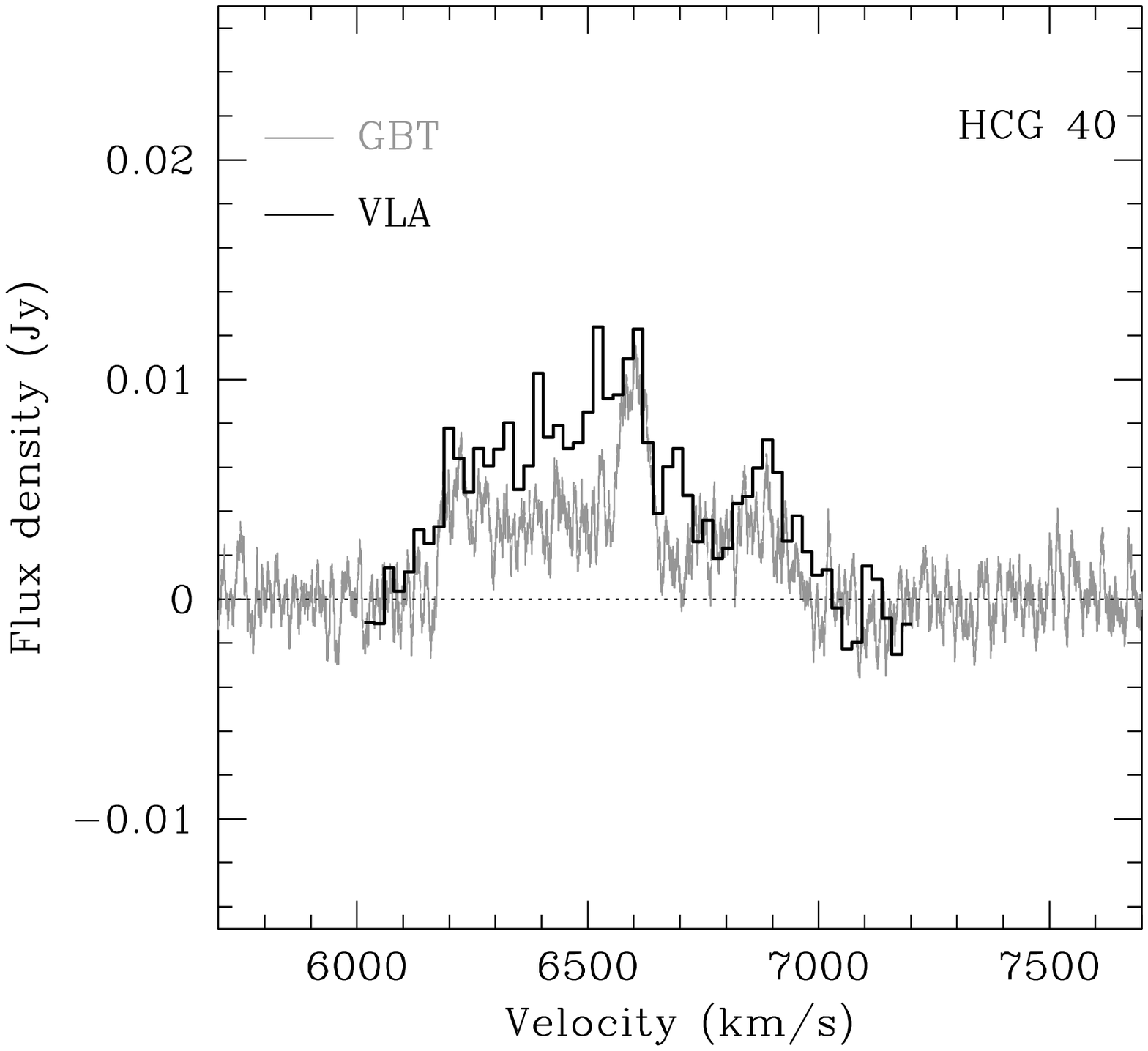} \\
\includegraphics[scale=0.35,angle=-0]{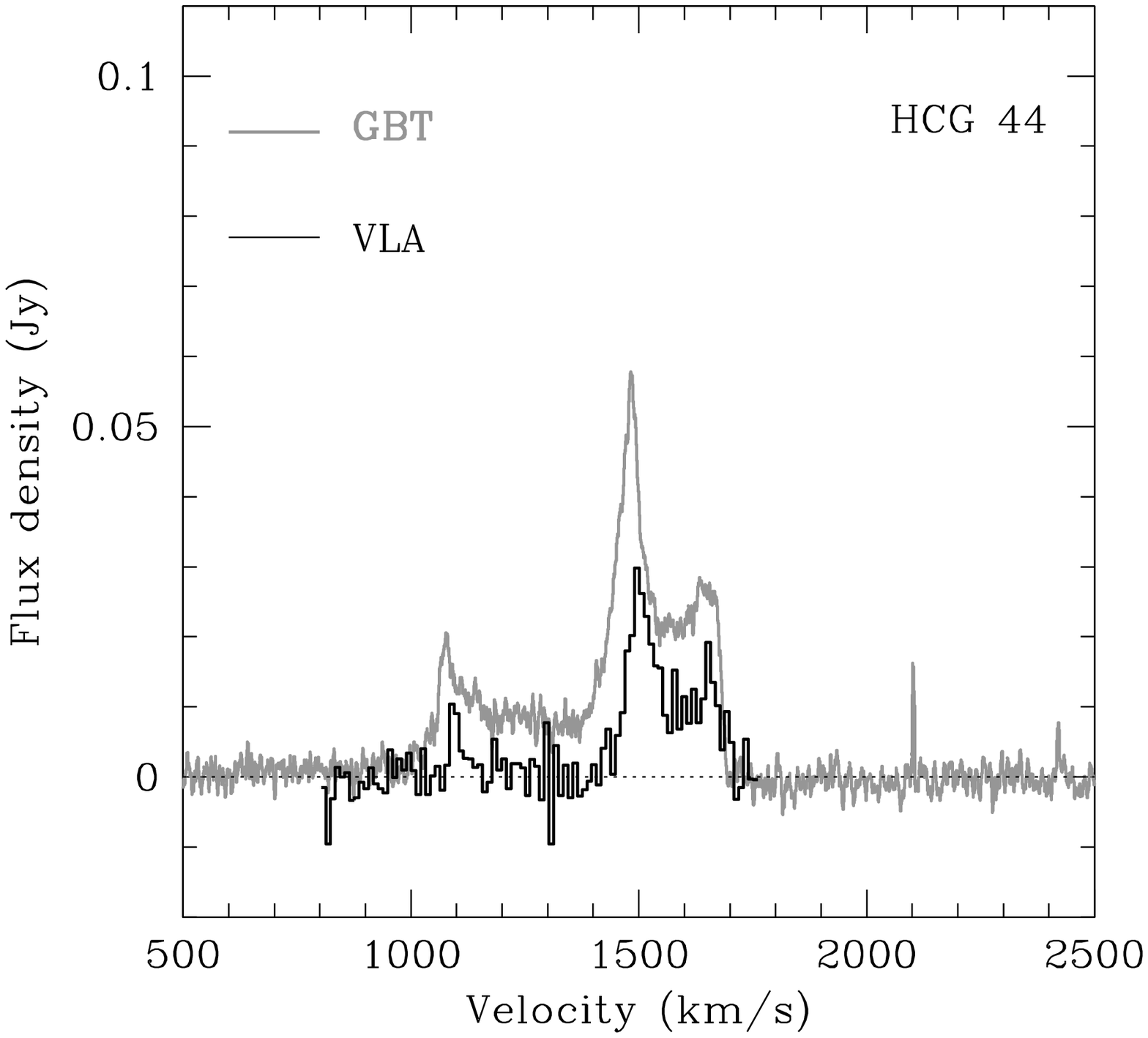} & \includegraphics[scale=0.35,angle=-0]{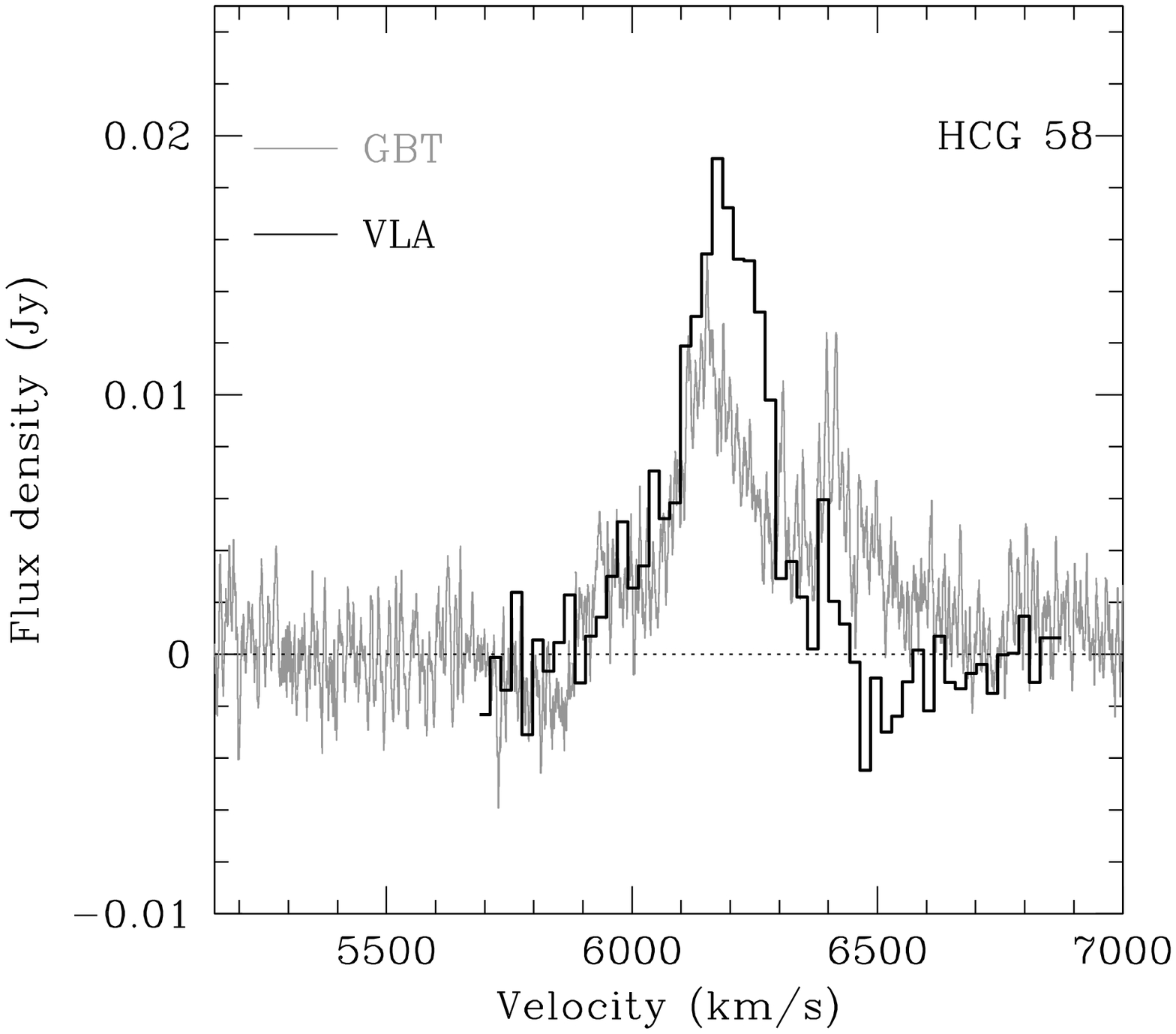} \\

\end{tabular}
{\caption{continued. }}
\end{figure}
\clearpage

 \begin{figure}
 \figurenum{4}

\begin{tabular}{c c}

\includegraphics[scale=0.35,angle=-0]{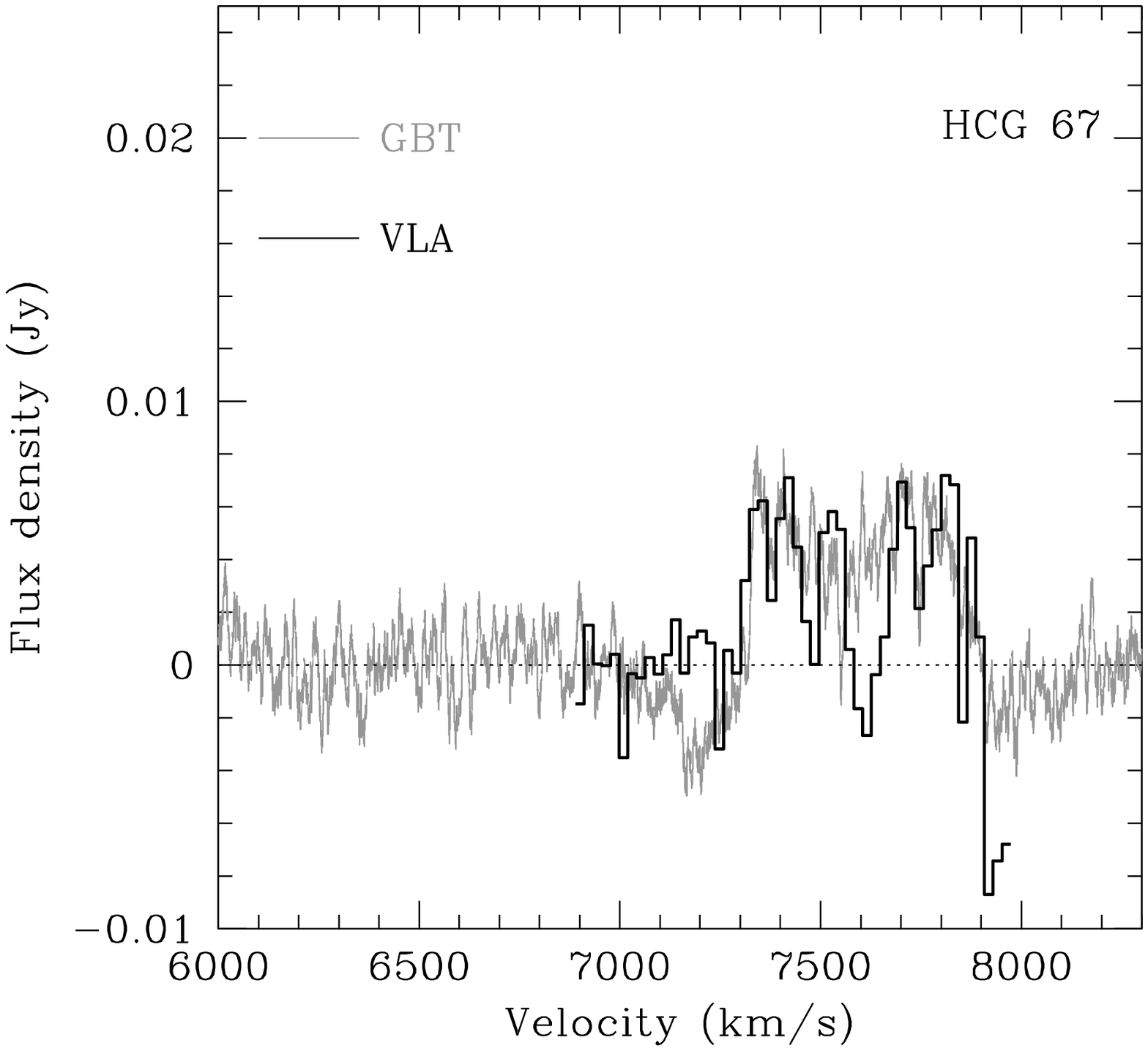} & \includegraphics[scale=0.35,angle=-0]{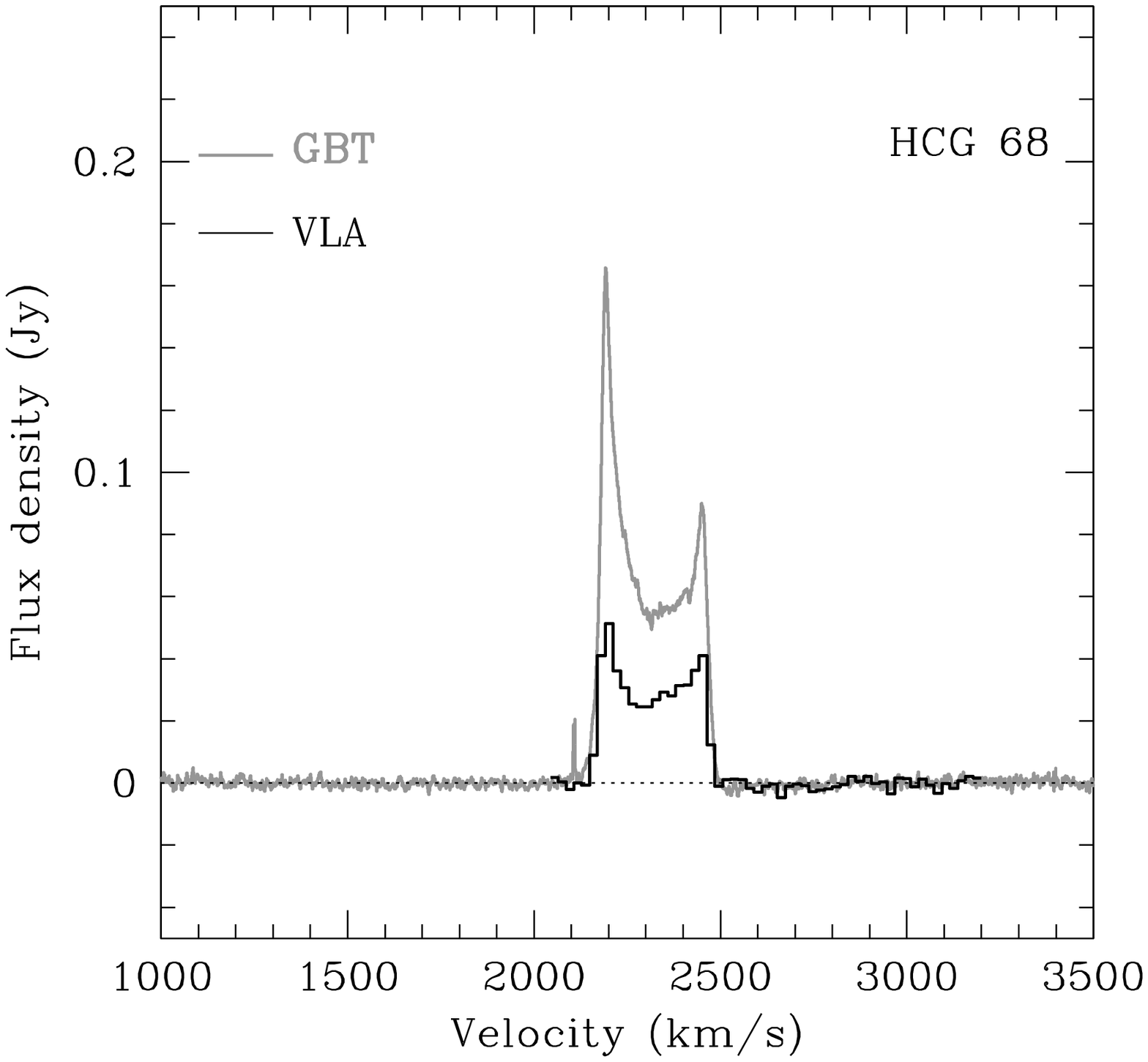} \\
\includegraphics[scale=0.35,angle=-0]{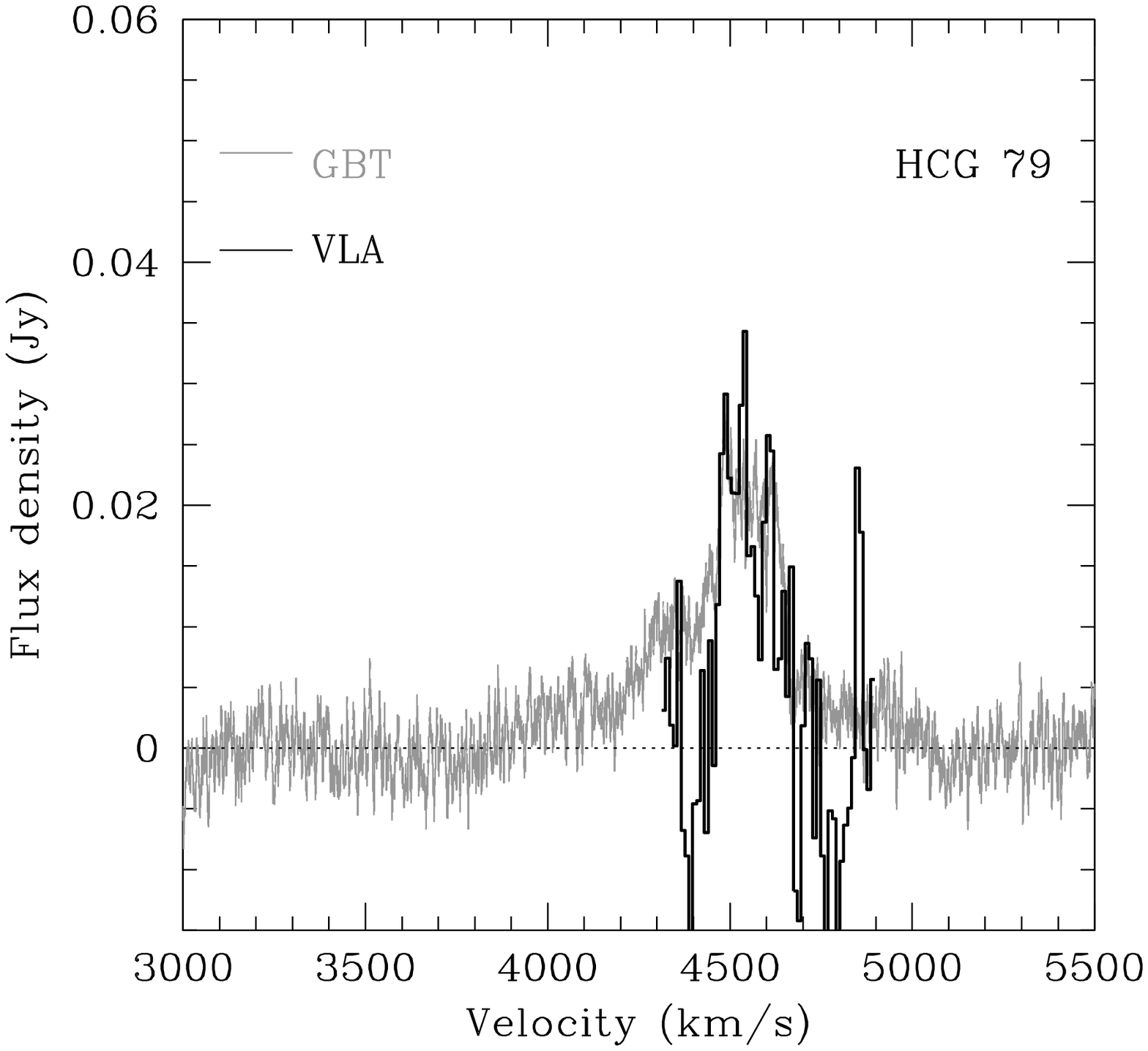} & \includegraphics[scale=0.35,angle=-0]{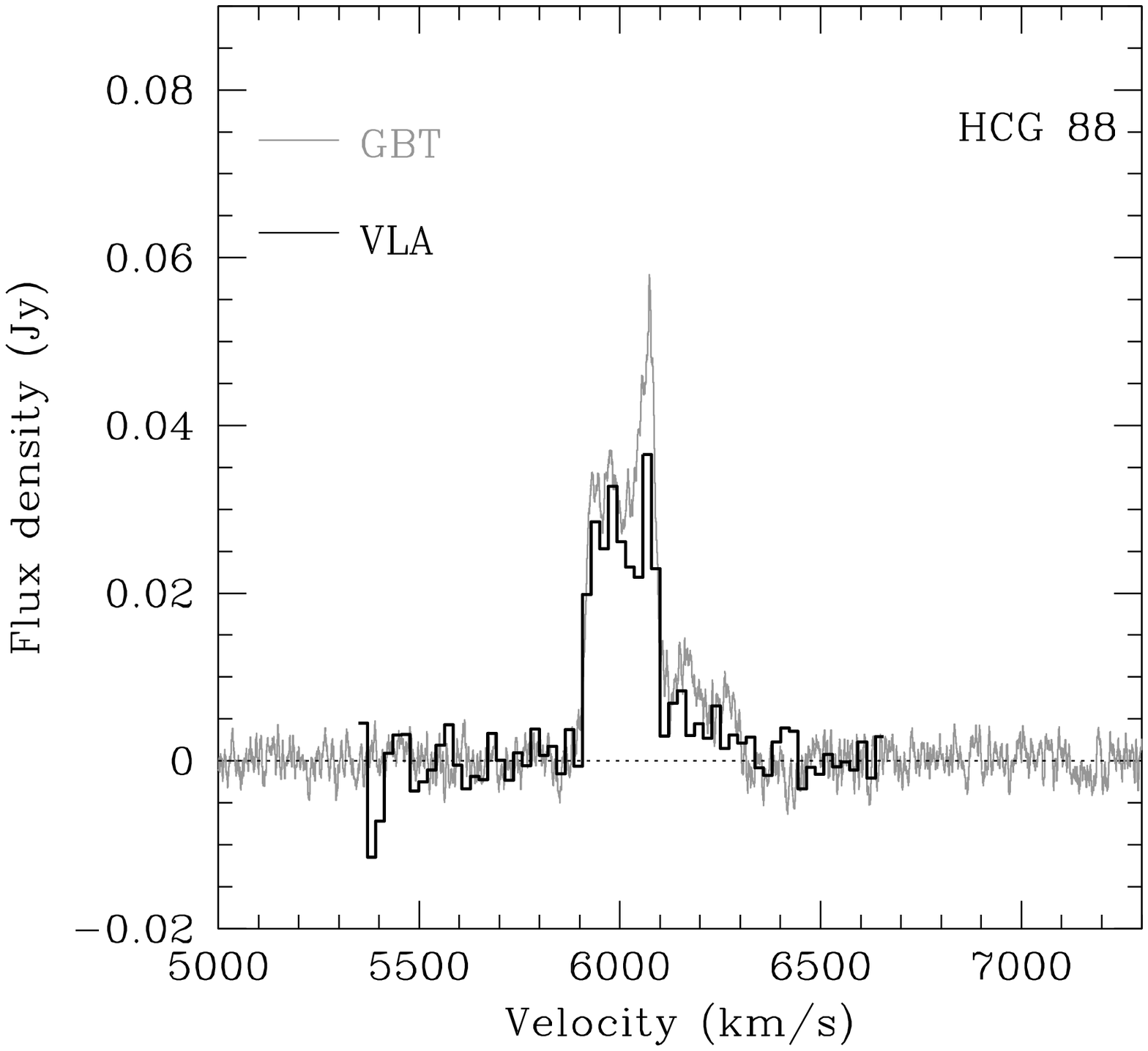} \\
\includegraphics[scale=0.35,angle=-0]{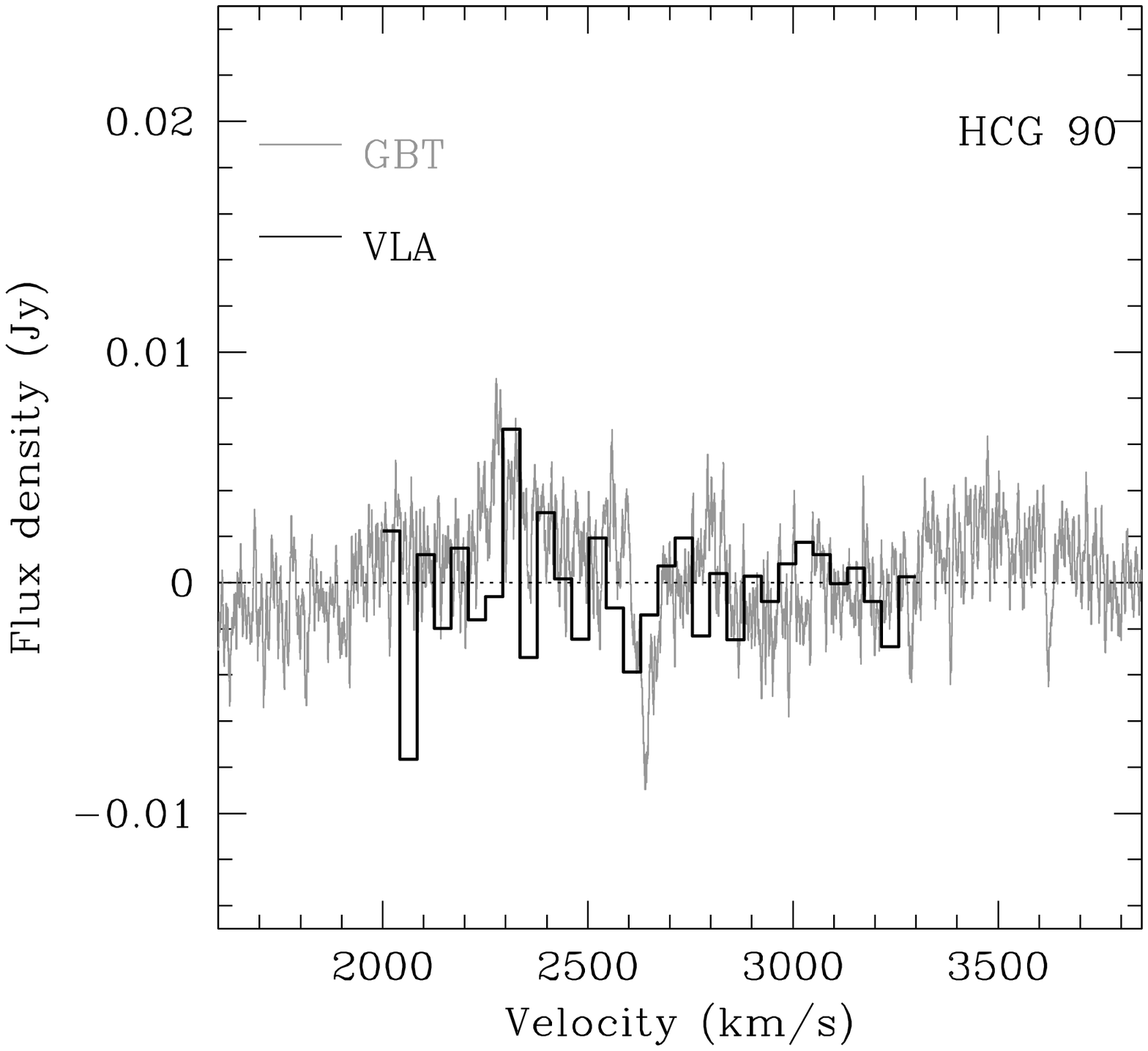} & \includegraphics[scale=0.35,angle=-0]{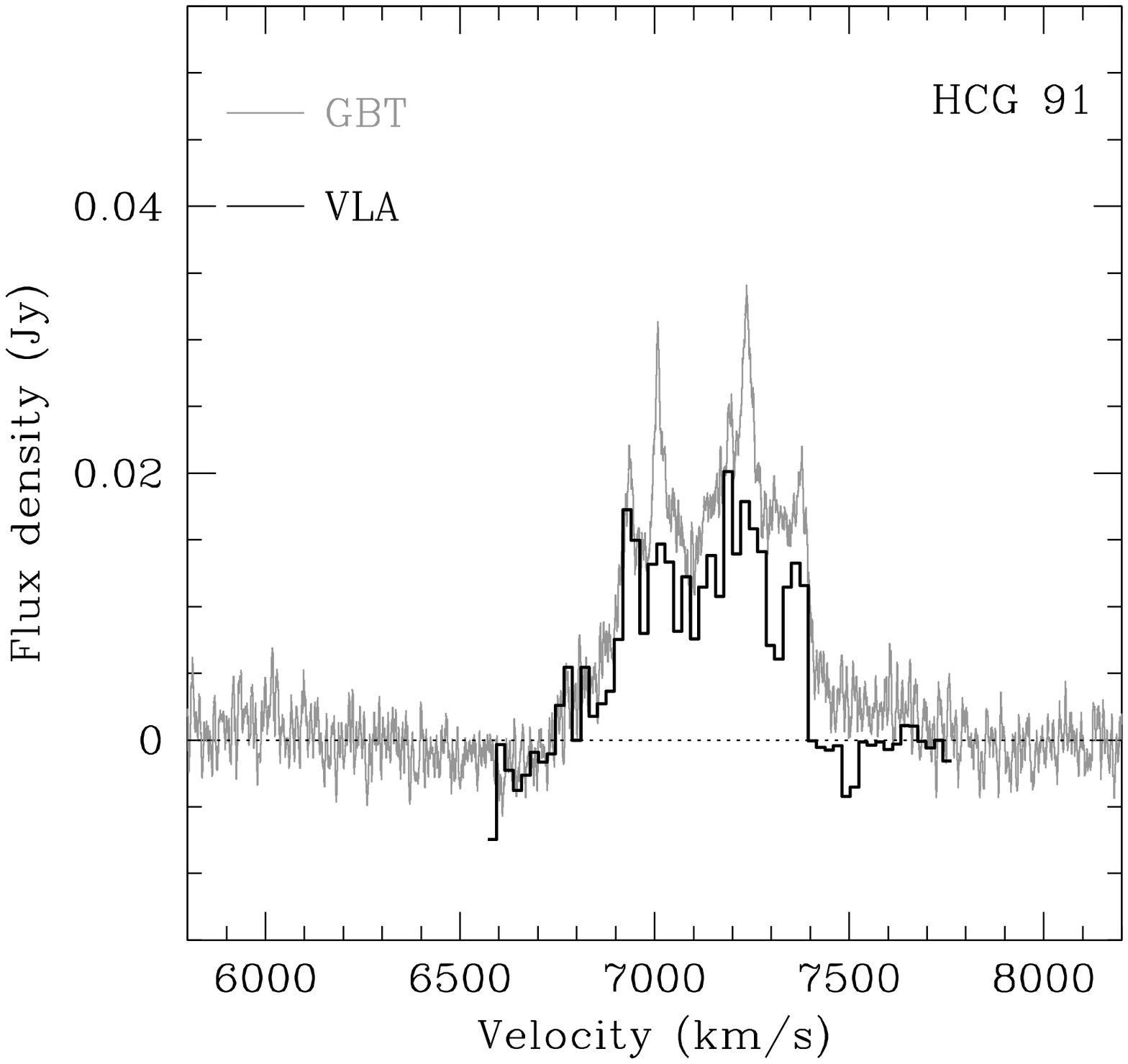} \\

\end{tabular}
{\caption{continued. }}
\end{figure}

\clearpage

 \begin{figure}
 \figurenum{4}

\begin{tabular}{c c}

\includegraphics[scale=0.35,angle=-0]{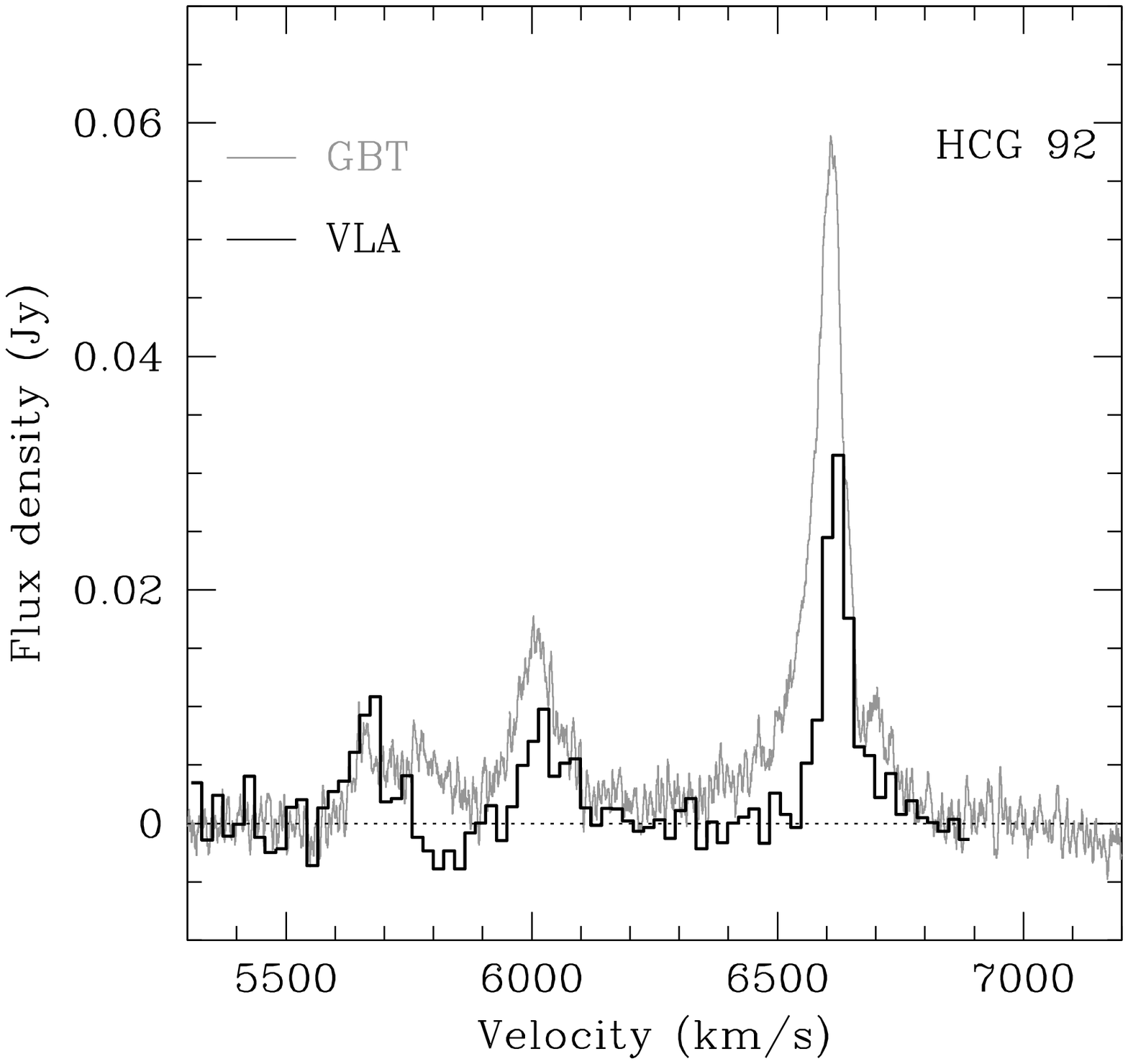} & \includegraphics[scale=0.35,angle=-0]{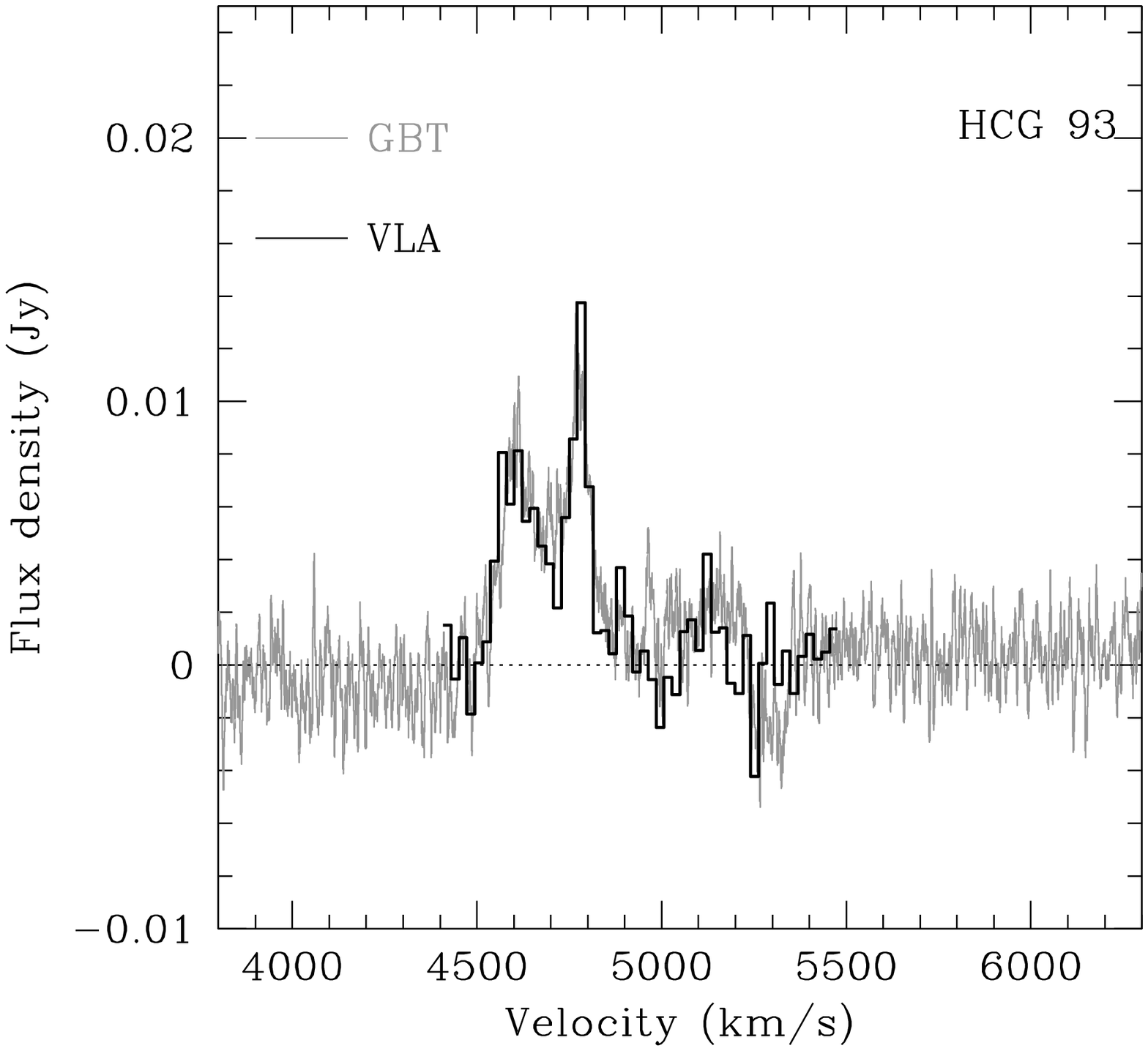} \\
\includegraphics[scale=0.35,angle=-0]{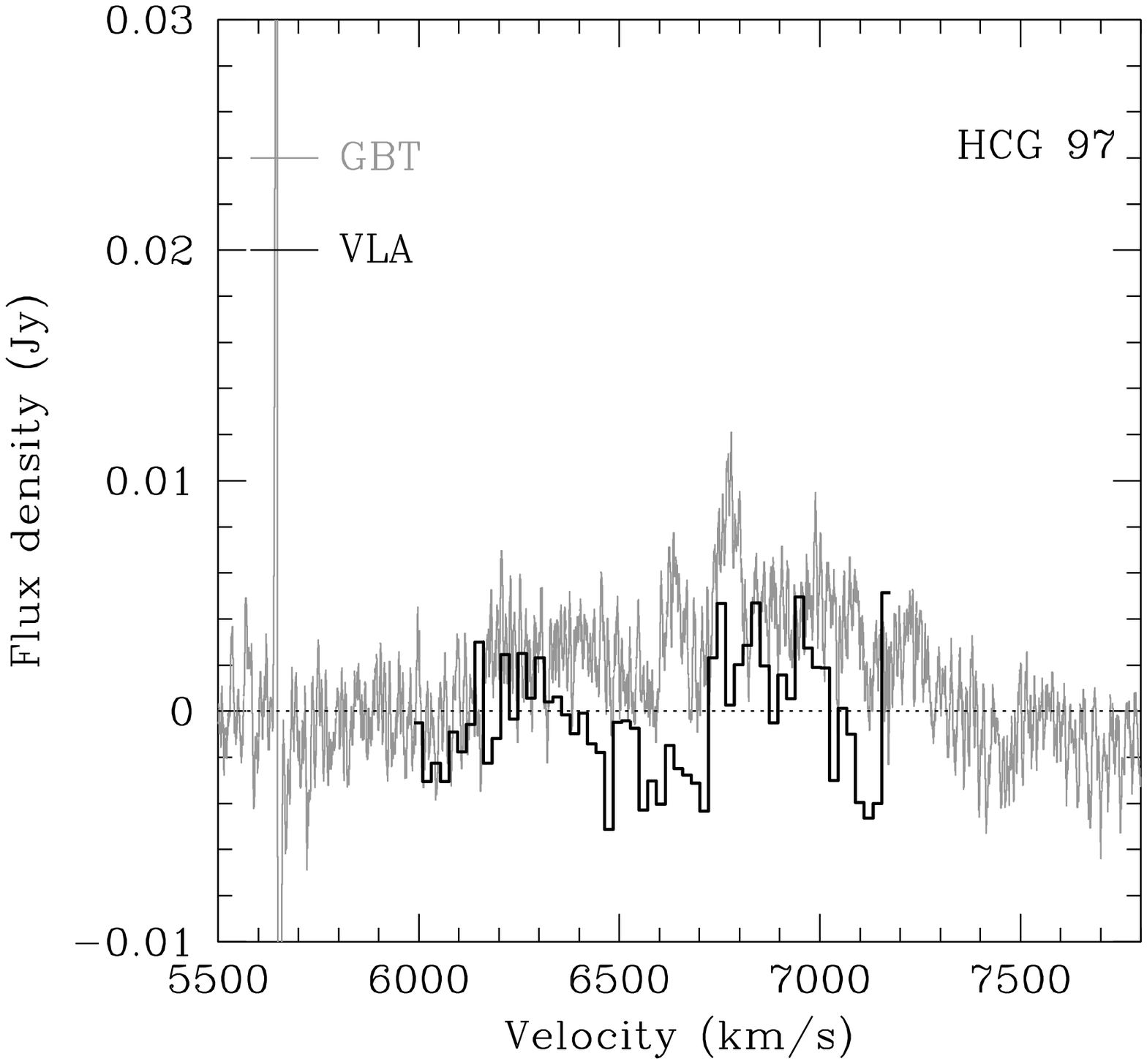} & \includegraphics[scale=0.35,angle=-0]{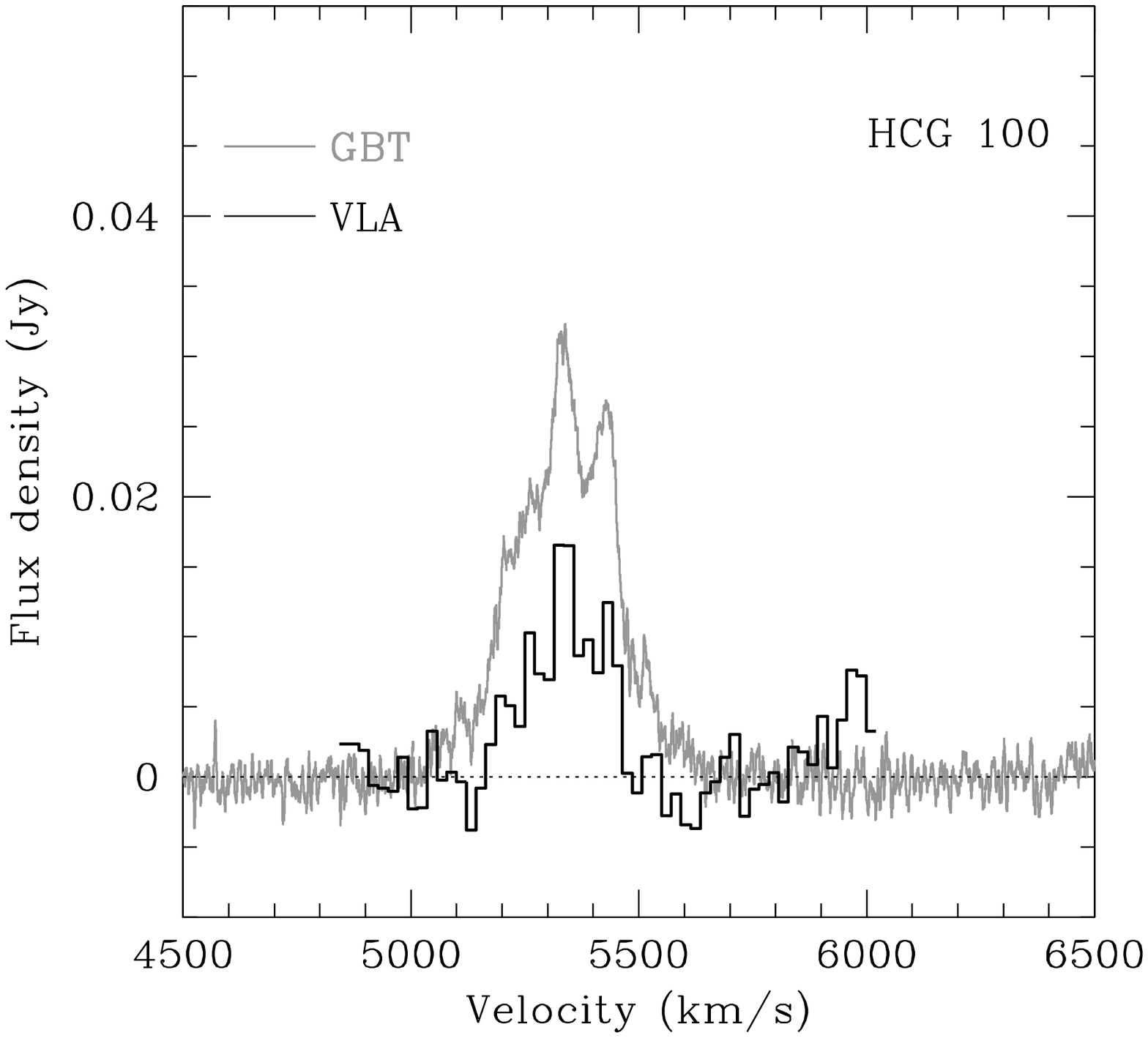} \\
\includegraphics[scale=0.35,angle=-0]{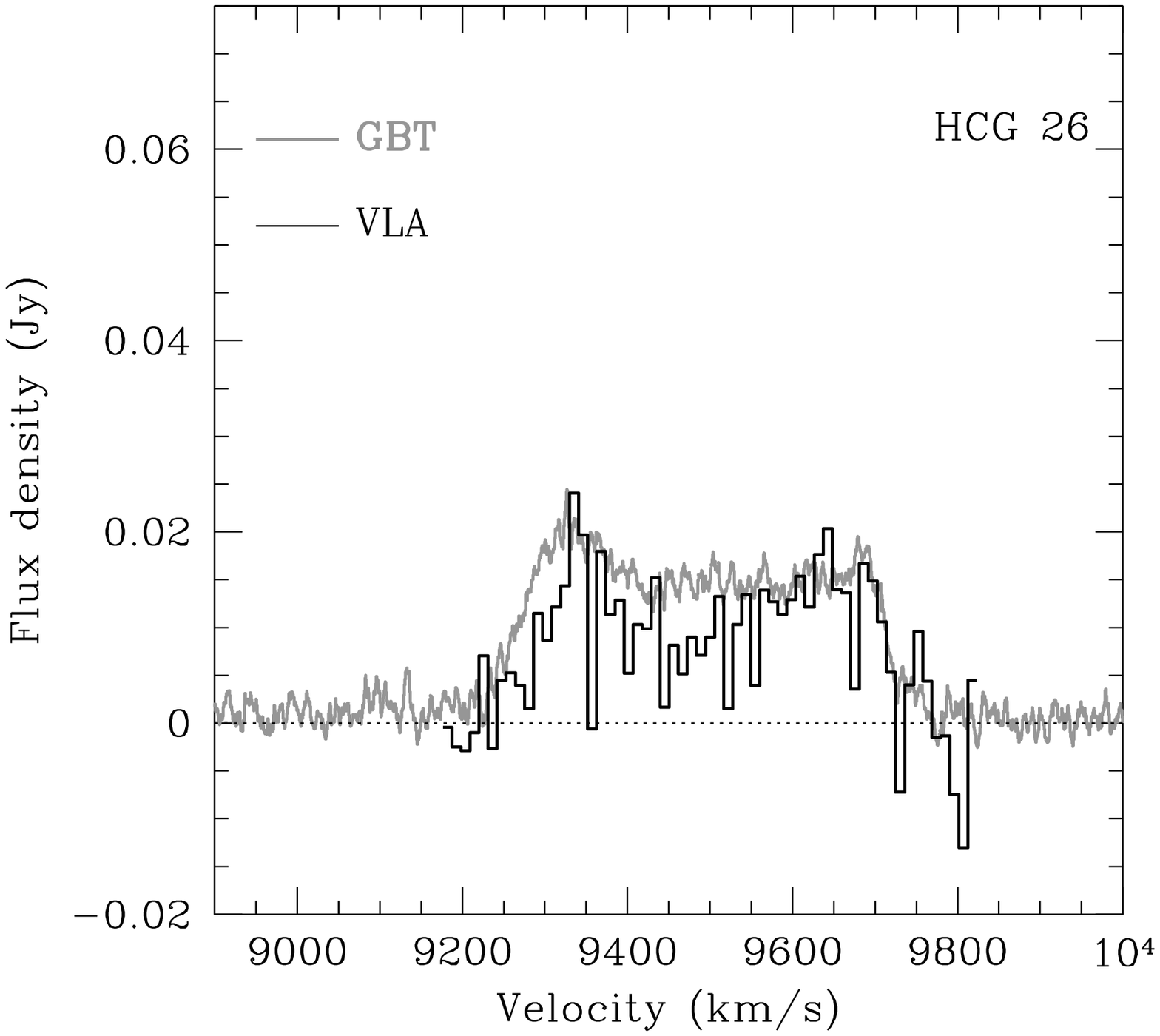} & \includegraphics[scale=0.35,angle=-0]{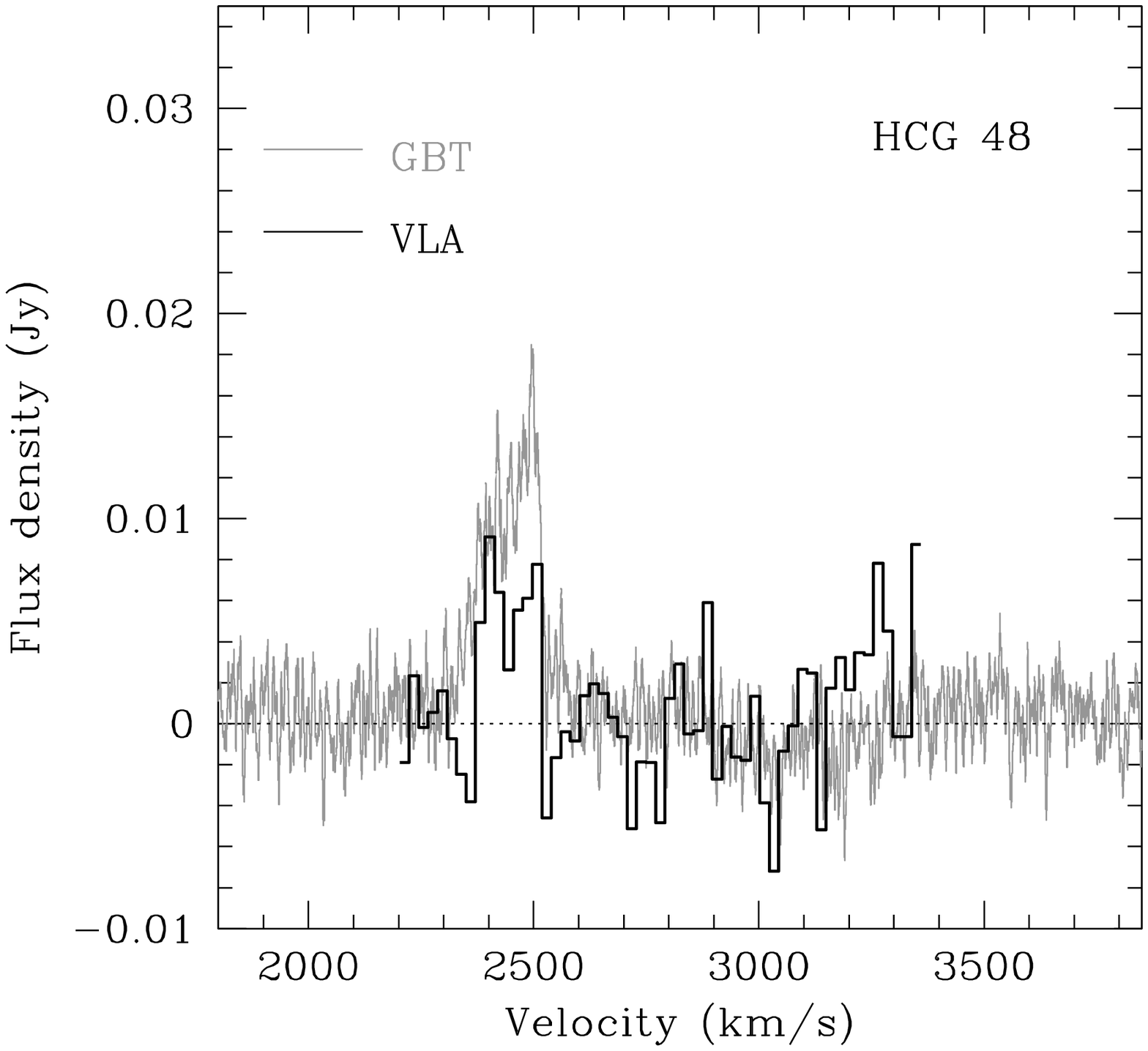} \\
\end{tabular}
{\caption{continued. }}
\end{figure}

\clearpage

\clearpage

\clearpage

 \clearpage

 \clearpage
 \clearpage

\clearpage

\clearpage

\clearpage

\clearpage

\begin{deluxetable}{ccccccccccl}
\tabletypesize{\scriptsize}
\rotate
\tablecaption{Information on the GBT Sample.\label{tbl-Sample}}
\tablewidth{0pt}
\tablehead{
\colhead{Source} & \colhead{$\theta$\tablenotemark{a}} & \colhead{Velocity\tablenotemark{b}} & \colhead{Lum. Dist.\tablenotemark{c}} & \colhead{Members} &\colhead{Velocities\tablenotemark{b}} & \colhead{Morphology\tablenotemark{d}} &   \\
\colhead{} & \colhead{arcmin} & \colhead{\kms\ } & \colhead{Mpc} & \colhead{} & \colhead{\kms\ } &\\
}

\startdata

HCG~~7	&	5.7	&	4227	&	56.6	& 	A, B, C, D & 4133, 4255, 4415, 4121 & 	Sb, SB0, SBc, SBc \\
HCG~10	&	10.9	&	4827	&	64.6	&	A, B, C, D &  5189, 4822, 4660, 4662 & 	SBb, E1, Sc, Scd\\
HCG~15	&	7.7	&	6835	&	91.7	& 	A, B, C, D, E, F & 6967, 7117, 9687, 6244, 7197,6242 & 	Sa, E0, E0, E2, Sa, Sbc \\
HCG~16	&	6.4	&	3634& 	52.9&	A, B, C, D, 16-3\tablenotemark{e}, 16-6\tablenotemark{e} &  4073, 3864, 3851, 3874, 4001, 3972 &	SBab, Sab, Im, Im, Sab, Sb\\
HCG~23	&	7.1	&	4827	&	64.6	&	A, B, C, D &  4869, 4877, 5373, 4467 &	Sab, SBc, S0, Sd \\
HCG~25	&	6.4	&	6318 &	85.2 &	A, B, D, F & 6285, 6331, 6322, 6244 &   SBc, SBa, S0, S0\\

HCG~30	&	4.5	&	4617	&	61.8	&	A, B, C, D & 4697, 4625,4508, 4666 &	SBa, Sa, SBbc, S0\\
HCG~31	&	4.2\tablenotemark{f} 	&	4037	&	54.5	&	A, B, C, G, Q & 4074, 4136, 4019, 3991, 4090 & Sdm, Sm, Im, cI, Im \\
HCG~37	&	3.2	&	6685	&	89.6	&	A, B, C, D, E & 6745, 6758, 7356, 6131, 6469 & E7, Sbc, S0a, SBdm, E0\\
HCG~40	&	1.7	&	6685	&	89.6	& 	A, B, C, D, E &  6628, 6842, 6406, 6492, 6633 & E3, S0, Sbc, SBa, Sc \\
HCG~44	&	16.4 &	1379	&	18.4	&	A, B, C, D, E\tablenotemark{g}  & 1271, 1399, 1217, 1521, 2100 & Sa, E2, SBc, Sd, Sc \\
HCG~58	&	8.8	&	6158	&	83.2	&	A, B, C, D, E &  6138, 6503, 6103, 6274, 6102 & Sb, SBab, SB0a, E1, Sbc\\
HCG~67	&	3.3	&	7587	&	98.5	&	A, B, C, D &  7213, 7588, 7528, 7101 & E1, Sc, Scd, S0\\
HCG~68	&	9.2	&	2310	&	32.0 &	A, B, C, D, E & 2325, 2579, 2321, 2344, 2401 & S0, E2, SBbc, E3, S0 \\
HCG~79	&	1.3	&	4347	&	58.2	&	A, B, C, D & 4560, 4447, 4053, 4620 & E0, S0, S0, Sdm \\
HCG~88	&	5.2	&	6026	&	80.7	&	A, B, C, D & 6033, 6192, 5956, 6041 & Sb, SBb, Sc, Sc\\
HCG~90	 &	7.4	 &	2638	 &	35.3	 &	A, B, C, D & 2603, 2511, 2497, 2659 & Sa, E0, E0, Im\\
HCG~91	&	5.2	&	7141	&	95.7	&	A, B, C, D & 6934, 7189, 7190, 7315 & SBc, Sc, Sc, SB0\\
HCG~92	&	3.2	&	6446	&	86.4	&	B, C, D, E, F & 5774, 6747, 6630, 6599, 5985 & Sbc, SBc, S0, E1, Sa \\
HCG~93	&	9.0	&	5036	&	67.4	&	A, B, C, D & 5072, 4736, 4772, 5173 &  E1, SBd, SBa, SB0 \\
HCG~97	&	5.2	&	6535	&	87.6	&	A, B, C, D, E & 6932, 6666, 6003, 6328, 6665 &  E5, Sc, Sa, E1, S0a\\
HCG~100	&	3.6	&	5336	&	71.5	& 	A, B, C, D & 5366, 5256, 5461, 5590 & Sb, Sm, SBc, Scd\\
HCG~18	&	2.0	&	4107	&	54.1	& 	B, C, D &  4105, 4143, 4067  &  Im, Im , Im \\
HCG~26	&	1.9	&	9437	&	127.3&	A, B, C, D, E, F, G & 9678, 9332, 9618, 9133, 9623, 9626, 9293 & Scd, E0, S0, cI, Im, cI, S0\\
HCG~35	&	2.2	&	16249	&	219.5&	A, B, C, D, E, F & 15919, 16338, 16249, 15798, 16711, 16330 &S0, E1, E1, Sb, S0, E1 \\
HCG~48	&	5.0	&	2818	&	37.7	&	A, B, C, D &  2267, 2437, 4381, 4361 & E2, Sc, S0a, E1 \\

\enddata
\tablenotetext{a}{\citet{hick82}}
\tablenotetext{b}{NED}
\tablenotetext{c}{Evaluated using same cosmology as VM2001 - $\Omega_{m} =1$ and $\Omega_{\lambda} =0$ }
\tablenotetext{d}{\citet{hick89}}
\tablenotetext{e}{\citet{decar97} }
\tablenotetext{f}{\citet{VM2005} }
\tablenotetext{g}{\citet{will91} }
\end{deluxetable}
\clearpage

\begin{deluxetable}{cccccccccccl}
\tabletypesize{\scriptsize}
\rotate
\tablecaption{Results from our GBT observations and comparison with previously published data.\label{tbl-GBT}}
\tablewidth{0pt}
\tablehead{
\colhead{Source} & \colhead{HI mass} &
\colhead{Uncertainty, $\sigma_{M(HI)}$\tablenotemark{c}} &\multicolumn{2}{c}{Line-width at zero flux} & \colhead{ Log[M(HI)$_{pred}$\tablenotemark{d}]} & \colhead{ Log[M(HI)$_{obs}$]} &
\multicolumn{2}{c}{Deficiency}  \\
& & & \colhead{Range}  & \colhead{$\Delta$ V}& & &\colhead{GBT} & \colhead{VM2001} \\
\colhead{}  & \colhead{$\times~10^{9} M_{\odot}$ } &
\colhead{$\times~10^{8} M_{\odot}$  } &\colhead{\kms\ }&\colhead{\kms\ }& \colhead{Log[M$_{\odot}$]} & \colhead{Log[M$_{\odot}$]} & &\\

}

\startdata

HCG~~7	&	5.8	&	0.4	&	3970-4550 	&     580		&	10.37	&	9.76	&	0.61	&	 0.69\\
HCG~10	&	9.9	&	0.6	&	4750-5500	&	750		&	10.22\tablenotemark{e}	&	10.00	&	0.22	&	0.24\\
HCG~15	&	3.7	&	1.7	&	6100-7000	& 	900		&	10.03	&	9.57	&	0.46	&	 0.62\\
HCG~16\tablenotemark{f}	&	14.2    &       0.5		&	3600-4180&	580	&	10.57	&	10.16	&	0.41	&	0.15 \\
HCG~23	&	10.6	&	0.9	&	4330-5200	&	870		&	10.00	&	10.03	&	-0.03	&	 -0.03\\
HCG~25\tablenotemark{f}	&	13.6    &       1.5 	&	6010-6670 	&	660 &	10.16	&	10.14	&	0.02	&	 0.26\\

HCG~30	&	0.6	&	0.6	&	4250-4760	&	510		&	10.18	&	8.79	&	1.39	&	 1.56\\
HCG~31	&	18.6	&	0.9	&	3870-4270	&	400		&	10.53	&	10.27	&	0.26	&	 0.18\\

HCG~37	&	5.4	&	1.5	&	6200-7400	&	1200		&	10.07	&	9.75	&	0.32	&	 0.88\\
HCG~40	&	6.6	&	1.5	&	6150-7000 	&	850		&	10.11	&	9.82	&	0.29	&	0.97 \\
HCG~44	&	0.9	&	0.1 	&	1010-1750	&	740		&	9.73\tablenotemark{e}	&	8.97	&	0.76	&	 0.69\\
HCG~58	&	8.7	&	1.6	&	5880-6990	&	1110		&	10.40	&	9.94	&	0.46	&	 0.57\\
HCG~67	&	4.7	 & 	2.7	&	7310-7900	&	590		&	10.30	&	9.68	&	0.62	&	 0.17\\
HCG~68\tablenotemark{f}	&	5.6	&	0.2	&	2100-2500 & 400	&	9.87\tablenotemark{e}	&	9.75	&	0.12	&	 0.48\\
HCG~79	&	4.2	&	1.3	&	3910-4940	&	1030		&	9.71	&	9.63	&	0.08	&	 0.41\\
HCG~88	&	11.9	&	1.9	&	5890-6290	&	400		&	10.55	&	10.08	&	0.47	&	 0.27\\
HCG~90	 &	0.5	 &	0.3	&	2000-2500	&	500	 	&	9.85\tablenotemark{e} &	8.73	 &	1.12	&	 -\\
HCG~91	&	22.9	&	3.0 	&       6750-7450	&	700 		&     10.56 		 &	10.36	&	0.20	&	 0.24\\
HCG~92	&	16.9	&	1.4	&	5610-6140, 6370-6770 		& 	930		&	10.51	&	10.23	&	0.28	&	 0.49\\
HCG~93	&	2.6	&	0.8	&	4500-5200	&	700		&	9.94\tablenotemark{e}	&	9.42	&	0.52	&	 0.80\\
HCG~97	&	4.3	&	1.8	&	6150-7300	&	1150		&	9.99	&	9.62	&	0.35	&	 0.89\\
HCG~100\tablenotemark{f}	&	9.0	&	0.9	&	5000-5600 &	600	&	10.24	&	9.96	&	0.28	&	 0.50 \\

HCG~18	&	10.7	&	0.5	&	3950-4200	&	250		&	9.89	&	10.03 & -0.14 &  -0.14\\
HCG~26	&	28.5	&	3.5	&	9220-9750	&	530		&	10.23	&	10.45	&	-0.22	&	 -0.19\\
HCG~35	&    $<$28.2\tablenotemark{g}&	15.7	&	- 	&	-	&	10.01	& $<$10.45		&	-	&	 -\\
HCG~48	&	1.2	&	0.5	&	2200-2600	& 	300		&	9.34	&	9.07	&	0.27	&	 0.82\\					
\enddata

\tablenotetext{a}{\citet{hick82}}
\tablenotetext{b}{Evaluated using same cosmology as VM2001 - $\Omega_{m} =1$ and $\Omega_{\lambda} =0$ }
\tablenotetext{c}{Noise measured for smoothed data of resolution 10~\kms\. \HI\ mass estimated assuming a line width $\Delta$~V=100~\kms.}
\tablenotetext{d}{Within the GBT beam. Values have been borrowed from VM2001.}
\tablenotetext{e}{Calculated by the authors using the same procedure as VM2001 taking into account the GBT beam ($\sim$~9.1$^{\prime}$) coverage.}
\tablenotetext{f}{GBT beam is smaller than the extent of \HI\ in VLA maps and hence values presented are significantly lower than total \HI\ content.}
\tablenotetext{g}{No \HI\ was detected. Values refer to mass corresponding to 3$\sigma$ noise level assuming a line-width of 600~\kms.}
\end{deluxetable}

\clearpage

\begin{deluxetable}{ c c c  c c c c c}
\tabletypesize{\scriptsize}
\rotate
\tablecaption{A summary of our VLA data (used for comparison with GBT spectra).\label{tbl-VLA}}
\tablewidth{0pt}
\tablehead{

\colhead{Source} & \colhead{Array} & \colhead{Synth. Beam} & \colhead{Channel Width}&  \colhead{Line width}& \colhead{Reference}\\
\colhead{} & \colhead{} &\colhead{}  & \colhead{\kms}&  \colhead{}\\
}

\startdata
 HCG 7 &D&64$^{\prime \prime} \times 50^{\prime \prime}$      &21.2& 600&VLA archive \\
 HCG 10  & D&64$^{\prime \prime} \times 53^{\prime \prime}$	&21.3 &600 &\citet{bor}\\
HCG 15  &D&65$^{\prime \prime} \times 55^{\prime \prime}$  	&  21.6&- &\citet{bor}\\
HCG 16 &CD&42$^{\prime \prime} \times 34^{\prime \prime}$&21.1&500 &\citet{VMpriv}\\
HCG 23&C&26$^{\prime \prime} \times 18^{\prime \prime}$ 	&21.2&200\tablenotemark{a}&\citet{willvan95}\\
HCG 25&D& $70^{\prime \prime} \times 51^{\prime \prime}$	&21.5&550&\citet{bor}\\
HCG 30  &D&52$^{\prime \prime} \times 44^{\prime \prime}$ &  21.3 &&\citet{Yunprep} \\
HCG 31&CD&$16^{\prime \prime} \times 15^{\prime \prime}$	&10.6& 200 &\citet{VM2005}\\
HCG 37  &D&44$^{\prime \prime} \times 43^{\prime \prime}$ & 21.5& &\citet{Yunprep} \\
HCG 40  &CD&47$^{\prime \prime} \times 36^{\prime \prime}$ & 21.6 &950&\citet{Yunprep} \\
HCG 44  &D&58$^{\prime \prime} \times 52^{\prime \prime}$ 	 & 10.4 &400\tablenotemark{a}& VLA archive \\
HCG 58  &D&62$^{\prime \prime} \times 59^{\prime \prime}$ 	&  21.5 &550 &\citet{Yunprep}\\
HCG 67 &D&72$^{\prime \prime} \times 59^{\prime \prime}$ 	&21.2&600&\citet{Yunprep}\\
HCG 68 &D&59$^{\prime \prime} \times 52^{\prime \prime}$ 	&21.0&350&VLA archive \\
HCG 79 &CD&31$^{\prime \prime} \times 18^{\prime \prime}$ &10.6&300&\citet{will91}\\
HCG 88 &C&24$^{\prime \prime} \times 17^{\prime \prime}$	&21.4&400&\citet{will98}\\
HCG 90  &CD& 77$^{\prime \prime}\times 40^{\prime \prime}$ & 42.0 &-& VLA archive \\ 
HCG 91  &D&74$^{\prime \prime} \times 47^{\prime \prime}$	&21.6&650&\citet{bor}\\
HCG 92 &CD&20$^{\prime \prime} \times 19^{\prime \prime}$ &21.4&550\tablenotemark{b} &\citet{willyun02}\\
HCG 93 &D& 60$^{\prime \prime} \times 56^{\prime \prime}$ 	& 21.4&600&\citet{bor} \\
HCG 97  &D&69$^{\prime \prime} \times 55^{\prime \prime}$ 	& 21.6&300&\citet{bor} \\
HCG 100  &D&60$^{\prime \prime} \times 56^{\prime \prime}$ & 21.4&300&\cite{bor}\\
HCG 26&C & 26$^{\prime \prime} \times 18^{\prime \prime}$	&11.0& 500 &\citet{willvan95}\\
HCG 48  &CD&53$^{\prime \prime} \times 39^{\prime \prime}$  & 21.0 &150&\citet{Yunprep} \\
\enddata
\tablenotetext{a}{Only part of the single dish emission was covered by the VLA spectral range}
\tablenotetext{b}{Sum of multiple emission regions}

\end{deluxetable}

\clearpage

\begin{deluxetable}{ c c cccc c c c c}
\tabletypesize{\scriptsize}
\rotate
\tablecaption{Filling factor calculation for the excess gas seen in the GBT spectra.\label{tbl-filling_factor}}
\tablewidth{0pt}
\tablehead{

\colhead{Source} & \colhead{Excess \HI\  frac.\tablenotemark{a}}&\colhead{Excess \HI\ mass } & \colhead{Spec. Distribut.}&\colhead{VLA Noise, $\sigma$}  & \colhead{Intrinsic $\Delta V$}& \colhead{3$\sigma$ Column Density}&\colhead{ $A_{ff}$}& \colhead{}\\
\colhead{} & \colhead{$\% $} & \colhead{$\times10^8~$M$_\odot$} &\colhead{(Categories)\tablenotemark{b}}&\colhead{~mJy/Beam}& \colhead{\kms\ } & \colhead{$\times~10^{19}~$cm$^{-2}$}& \colhead{\% }\\
}

\startdata
HCG~~7	&	16	&	9.2	& I		&	0.29	&	42	&	1.09	&	63		\\
HCG~10	&	17	&	17.1	& I		&	0.34	&	42	&	1.25	&	78		\\
HCG~15	&	56	&	20.3	& III		&	0.38	&	42	&	1.40	&	41		\\
HCG~16	&	5	&	6.7	& I		&	0.21	&	42	&	1.76	&	32		\\
HCG~23	&	30	&	31.9	& I\tablenotemark{c}	&	0.34	&	42	&	8.93	&	20		\\

HCG~25	&	15	&	20.7	& I		&	0.44	&	42	&	1.54	&	44		\\
HCG~30	&	23	&     1.4	& I		&	0.45	&	42	&      2.39	&	 4		\\
HCG~31	&	70	&     130.0	& I, II		&	0.41	&	42	&    21.68  &       48              \\
HCG~37	&	81	&     45.1	& III		&	0.91	&	42	&      6.30  &       21              \\
HCG~44	&	60	&	5.5	& I, II, III	&	0.44	&	42	&	1.88	&	205\tablenotemark{d}\\

HCG~58	&	13	&	11.4	& I, III	&	0.26	&	42	&	0.87	&	45		\\
HCG~67	&	15	&	6.9	& I		&	0.53	&	42	&	1.50	&	11		\\
HCG~68	&	54	&	30.3	& I		&	0.31	&	42	&	1.24	&	566\tablenotemark{d}\\
HCG~79	&	21	&	9.0	& II,  III 	&	0.90	&	42	&    19.64	&	3		\\
HCG~88	&	21	&	24.6	& I		&	0.31	&	42	&	9.38	&	10		\\

HCG~90	&	44	&	2.3	& III		&	0.34	&	42	&	1.41	&	32		\\
HCG~91	&	38	&	86.9	& I		&	0.51	&	42	&	0.89	&	254\tablenotemark{d}\\
HCG~92	&	51	&	86.3	& I, II, III	&	0.21	&	42	&	7.09	&	39		\\
HCG~93	&	8	&	2.2	& I		&	0.33	&	42	&	1.20	&	9		\\
HCG~97	&	55	&	23.5	& I, III	&	0.52	&	42	&	1.60	&	45		\\

HCG~100	&	55	&	49.3	& I, II	&	0.30	&	42	&	1.10	&	209\tablenotemark{d}\\

HCG~26	&	28	&	79.9	& I		&	0.54	&	42	&    14.70	&	8		\\
HCG~48	&	44	&	5.1	& I		&	0.42	&	42	&	2.50	&	34		\\

\enddata
\tablenotetext{a}{Excess \HI\ fraction, M$_{excess}$/M$_{GBT}$}
\tablenotetext{b}{See \S~\ref{sec:comparison} for details. }
\tablenotetext{c}{Spectrum used for comparison is incomplete in its velocity coverage.}
\tablenotetext{d}{Significant spatial filtration of \HI\ by the VLA in groups with the \HI\ structures are of the order 10-15$^{\prime}$.}
\tablecomments{HCG~40 is not included in the above table as excess gas mass could not be determined precisely due to uncertainty in calibration of the GBT data for this group. }
\end{deluxetable}

\clearpage

\begin{table}
\begin{center}
\caption{The excess gas mass fraction for the three \HI\ deficiency classes defined in \S~\ref{sec:HIdef} .\label{hi_def}}
 \begin{tabular}{ c c c c c c}
 \tableline\tableline
 \multicolumn{2}{c}{Normal \HI\ content} & \multicolumn{2}{c}{Slightly \HI\ deficient} & \multicolumn{2}{c}{Highly  \HI\ deficient}\\
\tableline
Source & \% detection &Source & \% detection &Source & \% detection \\
 & M$_{excess}$/M$_{GBT}$& & M$_{excess}$/M$_{GBT}$&& M$_{excess}$/M$_{GBT}$\\
 \tableline
 \\
 
 HCG 25 & 15 \% 	& 	HCG 16 & 5 \%  	& 	HCG 93 & 8 \%\\
 HCG 79 & 21 \%  	&	HCG 58& 13 \% 	& 	HCG 67 & 15 \%\\
 HCG 23\tablenotemark{a} & 30 \%& HCG 10 & 17 \% &HCG~~7& 16 \%\\
 HCG 68 & 54 \% 	&	HCG 88 & 21 \% 	& 	HCG 30& 23 \%\\ 
       		&                &	HCG 91& 38 \%  	&	HCG 90 & 44 \%\\
       		& 	       	 &	HCG 92 & 51 \%	&       HCG 44 & 60 \%\\      
		&                & 	HCG 97 & 55 \% &&\\	
       		&                &	HCG 100&55 \% &&\\	 
		& 		&	HCG 15& 56 \% 	&& \\
		& 		&	HCG 31& 70 \% 	&& \\
		&		&	HCG 37 &  81 \% &&\\

 \tableline
   Average&30 \%&Average&42 \%&Average & 28 \%\\
   Median &26 \%&Median &51 \%&Median & 20 \%\\
  
 \end{tabular}
\tablenotetext{a}{Spectrum used for comparison is incomplete in its velocity coverage.}
\tablecomments{HCG~40 is not included in the analysis (see \S~\ref{sec:trends} for details).}
\end{center}
\end{table}

\clearpage

\begin{table}
\begin{center}
\caption{The excess gas mass fraction for the three phases of group evolution \tablenotemark{a}.\label{evolution}}
 \begin{tabular}{ c c c c c c}
 \tableline\tableline
 \multicolumn{2}{c}{ Phase 1}&  \multicolumn{2}{c}{Phase 2} & \multicolumn{2}{c}{Phase 3} \\
\tableline
Source & \% detection &Source & \% detection &Source & \% detection \\
 & M$_{excess}$/M$_{GBT}$& & M$_{excess}$/M$_{GBT}$&& M$_{excess}$/M$_{GBT}$\\
  \tableline
 &&&&&\\
 HCG 67 & 15 \%	&      	HCG 16 & 5 \%    	&   	HCG 93 & 8 \%\\
 HCG~~7& 16 \%      & 	HCG  58& 13 \%  	&	HCG 30 & 23 \%\\
HCG 88 & 21 \%	&  	HCG 25 & 15 \%  	& 	 HCG 90 & 44 \%\\
HCG 79 & 21 \%  	& 	HCG 10 & 17 \% 	& 	HCG 92 & 51 \%\\
 HCG 23\tablenotemark{b} & 30 \% & HCG 91& 38 \%   & HCG 68 & 54 \%\\ 
 		 &		&	HCG 100& 55 \% 	& 	HCG 97 & 55 \%\\
 		&		&	HCG 31 & 70 \%	&	HCG 15& 56 \%\\ 
                  &		&		& 		         &	 HCG 44 & 60 \%\\	
  		&		&		& 			&	 HCG 37 & 81 \%\\					

   \tableline
   Average&21 \%&Average&30 \%&Average & 48 \%\\
   Median &21 \%&Median &17 \%&Median & 54 \%\\
\end{tabular}

\tablenotetext{a}{VM2001} 
\tablenotetext{b}{Spectrum used for comparison is incomplete in its velocity coverage.}
\tablecomments{HCG~ 40 is not included in the analysis (see \S~\ref{sec:trends} for details).}
\end{center}
 \end{table}

\clearpage

\begin{table}
\begin{center}
\caption{The excess gas mass fraction for the spiral dominated and non-spiral dominated HCGss \tablenotemark{a}.  \label{hubbletype}}
 \begin{tabular}{ c c c c }
 \tableline\tableline
\multicolumn{2}{c}{Spiral}&\multicolumn{2}{c}{Non-spiral}\\
\tableline
Source & \% detection &Source & \% detection \\
 & M$_{excess}$/M$_{GBT}$& & M$_{excess}$/M$_{GBT}$\\
 \tableline
 &&&\\
HCG 16 &   5 \%	&	HCG 93 &     8 \%\\
HCG 58 & 13 \%	&	HCG 67 &   15 \% \\
HCG 25 & 15 \%	&	HCG 79 &   21 \%\\ 
HCG   7 & 16 \% 	& 	HCG 90 &   44 \%	\\
HCG 10 & 17 \%	& 	HCG 68 &   54 \%\\
HCG 88 & 21 \%	&	HCG 97 &   55 \%\\
HCG 30 & 23 \%	&	HCG 15 &   56 \%\\
HCG 23\tablenotemark{b}&30 \%	&	HCG 37 &  81 \%\\
HCG 91 & 38 \%&&\\
HCG  92&51 \%&&\\
HCG 100& 55 \%&&\\	
HCG 44& 60 \%		&&\\
HCG 31& 70 \%		&&\\

   \tableline
   Average&32 \%&Average & 42 \%\\
   Median & 23 \% & Median & 49 \%\\

\end{tabular}

\tablenotetext{a}{\cite{hick82}}
\tablenotetext{b}{Spectrum used for comparison is incomplete in its velocity coverage.}
\tablecomments{HCG~40 is not included in the analysis (see \S~\ref{sec:trends} for details).}
\end{center}
\end{table}

 \end{document}